\documentclass[11pt]{article}
\usepackage{amsmath, amssymb, amsfonts, amsthm}

\newtheorem{theorem}{Theorem}[section]
\newtheorem{proposition}[theorem]{Proposition}

\newtheorem{lemma}[theorem]{Lemma}

\newcommand{\tri}{| \! | \! |}
\newcommand{\rd}{{\rm d}}
\newcommand{\be}{\begin{equation}}
\newcommand{\ee}{\end{equation}}
\newcommand{\bey}{\begin{eqnarray}}
\newcommand{\eey}{\end{eqnarray}}

\newcommand{\eps}{\varepsilon}

\newcommand{\bp}{{\bf p}}

\newcommand{\bx}{{\bf x}}
\newcommand{\by}{{\bf y}}

\newcommand{\ph}{\varphi}

\newcommand{\la}{\langle}
\newcommand{\ra}{\rangle}

\newcommand{\e}{\varepsilon}

\newcommand{\cU}{{\cal U}}

\newcommand{\bR}{{\mathbb R}}

\newcommand{\bN}{{\mathbb N}}

\newcommand{\tr}{\mbox{Tr}}

\newcommand{\wt}{\widetilde}
\newcommand{\wh}{\widehat}

\newcommand{\const}{\mathrm{const}}

\newcommand{\cE}{{\cal E}}

\newcommand{\cK}{{\cal K}}

\newcommand{\cL}{{\cal L}}

\newcommand{\cJ}{{\cal J}}

\input epsf

\newcommand{\fh}{{\frak h}}

\newcommand{\donothing}[1]{}

\begin{document}

\title{Rigorous Derivation of the Gross-Pitaevskii Equation \\ with a Large Interaction Potential}
\author{L\'aszl\'o Erd\H os${}^1$\thanks{Partially supported by SFB/TR12 Project from DFG} \; ,
Benjamin Schlein${}^1$\thanks{Supported by a Kovalevskaja Award from the Humboldt Foundation.
On leave from Cambridge University}, \, and Horng-Tzer
Yau${}^2$\thanks{Partially supported by NSF grant
DMS-0602038} \\
\\
Institute of Mathematics, University of Munich, \\
Theresienstr. 39, D-80333 Munich, Germany${}^1$ \\
Department of Mathematics, Harvard University\\ Cambridge, MA 02138, USA${}^2$\\
}

\maketitle

\begin{abstract}
Consider a system of $N$ bosons in three dimensions interacting via
a repulsive short range pair potential $N^2V(N(x_i-x_j))$, where
$\bx=(x_1, \ldots, x_N)$ denotes the positions of the particles. Let
$H_N$ denote the Hamiltonian of the system and let $\psi_{N,t}$ be
the solution to the Schr\"odinger equation. Suppose that the
initial data $\psi_{N,0}$ satisfies the energy condition
\[ \langle \psi_{N,0},  H_N \psi_{N,0} \rangle \leq C N \, . \]
and that the one-particle density matrix converges to a projection as $N \to \infty$.
Then, we prove that the $k$-particle density matrices of $\psi_{N,t}$
factorize in the limit $N \to \infty$. Moreover, the one particle orbital
wave function solves the time-dependent Gross-Pitaevskii equation, a cubic non-linear Schr\"odinger
equation with the coupling constant proportional to the scattering length of the
potential $V$. In \cite{ESY}, we proved the same statement under the condition that the interaction
potential $V$ is sufficiently small; in the present work we develop a new approach
that requires no restriction on the size of the potential.
\end{abstract}

\section{Introduction}
\setcounter{equation}{0}

We consider a bosonic system of $N$ particles with a repulsive
interaction. The states of the system are given by elements
of  the Hilbert space $L^2_s (\bR^{3N})$, the subspace of $L^2 (\bR^{3N})$ consisting of permutation
symmetric wave functions. We are interested in describing the time evolution of special initial wave
 functions $\psi_N \in L^2_s (\bR^{3N})$ that exhibit complete Bose-Einstein condensation.
 
\medskip

For a given wave function $\psi_N$, we define the density matrix
$\gamma_N = |\psi_N \rangle \langle \psi_N|$ associated with $\psi_N$
 as the orthogonal projection onto $\psi_N$. Moreover, for
 $k =1, \dots, N$, we define the
$k$-particle marginal density $\gamma^{(k)}_N$,
 associated with $\psi_N$, by taking the partial trace of $\gamma_N$
over the last $(N-k)$ variables. In other words, $\gamma^{(k)}_N$
 is defined as a positive trace-class operator on $L^2 (\bR^{3k})$ with kernel given by
\[ \gamma^{(k)}_N (\bx_k; \bx'_k) = \int \rd \bx_{N-k} \, \psi_N (\bx_k, \bx_{N-k})
\overline{\psi}_N (\bx'_k, \bx_{N-k})\,. \] Here and in the following we use the notation
$\bx_k = (x_1, \dots ,x_k)$, $\bx'_k = (x'_1, \dots, x'_k)$, and $\bx_{N-k} = (x_{k+1}, \dots, x_N)$;
we will also denote $\bx = (x_1, x_2, \dots ,x_N)$. A sequence $\{ \psi_N \}_{N \in \bN}$, with
$\psi_N \in L^2_s (\bR^{3N})$ for all $N$, is said to exhibit complete Bose-Einstein
condensation in $\ph \in L^2 (\bR^3)$ if
\begin{equation}\label{eq:1tok}
\gamma^{(1)}_N \to |\ph \rangle \langle \ph|
 \qquad \text{as } N \to \infty \end{equation}
 in the trace-norm topology (here and in the following $|\ph \rangle \langle \ph|$ indicates the orthogonal projection onto $\ph$). Physically, complete Bose-Einstein
condensation means that all particles in the system, apart from a fraction vanishing
 as $N \to \infty$,
are described by the same one-particle wave function $\ph$, known as the condensate
 wave function.
Note that (\ref{eq:1tok}) implies that \begin{equation}
\label{eq:1tok2}
\gamma^{(k)}_N \to |\ph \rangle \langle
 \ph |^{\otimes k} \end{equation}
for all $k \geq 1$, as was first proven by Lieb and Seiringer in \cite{LS}.

The time-evolution of $N$ boson systems is governed by the Schr\"odinger equation
\begin{equation}\label{eq:schr}
i \partial_t \psi_{N,t} = H_N \psi_{N,t} \,
\end{equation}
with the  Hamiltonian operator
\begin{equation}\label{eq:ham1}
 H_N = \sum_{j=1}^N -\Delta_{j} + \sum_{i < j} V_N (x_i -x_j). \end{equation}
Here and in the following we are going to use the convention $\Delta_j = \Delta_{x_j}$ and
$\nabla_{j} = \nabla_{x_j}$. We consider the scaling introduced by Lieb, Seiringer and Yngvason
in \cite{LSY} for the interaction potential $V_N$, i.e. we fix a non-negative potential $V$,
and then we rescale it by defining $V_N (x) = N^2 V (Nx)$. This scaling is chosen
so that the scattering length of $V_N$ is of the order $1/N$.

We recall that the scattering length associated with a potential $V$
decaying sufficiently fast at infinity ($V$ has to be integrable at infinity)
is defined through the solution of the zero-energy scattering equation
 \be
 \left(-\Delta + \frac{1}{2} V \right) f = 0
\label{eq:scatt}
\ee
with boundary condition $f(x) \to 1$ as $|x| \to \infty$. We usually write $f=1-w$.
 The scattering length of $V$, which is a measure of the effective range
 of the interaction, is defined by
\[ a_0 = \lim_{|x| \to \infty} |x| w(x). \]
 An equivalent definition of the scattering length is given by the
formula
\begin{equation}\label{eq:a0} \int \rd x \, V(x) (1-w(x)) = 8 \pi a_0\, .
\end{equation}
By these definitions, it is clear that, if $a_0$ denotes the
scattering length of the potential $V$, then the scattering length
of the scaled potential $V_N$ is given by $a= a_0/N$.

Our main result is as follows; suppose that the family of wave functions
$\{ \psi_N \}_{N \in \bN}$ exhibits complete Bose-Einstein condensation
\eqref{eq:1tok} with some $\ph \in H^1 (\bR^3)$ and assume its energy
per particle to be bounded (in the sense that
$\langle \psi_N, H_N \psi_N \rangle \leq C N$ for all $N$).
Denote by $\psi_{N,t}$ the solution of the Schr\"odinger equation
(\ref{eq:schr}) with initial data $\psi_{N,0}= \psi_N$. Under appropriate conditions on the
 potential $V$, we show that, for every time $t\in \bR$, the family $\{ \psi_{N,t} \}_{N \in \bN}$
still exhibits complete condensation, and that the condensate wave function evolves according
to a cubic nonlinear Schr\"odinger equation known as the Gross-Pitaevskii equation.
In other words, if $\gamma^{(1)}_{N,t}$ denotes the one-particle marginal density associated with
$\psi_{N,t}$, we prove that \begin{equation}\label{eq:gammaNT1} \gamma_{N,t}^{(1)} \to
 |\ph_t \rangle \langle \ph_t| \end{equation} as $N \to \infty$, where $\ph_t$ is determined
by the nonlinear Gross-Pitaevskii equation
\begin{equation}
\label{eq:GP1}
i \partial_t \ph_t
= -\Delta \ph_t + 8 \pi a_0 |\ph_t|^2 \ph_t \,
\end{equation}
with initial data $\ph_0 = \ph$.
The cubic non-linear term expresses the local on-site self-interaction of the
condensate wave function. Due to the strongly localized interaction, the
many-body wave function develops a singular correlation structure on the scale $1/N$. As
$N$ goes to infinity, this short scale structure produces  the scattering length as a coupling
constant in the limiting Gross-Pitaevskii equation.

\medskip

This result gives a mathematical description of the dynamics of initial data exhibiting
 complete Bose-Einstein condensation. The simplest example of such initial data are product
 states $\psi_N = \ph^{\otimes N}$, for arbitrary $\ph \in L^2 (\bR^{3})$. Physically more
interesting examples of complete Bose-Einstein condensates are the ground states of trapped
 Bose gases. A trapped Bose gas in the Gross-Pitaevskii scaling is an $N$-boson system
described by the Hamiltonian
\begin{equation}\label{eq:hamtrap} H^{\text{trap}}_N = \sum_{j=1}^N
\left(-\Delta_{x_j} + V_{\text{ext}} (x_j) \right) + \sum_{i<j}^N V_N (x_i -x_j)\, ,
 \end{equation}
where $V_{\text{ext}} (x) \geq 0$ with $V_{\text{ext}} (x) \to \infty$ as
 $|x| \to \infty$ is a confining potential.
 The Hamiltonian (\ref{eq:hamtrap}) describes therefore a system of $N$ particles confined
by the external potential $V_{\text{ext}}$ into a volume of order one, interacting through a potential with effective range of the order $1/N$. Since the typical distance
between neighboring particles is of order $N^{-1/3}$, and thus much bigger than the effective range of the interaction, (\ref{eq:hamtrap}) describes a very dilute system.

\medskip

In \cite{LSY}, Lieb, Seiringer and Yngvason studied the ground
state energy $E_N$ of the Hamiltonian (\ref{eq:hamtrap})
with a spherically symmetric interaction, $V(x)=V(|x|)$,
 and they proved that
\[ \lim_{N \to \infty} \frac{E_N}{N} =
\min_{\ph \in L^2 (\bR^3)} \cE_{\text{GP}} (\ph) ,
\]
 where $\cE_{\text{GP}}$ denotes the so-called Gross-Pitaevskii energy
 functional
\[ \cE_{\text{GP}} (\ph) = \int \rd x \left( |\nabla \ph|^2 +
V_{\text{ext}} (x) |\ph (x)|^2 + 4\pi a_0 |\ph (x)|^4 \right)\,. \]
 In \cite{LS}, Lieb and Seiringer proved then that the ground state of (\ref{eq:hamtrap})
 exhibits complete Bose-Einstein condensation in the minimizer
$\phi_{GP}$ of the Gross-Pitaevskii energy functional $\cE_{\text{GP}}$. In other words,
they proved that, if $\psi^{\text{trap}}_N$ denotes the ground state vector of (\ref{eq:hamtrap}),
then the corresponding one-particle marginal density $\gamma^{(1)}_{N,\text{trap}}$ satisfies
\[ \gamma^{(1)}_{N,\text{trap}} \to |\phi_{\text{GP}} \rangle \langle \phi_{\text{GP}} | \]
as $N \to \infty$, in the trace-norm topology. A survey of results concerning the ground
state properties of the dilute bosonic gases can be found in \cite{LSSY}.

\medskip

Since $\psi^{\text{trap}}_N$ describes a complete Bose-Einstein condensate, we can apply
(\ref{eq:gammaNT1}) to study its time evolution with respect to the Hamiltonian (\ref{eq:ham1});
it follows that, for every fixed $t \in \bR$, $e^{-iH_N t} \psi^{\text{trap}}_N$ exhibits
complete Bose-Einstein condensation in the one-particle state described by the solution $\ph_t$
of the Gross-Pitaevskii equation (\ref{eq:GP1}) with initial data $\phi_{\text{GP}}$.
This result gives the mathematical description of recent experiments (initiated in \cite{Kett,CW})
 where the dynamics of Bose-Einstein condensates has been observed.


\medskip

We already proved (\ref{eq:gammaNT1}) in \cite{ESY} (some partial results were previously
obtained in \cite{ESY0}), under the assumption of a sufficiently weak potential $V$. More
 precisely, in \cite{ESY} we required the dimensionless parameter
\begin{equation}\label{eq:alpha} \rho = \sup_{x \in \bR^3} |x|^2 V (x) +
\int \frac{\rd x}{|x|} V(x) \end{equation}
to be sufficiently small. In the present paper, we remove this condition and consider
arbitrary repulsive potentials $V \geq 0$, with the fast decay property $V (x) \leq
C \langle x \rangle^{-\sigma}$, for some $\sigma >5$ (here $\langle x \rangle = (1 + x^2)^{1/2}$).

\medskip

 The removal of the smallness condition requires completely new ideas.
The main challenge in the derivation of the Gross-Pitaevskii equation is
to identify and resolve the short scale correlation structure in the $N$-body wave function.
In  \cite{ESY} we achieved this by locally factoring out the solution of the zero energy scattering
equation \eqref{eq:scatt}.
This approach, however, is not sufficiently precise for a large
interaction potential. In the present paper we propose an {\it intrinsic} characterization
of the correlation structure in terms of the two-particle scattering wave operator. More precisely,
we prove that the action of the wave operator in the relative coordinate $x_i-x_j$
regularizes $\psi_{N,t}$ in this variable. This regularity
is essential to control the convergence of the many-body interaction
to the local on-site nonlinearity in the limiting equation \eqref{eq:GP1}.
An a-priori estimate leading to this regularity will be
 obtained  from the conservation of the second moment
of the energy, i.e. from  $\langle \psi_{N,t}, H_N^2 \psi_{N,t}\rangle = \langle \psi_N , H_N^2 \psi_N \rangle$. This a-priori bound, however, only controls a specific combination of two derivatives,
$\nabla_{x_i}\cdot \nabla_{x_j}$, acting on the regularized wave function.
We thus need to establish a new Poincar\'e-type inequality involving
this combination of derivatives only. In the next section
we discuss the main ideas of our  new  approach.

\section{Resolution of the correlation structure for large potential}
\setcounter{equation}{0}

\medskip

As in \cite{ESY}, the general strategy of our proof is based on the study of solutions of the
BBGKY hierarchy of equations for the marginal densities $\gamma^{(k)}_{N,t}$ associated with
the solution of the $N$-particle Schr\"odinger equation (\ref{eq:schr}):
\begin{equation}\label{eq:BBGKY}
\begin{split}
i\partial_t \gamma_{N,t}^{(k)} = \; &\sum_{j=1}^k \left[-\Delta_j , \gamma_{N,t}^{(k)} \right]
+ \sum_{i<j}^k \left[ V_N (x_i -x_j), \gamma^{(k)}_{N,t} \right] \\ &+ \left( N - k \right)
\sum_{j=1}^k \tr_{k+1} \; \left[ V_N (x_j - x_{k+1}) , \gamma^{(k+1)}_{N,t} \right]
\end{split}
\end{equation}
for $k=1, \dots ,N$. Here $\tr_{k+1}$ denotes the partial trace over the $(k+1)$-th variable,
 and we use the convention that $\gamma^{(k)}_{N,t}=0$ for $k = N+1$. The main observation is
 the fact that limit points $\{ \gamma^{(k)}_{\infty,t} \}_{k\geq 1}$ of the families
$\{ \gamma_{N,t}^{(k)} \}_{k=1}^N$ (with respect to an appropriate topology) are solution
of the infinite hierarchy of equations
\begin{equation}\label{eq:infhier}
\begin{split}
i\partial_t \gamma_{\infty,t}^{(k)} = \; &\sum_{j=1}^k \left[ -\Delta_j , \gamma_{\infty,t}^{(k)}
 \right]  + 8\pi a_0 \sum_{j=1}^k \tr_{k+1} \;
\left[ \delta (x_j - x_{k+1}) , \gamma^{(k+1)}_{\infty,t} \right]\,.
\end{split}
\end{equation}
It is easy to check that the product ansatz $\gamma_{\infty,t}^{(k)} = |\ph_t \rangle
\langle \ph_t|^{\otimes k}$
 satisfies (\ref{eq:infhier}) if and only if $\ph_t$ solves the Gross-Pitaevskii
equation (\ref{eq:GP1}).
Therefore, to conclude the proof of (\ref{eq:gammaNT1}), it suffices to show that:
1) every limit point
 of the family $\{ \gamma_{N,t}^{(k)} \}_{k=1}^N$ is a solution of the infinite
hierarchy (\ref{eq:infhier})
 and 2) the solution to (\ref{eq:infhier}) is unique.
This strategy has already been used to derive the non-linear Hartree equations for the effective
dynamics of so called mean-field systems, see \cite{Sp,EY,BGM,ES}, to derive the cubic non-linear
 Schr\"odinger equation with different (and simpler) scalings of the interaction potential, see
\cite{EESY,ESY2}, and to derive the non-linear Schr\"odinger
equation in a one-dimensional setting see \cite{ABGT,AGT}.
We remark that the first derivation of the Hartree equation was obtained
with a different method in \cite{Hepp, GV1}. With this method the  speed
of convergence was estimated recently in \cite{RS}.

\medskip

In all works based on the BBGKY hierarchy, the key step
consists in finding an appropriate norm and space of density matrices to work with. On the one hand,
the topology has to be sufficiently strong to guarantee the convergence of \eqref{eq:BBGKY} to
(\ref{eq:infhier}) and the space has to be sufficiently small to guarantee the uniqueness of
the solution to (\ref{eq:infhier}). On the other hand the norm defining this space cannot be
too strong since we have to prove, via an a-priori bound, that limit points of the sequence
$\gamma_{N,t}^{(k)}$ belong to this space.

\medskip

In \cite{ESY}, we use an appropriate Poincar{\'e}-type
inequality to prove the convergence of \eqref{eq:BBGKY} to (\ref{eq:infhier}).
To do that, we would need a control on a mixed Sobolev norm on $\gamma_{N,t}^{(k+1)}$ and
$\gamma_{\infty, t}^{(k+1)}$ with at least two derivatives (note that the commutator
with the delta function in \eqref{eq:infhier} is even ill-defined unless some regularity
is known on $\gamma_{\infty, t}^{(k+1)}$). However, due to the singularity of the interaction
potential, it turns out that the solution of the Schr\"odinger
equation $\psi_{N,t}$ develops a short scale correlation structure, living on a length-scale
 $O(1/N)$, which causes the
Sobolev norms with two or more derivatives to blow up as $N \to \infty$.

Instead of considering derivatives of $\psi_{N,t}$,
we prove therefore an a-priori bound of the form $\int |\nabla_1 \nabla_2 \phi_{12,N} (t)|^2 \leq C$
 on the $N$-body function
$\phi_{12,N} (t) = \psi_{N,t}/(1-w_{12})$ obtained from the original wave function after factoring out the singular short scale structure. Here $1-w_{12}=1-w_N (x_1-x_2)$, where $f_N=1-w_N$ is the zero-energy scattering solution to $(-\Delta + (1/2) V_N) f_N = 0$. Note that, by simple scaling,
$w_N (x) = w (Nx)$, where $f(x) = 1- w (x)$ is the zero-energy scattering solution to the unscaled
equation $(-\Delta + (1/2) V) f = 0$. It turns out that $\int |\nabla_1\nabla_2 \psi_{N,t}|^2$
 grows with $N$, but $\int |\nabla_1\nabla_2 \phi_{12,N} (t)|^2$ remains bounded. Although
$\psi_{N,t}$ and $\phi_{12,N} (t)$ behave very differently in the mixed Sobolev norm,
their difference in $L^2$-norm  vanishes in the $N\to\infty$ limit due to the scaling
$w_N(x) = w(Nx)$. This allows us to obtain control on the mixed Sobolev norm of
$\gamma_{\infty, t}$ despite the fact that it is defined as a limit of
$\gamma_{N,t}$ only in the weak topology of trace class operators. Moreover,
the boundedness of $\int |\nabla_1\nabla_2 \phi_{12,N} (t)|^2$
explains the emergence of the scattering length in (\ref{eq:infhier}).

\medskip

The proof of $\int |\nabla_1 \nabla_2 \phi_{12,N} (t)|^2 \leq C$ relies on the conservation of
 $H_N^2$ along the time evolution and on the key inequality
\begin{equation}\label{eq:en2}
\langle \psi_N, (H_N+N)^2 \psi_N \rangle \geq C N^2 \int \left| \nabla_1 \nabla_2
\frac{\psi_N  }{1-w_{12}} \right|^2 \,
\end{equation}
valid for all $\psi_N \in L^2_s (\bR^{3N})$.

\medskip

To show the uniqueness of the solution of the infinite hierarchy (\ref{eq:infhier}),
on the other hand, more information on the limiting densities $\gamma^{(k)}_{\infty,t}$
 were needed; more precisely, uniqueness was proven in \cite{ESY2} under the assumption
that \begin{equation}\label{eq:aprik0} \tr \; (1-\Delta_1) \dots (1-\Delta_k)
\gamma^{(k)}_{\infty,t} \leq C^k \end{equation} for all $k \geq 1$. Because of the
singular short scale structure characterizing the solution of the Schr\"odinger equation
 for finite $N$, the densities $\gamma_{N,t}^{(k)}$ do not satisfy (\ref{eq:aprik0}).
To circumvent this problem, we derived in \cite{ESY} a higher order energy estimate of the form
\begin{equation}\label{eq:enk}
\langle \psi_N, (H_N+N)^k \psi_N \rangle \geq C^k N^k \int \rd \bx \, \Theta_{k-1} (\bx) \;
 \left| \nabla_1 \dots \nabla_k \psi_N (\bx) \right|^2 \,
\end{equation}
for all $k \geq 1$, where $\Theta_{k-1} (\bx)$ is a cutoff function vanishing (up to
exponentially small contributions) in regions where a second particle comes close
 to $x_j$, for some $j \leq k-1$ (see Section \ref{sec:hoee} for the precise definition
 of the cutoff). This higher order energy estimate provides a control on the $L^2$ norm
 of the mixed derivatives $\nabla_1 \dots \nabla_k \psi_{N,t}$ restricted on regions
with no other particle close to $x_1, \dots ,x_{k-1}$. Choosing $\Theta_{k-1}$ to vanish in
 a sufficiently small volume, it was possible to remove the cutoffs in the weak limit $N \to \infty$
and to obtain the a-priori bounds (\ref{eq:aprik0}) on the limit points
$\{ \gamma^{(k)}_{\infty,t} \}_{k \geq 1}$.

\medskip

The estimates (\ref{eq:en2}) and (\ref{eq:enk}) were therefore the two main ingredients
used in \cite{ESY} to control the singularity of the interaction potential $V_N$ and the
singular short scale structure of $\psi_{N,t}$.
In \cite{ESY}, both these estimates heavily relied on the smallness of the parameter $\rho$
 introduced in (\ref{eq:alpha}). Therefore, although the general strategy of the current
 paper is similar to the one used in \cite{ESY}, the removal of the smallness
condition on $\rho$ requires completely new ideas, which we now explain.

\medskip

In \cite{ESY}, the energy estimate (\ref{eq:en2}) was obtained from an identity of the form
\be
\langle \psi_{N}, H^2_N\psi_{N}\rangle = N(N-1)\int |\nabla_1\nabla_2\phi_{12}|^2
 (1-w_{12})^2 + N(N-1) \int \nabla_1\bar\phi_{12} \cdot g_{12} \nabla_2\phi_{12} + \Omega \, ,
\label{eq:id}
\ee
where $\phi_{12} = \psi_N / (1-w_{12})$. Here $g_{12}$ is an explicit matrix
 involving the Hessian of $w_{12}$ and squares of
its first derivatives and $\Omega$ contains irrelevant terms.
The following bound is essentially  optimal for the size of $g_{12}$:
$$
   |g_{12}|\leq \frac{C\rho}{|x_1-x_2|^2}\,.
$$
The best strategy is then to estimate
\be
  \int \nabla_1\bar\phi_{12} \cdot g_{12} \nabla_2\phi_{12}
 \ge -  C\rho \int \frac{1}{|x_1-x_2|^2} \left( |\nabla_1\phi_{12}|^2 + |\nabla_2 \phi_{12}|^2 \right)
 \ge -  C\rho \int |\nabla_1\nabla_2\phi_{12}|^2\, ,
\label{eq:neg}
\ee
where we used Hardy's inequality in the last step. This term can be absorbed
in the first (positive) term in \eqref{eq:id} only if $\rho$ is sufficiently small.
A closer inspection of the structure of the terms in $\Omega$ reveals that,
although some of them are positive,  they
cannot compensate for the negative term \eqref{eq:neg}.
After estimating them from below at the expense of adding lower order terms to $H_N^2$,
we get the desired bound
\be\label{eq:contr}
   \int |\nabla_1\nabla_2\phi_{12}|^2 \leq C N^{-2} \langle \psi_{N}, (H_N+N)^2\psi_{N}\rangle\,.
\ee

The first main idea of the current paper is to prove the following replacement for \eqref{eq:contr}
(see Proposition \ref{prop:energ2})
\be
  \label{eq:repl}
 \int |\nabla_1\cdot \nabla_2 W_{12}^*\psi_{N}|^2 \leq C N^{-2} \langle \psi_{N}, (H_N+N)^2\psi_{N}\rangle\, ,
\ee
where $W_{12}$ denotes the wave operator of the one particle Hamiltonian $-\Delta +
\frac{1}{2}V_N$ acting on the  difference variable $x_1-x_2$. Note two main differences
 between \eqref{eq:contr} and \eqref{eq:repl}. First, $W_{12}^*\psi_N$ replaces
the function $\phi_{12,N} = \psi_N /(1-w_{12})$  that can be considered as a first order
 approximation to  $W_{12}^*\psi_N$. Second, instead of controlling the full mixed second
 derivative, $\nabla_1\nabla_2$, as in (\ref{eq:contr}), the new bound controls only
 $\nabla_1\cdot \nabla_2$. Although the control on $\nabla_1 \cdot \nabla_2$ is in general
 much weaker than the control on $\nabla_1 \nabla_2$, in the radial direction of the
 relative coordinate
$x_1-x_2$ the new bound is as strong as the former one.

\medskip

Both differences cause substantial difficulties in proving that \eqref{eq:BBGKY}
indeed converges to  \eqref{eq:infhier}. First, instead of working with a relatively explicit
function $w_{12}$ and using its fairly straighforward properties summarized
in Lemma 5.1 of \cite{ESY}, now analogous properties have to established for the
wave operator. Second, the lack of the control on the full mixed second derivatives
impedes using the Sobolev-type operator inequality
$$
    V(x_1-x_2) \leq C\| V\|_1 (1-\Delta_1)(1-\Delta_2)
$$
that was crucial  in  controlling many error terms. We have found a replacement
for this inequality (Lemma \ref{lm:VL1}):
\be \label{eq:VL1}
    V(x_1-x_2) \leq C\| V\|_1 \Big( (\nabla_1\cdot\nabla_2)^2 -\Delta_1-\Delta_2+1\Big)
\ee
that uses only the $\nabla_1\cdot \nabla_2$ combination in the highest order term.
Similarly to  Lemma 7.2 of \cite{ESY}, we are also able to improve (\ref{eq:VL1}) to a
 Poincar{\'e}-type inequality $\langle \varphi, (h_\alpha (x_1 -x_2) - \delta (x_1 -x_2))
 \psi \rangle\to 0$ as $\alpha\to 0$, where $h_\alpha (x) = \alpha^{-3} h (x/ \alpha)$
(with $\int h (x) \rd x = 1$) is an approximate Dirac delta function on scale $\alpha$
 (see Lemma \ref{lm:VL12}). This is necessary to control the convergence of
 (\ref{eq:BBGKY}) to (\ref{eq:infhier}).

\medskip

The second main novelty of this paper is a proof for the higher order
derivative estimates (\ref{eq:enk})  in the
large potential regime (Proposition \ref{prop:hk}).
Although the main conclusion is the same as Proposition~5.3 in \cite{ESY},
the proof does not require the smallness of $\rho$. The
proof in \cite{ESY} started from the trivial energy estimate
$\sum_{j=1}^N -\Delta_j \leq H_N$ and the estimate \eqref{eq:contr} on $H^2_N$
and it used an induction on the exponent to pass from $(H_N+N)^k$ to $(H_N+N)^{k+2}$.
The step two induction allowed us to start the induction estimate with
\be
    (H_N+N)^{k+2} \ge (H_N+N) \nabla_1^* \ldots \nabla_k^*\Theta \nabla_k \ldots \nabla_1 (H_N+N)\, ,
\label{eq:k+2}
\ee
and then we commuted the two $H_N+N$ factors through to the middle
and used the $(H_N+N)^2$ estimate (\ref{eq:contr}). The weaker $H_N^2$ estimate \eqref{eq:repl}
however does not allow us to regain a control on full derivatives, so
pursuing this path would, at best, yield a control on some complicated
combinations of partial derivatives.

In this paper we establish higher order derivative estimates by a step one induction,
i.e. passing from $(H_N+N)^k$ to $(H_N+N)^{k+1}$.
This eliminates using the $H_N^2$ estimate with incomplete derivatives,
but  the price is that instead of \eqref{eq:k+2} we have to work with
\be
    (H_N+N)^{k+1} \ge (H_N+N)^{1/2} \nabla_1^* \ldots \nabla_k^*\Theta \nabla_k \ldots \nabla_1 (H_N+N)^{1/2}
\label{eq:k+1}
\ee
in the induction step,
i.e. we have to commute the square root of the Hamiltonian through the derivatives.
The technical complications involved with the square root turn out to be reasonably shortly managable,
so that this new method actually provides a simpler proof than in \cite{ESY} for the higher
energy estimates even when $\rho$ is small.

\medskip

{\it Notation.} Throughout the paper we will use the notation $\bx = (x_1, \dots ,x_N) \in \bR^{3N}$, and, for $k = 1, \dots , N$, $\bx_k = (x_1, \dots ,x_k)$, $\bx'_k = (x'_1, \dots ,x'_k) \in \bR^{3k}$, $\bx_{N-k} = (x_{k+1}, \dots ,x_N) \in \bR^{3(N-k)}$. The notation $\| \cdot \|$ indicates the $L^2$-norm if the argument is a function and it denotes the operator norm (from $L^2$ to $L^2$) if the argument is an operator. For $1 \leq p \leq \infty$, $\| f \|_p$ indicates the $L^p$-norm of $f$. Moreover, we will use $\nabla_j$ and $\Delta_j$ as shorthand notations for $\nabla_{x_j}$ and,
respectively, $\Delta_{x_j}$. If $A$ is an operator acting on $L^2 (\bR^{3k})$, we will denote its kernel by $A (\bx_k ; \bx'_k)$. The letter $C$ denotes universal constants that may depend on $V$ and on the $H^1$-norm of the initial one particle wave function $\ph$, but is independent of $N$.

\section{Main Theorem}
\setcounter{equation}{0}

The following theorem is the main result of this paper.

\begin{theorem}\label{thm:main}
Suppose that $V \geq 0$, with $V(-x) = V(x)$ and $V(x) \leq C \langle x \rangle^{-\sigma}$,
for some $\sigma >5$, and for all $x \in \bR^3$.
Assume that the family $\psi_N \in L_s^2 (\bR^{3N})$, with $\| \psi_N \| =1$ for all $N$,
has finite energy per particle,
in the sense that \begin{equation}\label{eq:assH1} \langle \psi_N, H_N \psi_N \rangle
\leq C N \end{equation} and that it exhibits complete Bose-Einstein condensation in
the sense that the one-particle marginal $\gamma_N^{(1)}$ associated with $\psi_N$
satisfies \begin{equation}\label{eq:asscond} \gamma_N^{(1)} \to |\ph \rangle \langle \ph|
\qquad \text{in the trace norm topology as } \; N \to \infty \end{equation} for some
$\ph \in H^1 (\bR^3)$. Then, for every $k \geq 1$ and $t \in \bR$, we have
\[ \gamma^{(k)}_{N,t} \to |\ph_t \rangle \langle \ph_t|^{\otimes k} \] as $N \to \infty$,
in the trace norm. Here $\ph_t$ is the solution of the nonlinear Gross-Pitaevskii
equation \begin{equation}\label{eq:GPthm} i \partial_t \ph_t = -\Delta\ph_t +
 8\pi a_0 |\ph_t|^2 \ph_t \end{equation} with initial data $\ph_{t=0} = \ph$.
\end{theorem}

{\it Remark.} Note that the condition $V(x) \leq C \langle x \rangle^{-\sigma}$
for some $\sigma >5$ and for all $x \in \bR^3$ is only required to apply the result
 of Yajima in \cite{Ya}, which guarantees that the wave operator $W$ associated
with the Hamiltonian $\fh =-\Delta +(1/2) V$ maps $L^p (\bR^3)$ into itself,
for all $1 \leq p \leq \infty$ (see Proposition~\ref{prop:waveop}). If one knows,
by different means, that $\| W \|_{L^p \to L^p} < \infty$ (it suffices to know it
for $p=1$ and $p=\infty$), then it would be enough to assume that
$V \in L^1 (\bR^3, (1+|x|^2) \rd x) \cap L^2 (\bR^3, \rd x)$.

\medskip

{\it Remark.} Compared with our previous result in \cite{ESY}, we do not require
here the potential $V$ to be spherical symmetric (we only need that $V(-x) = V(x)$).

\medskip

{\it Remark.} The fact that $\ph \in H^1 (\bR^3)$ does not need to
 be assumed separately, since it already follows from the assumption (\ref{eq:assH1}).

\medskip

To prove this theorem we will use an approximation argument and the following
theorem, which proves Theorem \ref{thm:main} for a smaller class of initial $N$-particle wave functions.

\begin{theorem}\label{thm:main2}
Assume the same conditions on the potential $V$ as in Theorem \ref{thm:main}.
Suppose moreover that $|\nabla^{\alpha} V (x)| \leq C$ for all multi-indices $\alpha$ with $|\alpha| \leq 2$. Assume that the family $\psi_N \in L^2 (\bR^{3N})$, with $\| \psi_{N} \| =1$, is such that
\begin{equation}\label{eq:enerk} \langle \psi_N , H_N^k \psi_N \rangle \leq
C^k N^k \end{equation} for all $k \in \bN$, and that \begin{equation}\label{eq:init}
 \gamma_N^{(1)} \to |\ph \rangle \langle \ph| \qquad \text{in the trace norm topology as } N \to \infty \end{equation}
for some $\ph \in H^1 (\bR^3)$. Then, for every $k \geq 1$ and $t \in \bR$ \[ \gamma^{(k)}_{N,t}
\to |\ph_t \rangle \langle \ph_t|^{\otimes k} \] as $N \to \infty$, in the trace norm. Here $\ph_t$ is the solution of the nonlinear Gross-Pitaevskii equation
(\ref{eq:GPthm}) with initial data $\ph_{t=0} = \ph$.
\end{theorem}

\section{Proof of the main theorem}\label{sec:outline}
\setcounter{equation}{0}

In this section we present the proof of Theorem \ref{thm:main2} and we show how it implies Theorem \ref{thm:main}, making use of several key proposition, whose proof is deferred to subsequent sections.
\medskip

We start by defining an appropriate space of time-dependent density matrices. To use
Arzela-Ascoli compactness argument, we will need to establish the concept of uniform
continuity in this space, thus we have to metrize the weak* topology.

\medskip

Let $\cK_k \equiv \cK (L^2 (\bR^{3k}))$ denote the space of compact operators on $L^2 (\bR^{3k})$ equipped with the operator norm topology and let $\cL^1_k \equiv \cL^1 (L^2 (\bR^{3k}))$ denote the space of trace class operators on $L^2 (\bR^{3k})$ equipped with the trace norm. It is well known that $\cL^1_k$ is the dual of $\cK_k$. Since $\cK_k$ is separable, we can fix a dense countable subset of the unit ball of $\cK_k$; we denote it
by $\{J^{(k)}_i\}_{i \ge 1} \in \cK_k$, with $\| J^{(k)}_i \| \leq 1$ for all $i \ge 1$.
Using the operators $J^{(k)}_i$ we define the following metric on the space $\cL^1_k
\equiv \cL^1 (L^2 (\bR^{3k}))$: for $\gamma^{(k)},
\bar \gamma^{(k)} \in \cL^1_k$ we set
\begin{equation}\label{eq:etak}
\eta_k (\gamma^{(k)}, \bar \gamma^{(k)}) : = \sum_{i=1}^\infty
2^{-i} \left| \tr \; J^{(k)}_i \left( \gamma^{(k)} - \bar
\gamma^{(k)} \right) \right| \, .
\end{equation}
Then the topology induced by the metric $\eta_k$ and
the weak* topology are equivalent on the unit ball of $\cL^1_k$
(see \cite{Ru}, Theorem 3.16) and hence on
any ball of finite radius as well. In other words, a uniformly
bounded sequence $\gamma_N^{(k)} \in \cL^1_k$ converges to
$\gamma^{(k)} \in \cL^1_k$ with respect to the weak* topology, if
and only if $\eta_k (\gamma^{(k)}_N , \gamma^{(k)}) \to 0$ as $N \to
\infty$.

\medskip

For a fixed $T > 0$, let $C ([0,T], \cL^1_k)$ be the space of
functions of $t \in [0,T]$ with values in $\cL^1_k$ which are
continuous with respect to the metric $\eta_k$. On $C ([0,T],
\cL^1_k)$ we define the metric
\begin{equation}\label{eq:whetak}
\widehat \eta_k (\gamma^{(k)} (\cdot ) , \bar \gamma^{(k)} (\cdot ))
:= \sup_{t \in [0,T]} \eta_k (\gamma^{(k)} (t) , \bar \gamma^{(k)}
(t))\,.
\end{equation}
Finally, we denote by $\tau_{\text{prod}}$ the topology on the
space $\bigoplus_{k \geq 1} C([0,T], \cL^1_k)$ given by the product
of the topologies generated by the metrics $\wh \eta_k$ on $C([0,T],
\cL^1_k)$.

\medskip

\begin{proof}[Proof of Theorem \ref{thm:main2}] The proof is divided
in four steps.

\medskip

{\it Step 1. Compactness of $\Gamma_{N,t}=\{ \gamma^{(k)}_{N,t}
\}_{k\geq 1}$.} We fix $T>0$ and work on the interval $t\in [0, T]$.
Negative times can be handled analogously.

\medskip

In Theorem \ref{thm:compactness} we show that the sequence $\Gamma_{N,t}^{(k)} = \{
\gamma^{(k)}_{N,t} \}_{k \geq 1} \in \bigoplus_{k \geq 1} C([0,T], \cL^1_k)$ is compact
with respect to the product topology $\tau_{\text{prod}}$ defined above (we use the convention that
$\gamma^{(k)}_{N,t} =0 $ if $k >N$). It also follows from Theorem
\ref{thm:compactness} that any limit point $\Gamma_{\infty,t} =
\{\gamma_{\infty,t}^{(k)} \}_{k\geq 1} \in \bigoplus_{k \geq 1} C (
[0,T], \cL^1_k)$ is such that, for every $k \geq 1$,
$\gamma_{\infty,t}^{(k)} \geq 0$, and $\gamma_{\infty,t}^{(k)}$ is
symmetric w.r.t. permutations.

\medskip

Using higher order energy estimates from Proposition \ref{prop:hk},
we show in Theorem \ref{thm:aprik} that an arbitrary limit point
$\Gamma_{\infty,t} = \{ \gamma^{(k)}_{\infty,t} \}_{k \geq 1}$ of the sequence
$\Gamma_{N,t}^{(k)}$ (with respect to the topology $\tau_{\text{prod}}$) is such that
\begin{equation}\label{eq:apri}
\tr \; (1-\Delta_1) \dots (1-\Delta_k) \, \gamma^{(k)}_{\infty,t}
\leq C^k
\end{equation}
for every $t \in [0,T]$ and every $k \geq 1$.

\medskip

{\it Step 2. Convergence to the infinite hierarchy.} In Theorem
\ref{thm:conv} we prove that any limit point
$\Gamma_{\infty,t} = \{ \gamma_{\infty,t}^{(k)} \}_{k\geq 1} \in
\bigoplus_{k\geq 1} C([0,T], \cL^1_k)$ of $\Gamma_{N,t} = \{
\gamma_{N,t}^{(k)} \}_{k\ge1}$ with respect to the product topology
$\tau_{\text{prod}}$ is a solution of the infinite hierarchy of
integral equations ($k=1,2, \ldots$)
\begin{equation}\label{eq:BBGKYinf}
\gamma_{\infty,t}^{(k)}  = \; \cU^{(k)} (t) \gamma_{\infty,0}^{(k)}
- 8\pi i a_0 \sum_{j=1}^k \int_0^t \rd s \, \cU^{(k)} (t-s)
\tr_{k+1} \left[ \delta (x_j -x_{k+1}),
\gamma_{\infty,s}^{(k+1)}\right]\,
\end{equation}
with initial data $\gamma_{\infty,0}^{(k)} = |\ph\rangle
\langle\ph|^{\otimes k}$ (where $\ph \in H^1 (\bR^3)$ has been introduced in (\ref{eq:init})).
Here $\tr_{k+1}$ denotes the partial trace
over the $(k+1)$-th particle, and  $\cU^{(k)} (t)$ is the free
evolution, whose action on $k$-particle density matrices is given by
\begin{equation}\label{eq:Uk} \cU^{(k)} (t) \gamma^{(k)} := e^{it\sum_{j=1}^k \Delta_j}
\gamma^{(k)} e^{-it\sum_{j=1}^k \Delta_j}\,. \end{equation}

\medskip

We remark next that the family of factorized densities,
\begin{equation}
\label{eq:factsol} \gamma^{(k)}_t = |\ph_t \rangle \langle
\ph_t|^{\otimes k},
\end{equation}
is a solution of the infinite hierarchy (\ref{eq:BBGKYinf}) if
$\ph_t$ is the solution of the nonlinear Gross-Pitaevskii equation
(\ref{eq:GPthm}) with initial data $\ph_{t=0}= \ph$. The nonlinear
Schr\"odinger equation (\ref{eq:GPthm}) is well-posed in $H^1 (\bR^3)$
and it conserves the energy, $\cE (\ph) := \frac{1}{2} \int |\nabla
\ph|^2 + 4 \pi a_0 \int |\ph|^4$.
{F}rom $\ph \in H^1 (\bR^3)$, we thus
obtain that $\ph_t \in H^1 (\bR^3)$ for every $t \in \bR$, with a
uniformly bounded $H^1$-norm. Therefore
\begin{equation}\label{eq:phik}
\tr \; (1-\Delta_1) \dots (1-\Delta_k) |\ph_t \rangle \langle
\ph_t|^{\otimes k} \leq \| \ph_t \|_{H^1}^k \leq C^k
\end{equation}
for all $t \in \bR$, and a constant $C$ only depending on the
$H^1$-norm of $\ph$. For the well-posedness of the subcritical
nonlinear Schr\"odinger equation (\ref{eq:GPthm}) in $H^1$, see e.g. \cite{K}.
We remark that the well-posedness has been established even
for the critical (quintic) nonlinear  Schr\"odinger equation
in \cite{GV3,GV4,Str}  for small data and in \cite{Bour,CKSTT} for large data.

\medskip

{\it Step 3. Uniqueness of the solution to the infinite hierarchy.}
In Section 9 of \cite{ESY2} we proved the following theorem, which
states the uniqueness of solution to the infinite hierarchy
(\ref{eq:BBGKYinf}) in the space of densities satisfying the a
priori bound (\ref{eq:apri}). The proof of this theorem is based on
a diagrammatic expansion of the solution of (\ref{eq:BBGKYinf}). Remark that the uniqueness
of the infinite hierarchy in a different space of densities was proven in \cite{KM}.
\begin{theorem}\label{thm:uniqueness}[Theorem 9.1 of \cite{ESY2}]
Suppose $\Gamma = \{ \gamma^{(k)} \}_{k \geq 1} \in \bigoplus_{k
\geq 1} \cL^1_k$ is such that
\begin{equation}
\tr\; (1-\Delta_1) \dots (1-\Delta_k) \gamma^{(k)} \leq C^k \, .
\end{equation}
Then, for any fixed $T >0$, there exists at most one solution
$\Gamma_t = \{ \gamma^{(k)}_t \}_{k \geq 1} \in \bigoplus_{k \geq 1}
C([0,T], \cL^1_k)$ of (\ref{eq:BBGKYinf}) such that
\begin{equation}\label{eq:bougam}
\tr\; (1-\Delta_1) \dots (1-\Delta_k) \gamma^{(k)}_t \leq C^k
\end{equation}
for all $t \in [0,T]$ and for all $k \geq 1$.
\end{theorem}

\medskip

{\it Step 4. Conclusion of the proof.} {F}rom Step 2 and Step 3 it
follows that the sequence $\Gamma_{N,t} = \{ \gamma^{(k)}_{N,t}
\}_{k \geq 1} \in \bigoplus_{k \geq 1} C([0,T],\cL_k^1)$ is
convergent with respect to the product topology
$\tau_{\text{prod}}$; in fact a compact sequence with only one limit
point is always convergent. Since  the family of densities
$\Gamma_t = \{ \gamma^{(k)}_t \}_{k \geq 1}$ defined in
(\ref{eq:factsol}) satisfies (\ref{eq:phik}) and it
is a solution of (\ref{eq:BBGKYinf}), it follows that $\Gamma_{N,t} \to
\Gamma_t$ w.r.t. the topology $\tau_{\text{prod}}$. In particular
this implies that, for every fixed $k \geq 1$, and $t \in [0,T]$,
$\gamma_{N,t}^{(k)} \to |\ph_t \rangle \langle \ph_t|^{\otimes k}$ with respect to the
weak* topology of $\cL^1_k$, and thus, by a standard argument, also in the trace-norm
topology. This completes the proof of Theorem
\ref{thm:main2}.
\end{proof}

\medskip

Next we prove Theorem \ref{thm:main}; to this end we have to combine Theorem \ref{thm:main2}
 with an approximation argument for the initial $N$-particle wave function, which is needed
 to make sure that the energy condition (\ref{eq:enerk}) is satisfied. This argument was
already used in \cite{ESY}; we present it here for completeness.

\begin{proof}[Proof of Theorem \ref{thm:main}]

We assume here that, as in Theorem \ref{thm:main2}, the interaction potential $V$ is such that $|\nabla^{\alpha} V (x)| \leq C$ for all multi-indices $\alpha$ with $|\alpha| \leq 2$. We show how to remove this condition in Appendix \ref{app:nablaV}.

\medskip

Fix $\kappa >0$ and $\chi \in C_0^{\infty} (\bR)$, with $0\leq
\chi\leq 1$, $\chi (s) = 1$, for $0 \leq s \leq 1$, and $\chi (s)
=0$ if $s \geq 2$. We define the regularized initial wave function
\[
\wt \psi_N := \frac{ \chi (\kappa H_N /N) \psi_N }{ \| \chi (\kappa
H_N /N) \psi_N \|} ,
\]
and  we denote by $\wt\psi_{N,t}$ the solution of the Schr\"odinger
equation (\ref{eq:schr}) with initial data $\wt \psi_N$. Denote by
$\wt \Gamma_{N,t} = \{ \wt \gamma_{N,t}^{(k)} \}_{k=1}^\infty$ the
family of marginal densities associated with $\wt \psi_{N,t}$. By
convention, we set $\wt\gamma^{(k)}_{N,t}:=0$ if $k >N$. The tilde
in the notation indicates the dependence on the cutoff parameter
$\kappa$. In Proposition \ref{prop:initialdata}, part i), we prove
that
\begin{equation}\label{eq:thm2-1}
\langle \wt \psi_{N,t} , H_N^k \wt\psi_{N,t} \rangle \leq {\wt C}^k
N^k \end{equation} if $\kappa >0$ is sufficiently small (the
constant $\wt C$ depends on $\kappa$). Moreover, in
part iii) of Proposition \ref{prop:initialdata}, we show that, for every
$J^{(k)} \in \cK_k$,
\begin{equation}\label{eq:thm2-2}
\tr \; J^{(k)} \left( \wt \gamma_N^{(k)} - |\ph \rangle \langle
\ph|^{\otimes k} \right) \to 0
\end{equation}
as $N \to \infty$. {F}rom (\ref{eq:thm2-1}) and (\ref{eq:thm2-2}),
the assumptions (\ref{eq:enerk}) and (\ref{eq:init}) of Theorem \ref{thm:main2}
are satisfied by the regularized wave function $\wt \psi_N$ and by the regularized
marginal densities $\wt \gamma^{(k)}_{N,t}$. Applying Theorem~\ref{thm:main2},
we obtain that, for every $t \in \bR$ and
$k \geq 1$,
\begin{equation}\label{eq:conve} \wt \gamma_{N,t}^{(k)} \to |\ph_t \rangle \langle
\ph_t|^{\otimes k} \, , \end{equation} where $\ph_t$ is the solution of
(\ref{eq:GPthm}) with initial data $\ph_{t=0} = \ph$.

\medskip

It remains to prove that the densities $\gamma^{(k)}_{N,t}$
associated with the original wave function $\psi_{N,t}$ (without
cutoff $\kappa$) converge and have the same limit as the regularized
densities $\wt \gamma^{(k)}_{N,t}$. This follows from Proposition
\ref{prop:initialdata}, part ii), where we prove that
\[
\| \psi_{N,t} - \wt\psi_{N,t} \| = \| \psi_{N} -
\wt\psi_{N} \| \leq C \kappa^{1/2} \; ,
\]
for a constant
$C$ independent of $N$ and $\kappa$. This implies that, for every
$J^{(k)} \in \cK_k$, we have
\begin{equation}\label{eq:remove}
\Big| \tr \; J^{(k)} \left( \gamma^{(k)}_{N,t} - \wt
\gamma^{(k)}_{N,t} \right) \Big| \leq C \| J^{(k)} \| \, \kappa^{1/2}\,.
\end{equation}
Therefore, for every fixed $k \geq 1$, $t \in \bR$,
$J^{(k)} \in \cK_k$,  we have
\begin{equation}\label{eq:lastproof}
\begin{split}
\Big| \tr \; J^{(k)} \left( \gamma^{(k)}_{N,t} - |\ph_t \rangle
\langle \ph_t |^{\otimes k} \right) \Big| \leq &\; \Big| \tr \;
J^{(k)} \left( \gamma^{(k)}_{N,t} - \wt \gamma^{(k)}_{N,t} \right)
\Big| + \Big| \tr \; J^{(k)} \left( \wt \gamma^{(k)}_{N,t} - |\ph_t
\rangle \langle \ph_t |^{\otimes k} \right) \Big| \\
\leq & \; C\, \| J^{(k)} \| \kappa^{1/2} + \Big| \tr \; J^{(k)} \left( \wt
\gamma^{(k)}_{N,t} - |\ph_t \rangle \langle \ph_t |^{\otimes k}
\right) \Big| \, .
\end{split}
\end{equation}
Since $\kappa >0$ was arbitrary, it follows from (\ref{eq:conve})
that the l.h.s. of (\ref{eq:lastproof}) converges to zero as $N \to
\infty$. This implies that, for arbitrary $k \geq 1$ and $t \in \bR$,
$\gamma^{(k)}_{N,t} \to |\ph_t \rangle \langle \ph_t |^{\otimes k}$ in the weak*
topology of $\cL^1_k$, and thus also in the trace-norm topology. This completes
 the proof of Theorem \ref{thm:main2}.
\end{proof}

\section{The wave operator and a-priori bounds on $\gamma_{N,t}^{(k)}$}
\setcounter{equation}{0}

In order to derive a-priori bounds for the marginal densities $\gamma^{(k)}_{N,t}$,
 we need to introduce wave operators. We denote by $W$ and $W_N$ the wave operators
 associated with the one-particle Hamiltonian $\fh = -\Delta + (1/2) V (x)$ and,
respectively, $\fh_N = -\Delta + (1/2) V_N (x)$, with $V_N (x) = N^2 V(Nx)$. The
existence of these wave operators and their most important properties are stated
in the following proposition (we denote by $s-\lim$ the limit in the
strong operator topology).

\begin{proposition} \label{prop:waveop}
Suppose $V \geq 0$, with $V \in L^1 (\bR^3)$. Then:
\begin{itemize}
\item[i)] ({\it Existence of the wave operator}). The limit \[ W = s-\lim_{t\to \infty}
e^{i\fh t} e^{i\Delta t} \] exists.
\item[ii)] ({\it Completeness of the wave operator}). $W$ is a unitary operator on
 $L^2 (\bR^3)$ with \[ W^* = W^{-1} =  s-\lim_{t\to \infty} e^{-i\Delta t} e^{-i\fh t} \, . \]
\item[iii)] ({\it Intertwining relations}). On $D(\fh) = D (-\Delta)$, we have
\begin{equation}\label{eq:intertw} W^* \fh W  = - \Delta \end{equation} \, .
\item[iv)] ({\it Yajima's bounds}). Suppose moreover that $V(x)\leq C \langle x \rangle^{-\sigma}$,
 for some $\sigma >5$. Then, for every $1 \leq p \leq \infty$, $W$ and $W^*$ map $L^p (\bR^3)$
into $L^p (\bR^3)$, and \[ \| W \|_{L^p \to L^p} < \infty \qquad \text{for all }
\quad 1 \leq p \leq \infty /, . \]
\item[v)] ({\it Rescaled wave operator}). If $\fh_N = -\Delta + (1/2) V_N (x)$, with
$V_N (x) = N^2 V (Nx)$, then the limit \[ W_N = s-\lim_{t \to \infty} e^{i\fh_N t} e^{i\Delta t} \]
 exists and it defines a unitary operator $W_N$ on $L^2 (\bR^3)$ with \[ W_N^* =
W_N^{-1} = s-\lim_{t\to \infty} e^{-i\Delta t} e^{-i\fh_N t} . \]
The wave operator $W_N$ satisfies the intertwining relations \[ W_N^* \fh_N W_N = -\Delta \, .\]
 Moreover, the kernel of $W_N$ is given by
    \[ W_N (x;y) = N^3 W(Nx;Ny) \qquad \text{and} \qquad W^*_N (x;y) = N^3 W^* (Nx; Ny) \]
where $W (x;y)$ and $W^* (x;y)$ denote the kernels of $W$ and $W^*$. In particular,
it follows that, if for every $1 \leq p \leq \infty$, the norms \[ \| W_N \|_{L^p \to L^p}
 = \| W \|_{L^p \to L^p} < \infty \qquad  \text{and} \qquad \| W^*_N \|_{L^p \to L^p}
= \| W^* \|_{L^p \to L^p} < \infty \] are finite and independent of $N$.
\end{itemize}
\end{proposition}
\begin{proof}
The proof of i), ii), and iii) can be found in \cite{RS3}. Part iv) is proven in
\cite{Ya0,Ya}. Part~v) follows by simple scaling arguments.
\end{proof}

In the following we will denote by $W_{(i,j)}$ and, respectively, by $W_{N, (i,j)}$,
 the wave operators $W$ and $W_N$ acting only on the relative variable $x_j - x_i$.
In other words, the action of $W_{(i,j)}$ on a $N$-particle wave function $\psi_N
\in L^2 (\bR^{3N})$ is given by \begin{equation}\label{eq:Wij} \left( W_{(i,j)}
\psi_N \right) (\bx) = \int \rd v \; W (x_j - x_i; v) \, \psi_N \left( x_1, \dots ,
 \frac{x_i + x_j}{2} + \frac{v}{2}, \dots , \frac{x_i + x_j}{2} - \frac{v}{2},
\dots, x_N \right) \end{equation} if $j <i$ (the formula for $i >j$ is similar).
Here $W(x;y)$ is the kernel of the wave operator $W$. An analogous formula holds for
 the rescaled wave operator $W_N$. Similarly, we define $W^*_{(i,j)}$ and $W^*_{N,(i,j)}$.

Using the wave operators, we have the following energy estimate.

\begin{proposition}\label{prop:energ2}
Suppose $V \geq 0$, $V \in L^1 (\bR^3)$ and $V(x) = V(-x)$ for all $x\in \bR^3$.
Then we have, for every $i \neq j$,
\begin{equation}\label{eq:energ2}
\langle \psi_N , H_N^2 \psi_N \rangle \geq C N^2 \int \rd \bx \; \left|
\left(\nabla_i \cdot \nabla_j \right)^2 W^*_{N,(i,j)} \psi_N \right|^2\, ,
\end{equation}
where $W^*_{N,(i,j)}$ denotes the wave operator $W^*_N$ defined in Proposition
\ref{prop:waveop} acting on the variable $v = x_j -x_i$ (defined similarly to (\ref{eq:Wij})).
\end{proposition}
\begin{proof}
We define, for $j=1,\dots,N$,
\[ h_j = -\Delta_j + \frac{1}{2}\sum_{i \neq j} V_N (x_i -x_j). \]
Then we have $H_N = \sum_{j=1}^N h_j$ and thus
\begin{equation}\label{eq:h1}
\begin{split}
\langle \psi_N, H_N^2 \, \psi_N \rangle \geq \; &N(N-1) \langle
\psi_N, h_1 h_2 \psi_N \rangle \\ = \; &N(N-1) \left\langle \psi_N ,
\left( -\Delta_1 + \frac{1}{2} \sum_{i \neq 1} V_N (x_i -x_1)
\right)\left( -\Delta_2 + \frac{1}{2} \sum_{j\neq 2} V_N (x_j -x_2)
\right) \psi_N \right\rangle \\
\geq \; &N(N-1) \left\langle \psi_N, \left( -\Delta_1 + \frac{1}{2}
V_N (x_1-x_2)\right) \left( -\Delta_2 + \frac{1}{2} V_N (x_1
-x_2)\right) \psi_N \right\rangle\,.
\end{split}
\end{equation}
Now we define the new variables
\[ u= \frac{x_1+x_2}{2}, \quad \text{and } \quad v=x_1 -x_2 \,.\] Then we have \[ \nabla_1 =
\frac{1}{2} \nabla_u + \nabla_v \quad \text{and } \quad \nabla_2 = \frac{1}{2} \nabla_u
-\nabla_v \] and thus
\[ \Delta_1 = \frac{1}{4} \Delta_u + \Delta_v + \nabla_u \cdot \nabla_v, \quad \text{and } \quad
\Delta_2 = \frac{1}{4} \Delta_u +\Delta_v - \nabla_u \cdot \nabla_v \,.\] We
set
\[ h_v = -\Delta_v + \frac{1}{2} V_N (v) \,. \] Then
\begin{equation}\label{eq:H-2}
\begin{split}
\langle \psi_N, H_N^2 \, \psi_N \rangle \geq \; &N(N-1) \left\langle
\psi_N, \left( - \frac{1}{4} \Delta_u + h_v + \nabla_u \cdot \nabla_v \right)
\left( -\frac{1}{4} \Delta_u + h_v - \nabla_u \cdot \nabla_v \right) \psi_N
\right\rangle \\ = \; &N(N-1) \left\langle \psi_N, \left[ \left( - \frac{1}{4} \Delta_u +
h_v \right)^2  - \left(\nabla_u \cdot \nabla_v \right)^2 + \frac{1}{2} \,
\nabla_u \cdot \left(\nabla V_N (v)\right) \right] \psi_N \right\rangle\,.
\end{split}
\end{equation}
Next we note that
\begin{equation}\label{eq:inte}
\begin{split}
\langle \psi_N, \nabla_u &\cdot \nabla V_N (v) \psi_N \rangle \\ &= \int \rd u
\rd v \; \overline{\psi}_N (u+v/2,u-v/2, \bx_{N-2}) \nabla V_N
(v) \cdot \nabla_u \psi_N (u+v/2,u-v/2,\bx_{N-2}) = 0\,.
\end{split}
\end{equation}
In fact, by the permutation symmetry, $\psi_N (x_1,x_2,\bx_{N-2}) =
\psi_N (x_2,x_1,\bx_{N-2})$. This implies, in the $u,v$-coordinates,
that $\psi_N (u+v/2,u-v/2,\bx_{N-2}) = \psi_N (u-v/2,u+v/2,\bx_{N-2})$ and also
that $\nabla_u \psi_N (u+v/2,u-v/2,\bx_{N-2}) = \nabla_u \psi_N
(u-v/2,u+v/2,\bx_{N-2})$. On the other hand $\left(\nabla V_N \right)
(-v)=-\left(\nabla V_N \right) (v)$. Therefore, the
integrand in (\ref{eq:inte}) is antisymmetric w.r.t. the change of
variables $v \to -v$, and the integral vanishes.

\medskip

Using also that \[(\nabla_u \cdot \nabla_v)^2 \leq
\left(-\Delta_u\right) \left(-\Delta_v\right) \leq \left(-\Delta_u
\right) \, h_v \, , \] it follows from (\ref{eq:H-2}) that
\begin{equation}\label{eq:H-3}
\langle \psi_N, H_N^2 \, \psi_N \rangle \geq N(N-1) \left\langle \psi_N,
\left( -\frac{1}{4} \Delta_u - h_v \right)^2 \psi_N \right\rangle \,.
\end{equation}
Next we make use of the wave operator $W_N$ defined in Proposition \ref{prop:waveop},
acting on the variable $v=x_2 - x_1$. By the intertwining relations (\ref{eq:intertw}), we find
\begin{equation}\label{eq:H-4}
\langle \psi_N, H_N^2 \, \psi_N \rangle \geq N(N-1) \left\langle W_{N,(1,2)}^*
\psi_N, \left( \frac{1}{4} \Delta_u - \Delta_v \right)^2 \, W_{N,(1,2)}^* \psi_N \right\rangle\,.
\end{equation}
In terms of the coordinates $x_1$ and $x_2$, we have $\nabla_1 \cdot
\nabla_2 = (1/4) \Delta_u - \Delta_v$. Therefore, by the permutation symmetry,
the bound (\ref{eq:H-4}) implies (\ref{eq:energ2}).
\end{proof}

Proposition \ref{prop:energ2} implies strong a-priori bounds on the solution of
the $N$-particle Schr\"odinger equation.
\begin{proposition}\label{prop:apri2}
Suppose that $V \geq 0$, $V \in L^1 (\bR^3)$, and $V(-x) = V(x)$ for all $x\in \bR^3$.
 Let $\psi_{N,t}$ be the solution of the Schr\"odinger equation (\ref{eq:schr}),
with initial data satisfying the assumption (\ref{eq:enerk}) (with $k=2$) of
Theorem \ref{thm:main2}, and let $\{ \gamma^{(k)}_{N,t} \}_{k=1}^N$ be the
 marginals associated with $\psi_{N,t}$. Then, for every $1\leq j \leq N$,
we have \[ \langle \psi_{N,t}, (1-\Delta_j) \psi_{N,t} \rangle \leq C \qquad
\text{and thus} \qquad \tr \; (1-\Delta_j) \gamma_{N,t}^{(k)} \leq C \] for
every $1\leq j\leq k \leq N$ (and for a constant $C$ which only depends on the
 initial data $\psi_N$ through the constant on the r.h.s. of (\ref{eq:enerk})).
Moreover, for any $i \neq j$, \[ \left\langle W_{N,(i,j)}^* \psi_{N,t} ,
\left((\nabla_i \cdot \nabla_j)^2 -\Delta_i - \Delta_j + 1 \right) W_{N,(i,j)}^*
 \psi_{N,t} \right\rangle \leq C \] uniformly in $N \geq 1$ and in $t \in \bR$. Here
$W_{N,(i,j)}$ denotes the wave operator $W_N$ defined in Proposition~\ref{prop:waveop}
 acting on the variable $x_j - x_i$.  In terms of density matrices, we obtain the a-priori bounds
\[ \tr \; \left((\nabla_i \cdot \nabla_j)^2 -\Delta_i - \Delta_j + 1 \right) W^*_{N,(i,j)}
 \gamma^{(k)}_{N,t} W_{N,(i,j)} \leq C \] uniformly in $N \geq 1$ and in $t \in \bR$ and for all $1 \leq i < j \leq k$ (with a slight abuse of notation, we denote here by $W_{N,(i,j)}$ and $W^*_{N,(i,j)}$
 the operators acting on the $k$-particle space $L^2 (\bR^{3k})$).
\end{proposition}
\begin{proof}
The first bound follows simply by the symmetry of the wave function, by energy conservation,
and by the condition $V \geq 0$. To prove the second bound, we compute
\begin{equation}\label{eq:apri2-1}
\begin{split}
\langle W_{N,(i,j)}^* \psi_{N,t} , \left((\nabla_i \cdot \nabla_j)^2 -\Delta_i -
\Delta_j + 1 \right) &W_{N,(i,j)}^* \psi_{N,t} \rangle
\\ = \; & \langle W_{N, (1,2)}^* \psi_{N,t} ,(\nabla_1 \cdot \nabla_2)^2 W_{N,(1,2)}^*
\psi_{N,t} \rangle \\ &+\langle W_{N,(1,2)}^* \psi_{N,t} , \left(-\Delta_1 - \Delta_2 \right)
 W_{N,(1,2)}^* \psi_{N,t} \rangle +1 \,.
\end{split}
\end{equation}
The first term on the r.h.s. of the last equation can be bounded by
\begin{equation}
\begin{split}
\langle W_{N,(1,2)}^* \psi_{N,t} , (\nabla_1 \cdot \nabla_2)^2 W_{N,(1,2)}^* \psi_{N,t}
\rangle \leq \; & CN^{-2} \langle \psi_{N,t}, H_N^2 \psi_{N,t} \rangle
\\ = \; & CN^{-2} \langle \psi_{N,0}, H_N^2 \psi_{N,0} \rangle \leq C \,
\end{split}
\end{equation}
using Proposition
\ref{prop:energ2} and (\ref{eq:enerk}). The second term on the r.h.s. of (\ref{eq:apri2-1}) is estimated by
\begin{equation}\label{eq:rev6}
\begin{split}
\langle W_{N, (1,2)}^* \psi_{N,t} , & \left(-\Delta_1 - \Delta_2 \right)W_{N,(1,2)}^*
\psi_{N,t} \rangle \\ & = 2 \, \left\langle W_{N,(1,2)}^* \psi_{N,t} ,
 \left(-\Delta_{x_1-x_2} - \frac{1}{4} \Delta_{(x_1 + x_2)/2} \right)
W_{N,(1,2)}^* \psi_{N,t} \right\rangle  \\ & = 2\, \left\langle \psi_{N,t} ,
\left(-\Delta_{x_1-x_2} + \frac{1}{2} V_N  (x_1 -x_2) - \frac{1}{4} \,
\Delta_{(x_1 + x_2)/2} \right) \psi_{N,t} \right\rangle  \\ & = \left\langle \psi_{N,t} ,
\left(-\Delta_{x_1} - \Delta_{x_2} + V_N  (x_1 -x_2) \right) \psi_{N,t} \right\rangle
 \\ &\leq \frac{2}{N} \langle \psi_{N,t}, H_N \psi_{N,t} \rangle =
\frac{2}{N} \langle \psi_{N,0} , H_N \psi_{N,0} \rangle \leq C\,.
\end{split}
\end{equation}
\end{proof}

\section{Compactness}
\setcounter{equation}{0}

In this section we prove the compactness of the sequence $
\Gamma_{N,t} = \{ \gamma^{(k)}_{N,t} \}_{k\ge1}$ w.r.t. the topology
$\tau_{\text{prod}}$ (defined in Section \ref{sec:outline}).

\begin{theorem}\label{thm:compactness}
Let the assumptions of Theorem \ref{thm:main2} be satisfied and
fix an arbitrary $T>0$. Then the sequence
$\Gamma_{N,t} \in \bigoplus_{k \geq 1} C([0,T], \cL_k^1)$ is compact
with respect to the product topology $\tau_{\text{prod}}$ generated
by the metrics $\wh \eta_k$ (defined in Section \ref{sec:outline}).
For any limit point $ \Gamma_{\infty,t} = \{ \gamma_{\infty,t}^{(k)}
\}_{k \geq 1}$, $ \gamma^{(k)}_{\infty,t}$ is symmetric w.r.t.
permutations, $ \gamma^{(k)}_{\infty,t} \geq 0$, and
\begin{equation}\label{eq:bou} \tr \; \gamma^{(k)}_{\infty,t} \leq 1
\,\end{equation} for every $k \geq 1$.
\end{theorem}

\begin{proof}
By a standard argument it is enough to prove the compactness of
$\gamma_{N,t}^{(k)}$ for fixed $k \geq 1$ with respect to the metric $\wh \eta_k$. To
this end, it is enough to show the equicontinuity of $\gamma_{N,t}^{(k)}$ with respect to the metric $\eta_k$. A useful criterium for equicontinuity is given  by the following lemma, whose proof can be found in \cite[Proposition 9.2]{ESY2}.

\begin{lemma}\label{lm:equi}  Fix $k \in \bN$ and $T > 0$.
A sequence $\gamma_{N, t}^{(k)} \in \cL^1_k$, $N=k,k+1, \ldots$,
with $\gamma_{N,t}^{(k)} \geq 0$ and $\tr \; \gamma_{N,t}^{(k)} = 1$
for all $t \in [0,T]$ and $N \geq k$, is equicontinuous in $C([0,T],
\cL^1_k)$ with respect to the metric $\eta_k$, if and only if there
exists a dense subset $\cJ_k$ of $\cK_k$ such that for any
$J^{(k)}\in \cJ_k$ and for every $\eps
>0$ there exists a $\delta > 0$ such that
\begin{equation}\label{eq:equi02}
\sup_{N\ge 1}\Big| \tr \;  J^{(k)} \left( \gamma_{N,t}^{(k)} -
\gamma_{N,s}^{(k)} \right) \Big| \leq \eps
\end{equation}
for all $t,s \in [0,T]$ with $|t -s| \leq \delta$.
\end{lemma}

We prove (\ref{eq:equi02}) for all $J^{(k)} \in \cK_k$ such that $\tri J^{(k)} \tri < \infty$, where we introduced the norm
\begin{equation}\label{eq:tri}\begin{split} \tri J^{(k)} \tri = \sup_{\bx_k , \bx'_k} &\la x_1 \ra^4 \dots \la x_k \ra^4 \la x'_1 \ra^4 \dots \la x'_k \ra^4 \, \\ &\times \left( |J^{(k)} (\bx'_k; \bx_k)| + \sum_{j=1}^k \left( |\nabla_{x_j} J^{(k)} (\bx'_k; \bx_k)| + |\nabla_{x'_j} J^{(k)} (\bx'_k; \bx_k)| \right) \right). \end{split}\end{equation}
It is simple to check that the set of $J^{(k)} \in \cK_k$ for which $\tri J^{(k)} \tri < \infty$ is dense in $\cK_k$.

\medskip

Rewriting the BBGKY hierarchy (\ref{eq:BBGKY}) in integral form and multiplying with an arbitrary observable $J^{(k)} \in \cK_k$ with $\tri J^{(k)} \tri < \infty$, we obtain that, for any $r \leq t$,
\begin{equation}\label{eq:equi-1}
\begin{split}
\Big| \tr \, J^{(k)} \left(  \gamma_{N,t}^{(k)} -
\gamma_{N,r}^{(k)} \right) \Big| \leq \; &\sum_{j=1}^k \int_r^t \rd s \,
\Big| \tr \; J^{(k)} [ -\Delta_j , \gamma_{N,s}^{(k)}] \Big| + \sum_{i<j}^k
\int_r^t \rd s \, \Big| \tr \; J^{(k)} [ V_N (x_i -x_j) , \gamma^{(k)}_{N,s} ] \Big| \\
&+\left(1-\frac{k}{N} \right) \sum_{j=1}^k \int_r^t \rd s \, \Big|
\tr \; J^{(k)}\left[ N V_N (x_j - x_{k+1}), \gamma^{(k+1)}_{N,s}
\right]\Big|.
\end{split}
\end{equation}
To control the first term on the r.h.s. of the last equation, we observe that, using the notation $S_j = (1-\Delta_j)^{1/2}$, we have
\begin{equation}\label{eq:equi-2}
\begin{split}
\Big| \tr \; J^{(k)} [ -\Delta_j , \gamma_{N,s}^{(k)}] \Big| = \; &\Big| \tr \; \left( S_j^{-1} J^{(k)} S_j - S_j J^{(k)} S_j^{-1} \right) S_j \gamma_{N,s}^{(k)} S_j  \Big| \\
\leq \; & \left( \left\| S_j^{-1} J^{(k)} S_j \right\| + \left\| S_j J^{(k)} S_j^{-1} \right\| \right) \tr (1-\Delta_j) \gamma_{N,s}^{(k)}
\\ \leq \;& C \tri J^{(k)} \tri \, .
\end{split}
\end{equation} Here we used that, by Proposition \ref{prop:apri2},  \[ \sup_{s \in \bR} \; \tr (1-\Delta_j) \gamma_{N,s}^{(k)} \leq C \] uniformly in $N$.

\medskip

To bound the second term on the r.h.s. of (\ref{eq:equi-1}), we decompose $\gamma_{N,s}^{(k)} = \sum_{\ell} \lambda^{(k)}_{\ell} \, |\xi^{(k)}_{\ell} \rangle \langle \xi^{(k)}_{\ell}|$ for $\xi^{(k)}_{\ell} \in L^2 (\bR^{3k})$, with $\| \xi_{\ell}^{(k)} \| =1$, $\lambda^{(k)}_{\ell} >0$, and $\sum_{\ell} \lambda^{(k)}_{\ell} =1$ (here we omitted the dependence of $\xi^{(k)}_{\ell}$ and $\lambda^{(k)}_{\ell}$ on $N,s$ from the notation). Then we find, for example for the term with $i=1, j=2$,
\begin{equation}\label{eq:equi-2ndterm}
\begin{split}
\tr \; J^{(k)} \, & V_N (x_1 -x_2) \gamma_{N,s}^{(k)} = \sum_{\ell} \lambda^{(k)}_{\ell} \, \int \rd \bx_k \, \rd \bx'_k \; J^{(k)} (\bx'_k ; \bx_k) V_N (x_1 -x_2) \xi^{(k)}_{\ell} ( \bx_k) \overline{\xi}^{(k)}_{\ell} (\bx'_k).
\end{split}
\end{equation}
Denoting by $W_N$ the wave operator associated with the Hamiltonian $\fh_N = -\Delta + (1/2) V_N (x)$, and by $W_{N,(i,j)}$ the wave operator $W_N$ acting on the variable $x_j -x_i$ (as defined in (\ref{eq:Wij})), we can estimate (introducing the new variables $u= (x_1+ x_2)/2$ and $v = x_1-x_2$)
\begin{equation*}
\begin{split}
\Big| \int \rd \bx_k \,& \rd \bx'_k  J^{(k)} (\bx'_k ; \bx_k) V_N (x_1 -x_2) \xi^{(k)}_{\ell}
( \bx_k) \overline{\xi}^{(k)}_{\ell} (\bx'_k) \Big| \\ 
= \, & \Big| \int \rd u \rd v \rd x_3 \dots \rd x_k  \rd \bx'_k \; J^{(k)} (\bx'_k ; u+ v/2, u-v/2, x_3, \dots ,x_k) 
\\ &\hspace{1.5cm} \times 
V_N (v) \xi^{(k)}_{\ell} (u+v/2,u-v/2,x_3, \dots ,x_k) \, \overline{\xi}^{(k)}_{\ell} (\bx'_k) \Big|
\\  \leq \, &\int \rd u \rd v \rd x_3 \dots \rd x_k  \rd \bx'_k \; \Big| J^{(k)} (\bx'_k ; u+ v/2, u-v/2, x_3, \dots ,x_k) - J^{(k)} (\bx'_k; u,u,x_3, \dots ,x_k)\Big| 
\\ &\hspace{1.5cm} \times V_N (v) |\xi^{(k)}_{\ell} (u+v/2,u-v/2,x_3, \dots ,x_k)| \, |\xi^{(k)}_{\ell} (\bx'_k)| 
\\ & +   \int \rd u \rd v \rd x_3 \dots \rd x_k  \rd \bx'_k \;  |J^{(k)} (\bx'_k; u,u,x_3, \dots ,x_k)| \, | (W_{N}^* V_N) (v)| \\ & \hspace{1.5cm} \times \left| \left( W^*_{N, (1,2)}\xi^{(k)}_{\ell}\right) (u+v/2,u-v/2,x_3, \dots ,x_k) \right| \, |\xi^{(k)}_{\ell} (\bx'_k)| 
\end{split}\end{equation*}
where in the last line we used the $L^2$-unitarity of the wave operator in the $v$-variable (before taking the absolute value inside the integral). Hence 
\begin{equation*}
\begin{split}
\Big| \int &\rd \bx_k \, \rd \bx'_k  J^{(k)} (\bx'_k ; \bx_k) V_N (x_1 -x_2) \xi^{(k)}_{\ell}
( \bx_k) \overline{\xi}^{(k)}_{\ell} (\bx'_k) \Big| \\  \leq \, &\sum_{j=1}^2 \int \rd u \rd v \rd x_3 \dots \rd x_k  \rd \bx'_k \int_0^1 \rd \tau | \nabla_{x_j} J^{(k)} (\bx'_k ; u+ \frac{\tau v}{2}, u- \frac{\tau v}{2}, x_3, .. ,x_k)|  |v| V_N (v) |\xi^{(k)}_{\ell} (\bx'_k)|^2 
\\ &+  \sum_{j=1}^2 \int \rd u \rd v \rd x_3 \dots \rd x_k  \rd \bx'_k \int_0^1 \rd \tau \, | \nabla_{x_j} J^{(k)} (\bx'_k ; u+ \frac{\tau v}{2}, u- \frac{\tau v}{2}, x_3, .. ,x_k)|  |v| V_N (v) \\ &\hspace{3.5cm} \times |\xi^{(k)}_{\ell} (u+v/2,u-v/2,x_3, \dots ,x_k)|^2 
\\ &+  \int \rd u \rd v \rd x_3 \dots \rd x_k  \rd \bx'_k \;  |J^{(k)} (\bx'_k; u,u,x_3, \dots ,x_k)| \, | (W_{N}^* V_N) (v)|  \, |\xi^{(k)}_{\ell} (\bx'_k)|^2 
\\ &+ 
  \int \rd u \rd v \rd x_3 \dots \rd x_k  \rd \bx'_k \;  |J^{(k)} (\bx'_k; u,u,x_3, \dots ,x_k)| \, | (W_{N}^* V_N) (v)| \\ &\hspace{3.5cm} \times \left|\left( W^*_{N, (1,2)}\xi^{(k)}_{\ell}\right) (u+v/2,u-v/2,x_3, \dots ,x_k) \right|^2 \,. 
\end{split}\end{equation*}
Using the norm defined in (\ref{eq:tri}), we find 
\begin{equation}\label{eq:rev1}
\begin{split}
\Big| \int \rd &\bx_k \, \rd \bx'_k  J^{(k)} (\bx'_k ; \bx_k) V_N (x_1 -x_2) \xi^{(k)}_{\ell}
( \bx_k) \overline{\xi}^{(k)}_{\ell} (\bx'_k) \Big| \\ \leq \, & C_k \, \tri J^{(k)} \tri \, \| \xi^{(k)}_{\ell} \|^2  \, \int \rd v \, |v| \, V_N (v) \\ &+ C_k \, \tri J^{(k)} \tri \int \rd v \rd u \rd x_3 \dots \rd x_N \, | v| V_N (v) \,  |\xi^{(k)}_{\ell} (u+v/2,u-v/2,x_3, \dots ,x_k)|^2 
\\ &+ C_k \,  \tri J^{(k)} \tri \, \| \xi^{(k)}_{\ell} \|^2 \,   \int \rd v \, | (W_{N}^* V_N) (v)|  \, \\ &
+ C_k \, \tri J^{(k)} \tri \, 
  \int \rd u \rd v \rd x_3 \dots \rd x_k  \, |(W_{N}^* V_N) (v)| \, \left|\left( W^*_{N, (1,2)}\xi^{(k)}_{\ell} \right) (u+v/2,u-v/2,x_3, \dots ,x_k) \right|^2 .
\end{split}\end{equation}
Since, by scaling $\| \, |v| V_N \,\|_1 \leq C N^{-2}$ and $\|\,  |v| V_N \, \|_{3/2} \leq C N^{-1}$ 
and since $\| W_N^* V_N \|_{1} \leq C \| V_N \|_1 \leq C N^{-1}$ (by the Yajima's bounds in part v) of Proposition \ref{prop:waveop}), we conclude that
\begin{equation}\label{eq:rev2} 
\begin{split}
\Big| \int \rd \bx_k \, \rd \bx'_k  J^{(k)} (\bx'_k ; \bx_k) &V_N (x_1 -x_2) \xi^{(k)}_{\ell}
( \bx_k) \overline{\xi}^{(k)}_{\ell} (\bx'_k) \Big| \\ &\leq \,  \frac{C_k \, \tri J^{(k)} \tri}{N} \left\langle W_{N,(1,2)}^* \xi_{\ell}^{(k)}, \left( (\nabla_1 \cdot \nabla_2)^2 -\Delta_1 - \Delta_2 + 1\right) W_{N,(1,2)}^* \xi^{(k)}_{\ell} \right\rangle 
\end{split}
\end{equation}
for a $k$-dependent constant $C_k$. Here we used Lemma \ref{lm:VL1} to bound the last term on the r.h.s. of (\ref{eq:rev1}), and that, by the Sobolev inequality $\| \xi_{\ell}^{(k)} \|_{L^6_v} \leq C \| \nabla_v \xi^{(k)}_{\ell} \|_{L^2_v}
$,
\begin{equation}\label{eq:rev3}
\begin{split}
\int \rd u \rd x_3 \dots \rd x_k \rd v \, |v| V_N(v) &|\xi^{(k)}_{\ell} (u+v/2,u-v/2,x_3, \dots ,x_k)|^2 \\ \leq \; & \| |v| V_N \|_{3/2} \int \rd u \rd x_3 \dots \rd x_k  \rd v \, |\nabla_v \xi^{(k)}_{\ell} (u+\frac{v}{2},u-\frac{v}{2},x_3, .. ,x_k)|^2 \\ \leq \; & C N^{-1}\left\langle \xi^{(k)}_{\ell} , (-\Delta_v) \, \xi^{(k)}_{\ell} \right\rangle \\ \leq \; & C N^{-1} \left\langle \xi^{(k)}_{\ell} , \left( -\Delta_v +(1/2) V_N (v) \right) \xi^{(k)}_{\ell} \right\rangle \\ \leq \; & C N^{-1}\left\langle W_{N,(1,2)}^* \xi^{(k)}_{\ell}, \left(-\Delta_1 - \Delta_2 \right)  W_{N,(1,2)}^* \xi^{(k)}_{\ell} \right\rangle 
\end{split}
\end{equation}
to bound the second term on the r.h.s. of (\ref{eq:rev1}) (applying the intertwining relations (\ref{eq:intertw}) and adding a positive term $-\Delta_u$ as in (\ref{eq:rev6})). {F}rom (\ref{eq:equi-2ndterm}), (\ref{eq:rev2}),  and from Proposition \ref{prop:apri2} we obtain that
\begin{equation}\label{eq:equi-est12}
\begin{split}
\Big| \tr \; J^{(k)} \, V_N (x_1 -x_2) \gamma_{N,s}^{(k)} \Big| & \leq \frac{C_k \tri J^{(k)} \tri}{N} \, \tr \; \left( (\nabla_1 \cdot \nabla_2)^2 -\Delta_1 - \Delta_2 + 1\right) W_{N,(1,2)}^* \gamma_{N,s}^{(k)} W_{N,(1,2)} \\ &\leq \frac{C_k \tri J^{(k)} \tri}{N}
\end{split}
\end{equation}
for all $s\in \bR$ and a constant $C_k$ only depending on $k$ (and on the constant appearing on the r.h.s. of (\ref{eq:enerk})). Similarly to (\ref{eq:equi-est12}), we can also show that
\begin{equation}\label{eq:equi-est12b} \Big| \tr \; J^{(k)} \, \gamma_{N,s}^{(k)} V_N (x_1 -x_2) \Big| \leq \frac{C_k \, \tri J^{(k)} \tri}{N}\, . \end{equation}
Since (\ref{eq:equi-est12}) and (\ref{eq:equi-est12b}) remain valid for all summands in the second term on the r.h.s. of (\ref{eq:equi-1}), we obtain that, for all $k \in \bN$, for all $t \in [0,T]$ and for all $J^{(k)} \in \cK_k$ with $\tri J^{(k)} \tri < \infty$,
\begin{equation}\label{eq:equi-3}
\sum_{i<j}^k  \Big| \tr \; J^{(k)} \,  \left[ V_N (x_i -x_j), \gamma_{N,s}^{(k)} \right] \Big| \leq \frac{C_k \, \tri J^{(k)} \tri}{N}
\end{equation}
for all $s \in \bR$.

\medskip

Also the third term on the r.h.s. of (\ref{eq:equi-1}) can be bounded similarly. In fact,
using again the decomposition $\gamma_{N,s}^{(k+1)} = \sum_{\ell} \lambda^{(k+1)}_{\ell} \, |\xi^{(k+1)}_{\ell} \rangle \langle \xi^{(k+1)}_{\ell}|$ we have, for example considering the term with $j=1$,
\begin{equation}\label{eq:comp31}
\begin{split}
\tr \, &J^{(k)} \, N \, V_N (x_1 -x_{k+1})  \gamma_{N,s}^{(k+1)} \\ = \; &\sum_{\ell} \lambda^{(k+1)}_{\ell} \int \rd \bx_k \, \rd \bx'_k \rd x_{k+1} \, J^{(k)} (\bx'_k ; \bx_k) N V_N (x_1 -x_{k+1}) \xi^{(k+1)}_{\ell} (\bx_k,x_{k+1}) \overline{\xi}^{(k+1)}_{\ell} (\bx'_k,x_{k+1}) \,. 
\end{split}
\end{equation}
The absolute value of the $\ell$-th summand can be estimated by 
\begin{equation}\label{eq:rev4} \begin{split}
\Big| \int \rd \bx_k \,& \rd \bx'_k \rd x_{k+1} \, J^{(k)} (\bx'_k ; \bx_k) N V_N (x_1 -x_{k+1}) \xi^{(k+1)}_{\ell} (\bx_k,x_{k+1}) \overline{\xi}^{(k+1)}_{\ell} (\bx'_k,x_{k+1})  \Big| 
\\ \leq &\int \rd u \rd v \rd x_2 \dots \rd x_k  \rd \bx'_k \, N V_N (v) \, |\xi^{(k+1)}_{\ell} (u+v/2,x_2, \dots ,x_k,u-v/2)|\\ &\hspace{1.5cm} \times \left| J^{(k)} (\bx'_k ; u+v/2,x_2, \dots ,x_k) \overline{\xi}^{(k+1)}_{\ell} (\bx'_k, u-v/2) \right. \\ &\hspace{3cm} \left.- J^{(k)} (\bx'_k ; u,x_2, \dots ,x_k)  \overline{\xi}^{(k+1)}_{\ell} (\bx'_k, u)\right|   \, \\
&+ \int \rd u \rd v \rd x_2 \dots \rd x_k  \rd \bx'_k N |(W^*_N V_N )(v)| \, |J^{(k)} (\bx'_k ; u,x_2, \dots ,x_k)|  \\ &\hspace{1.5cm} \times  |\xi^{(k+1)}_{\ell} (\bx'_k, u)| \left|\left(W_{N,(1,k+1)}^* \xi^{(k+1)}_{\ell}\right) (u+v/2,x_2, \dots ,x_k,u-v/2)\right| \\ = & \; \text{I} + \text{II} \, .
\end{split}
\end{equation}
Here 
\begin{equation}\label{eq:rev5} \begin{split}
\text{I} =  \; &\int \rd u \rd v \rd x_2 \dots \rd x_k  \rd \bx'_k \, N V_N (v) \, |\xi^{(k+1)}_{\ell} (u+v/2,x_2, \dots ,x_k,u-v/2)|\\ &\hspace{1.5cm} \times \left| \int_0^1 \rd \tau \, \frac{\rd}{\rd \tau} \left[ J^{(k)} (\bx'_k ; u+\frac{\tau v}{2},x_2, \dots ,x_k) \overline{\xi}^{(k+1)}_{\ell} (\bx'_k, u-\frac{\tau v}{2}) \right] \right| \\
\leq \; &\int \rd u \rd v \rd x_2 \dots \rd x_k  \rd \bx'_k \int_0^1 \rd \tau \, N V_N (v) |v| \, |\xi^{(k+1)}_{\ell} (u+v/2,x_2, \dots ,x_k,u-v/2)|
\\ &\hspace{1.5cm} \times \left( | \nabla_{x_1} J^{(k)} (\bx'_k ; u+\frac{\tau v}{2},x_2, \dots ,x_k)| \, | \xi^{(k+1)}_{\ell} (\bx'_k, u-\frac{\tau v}{2})| \right. \\ &\hspace{3cm} \left.+ |J^{(k)} (\bx'_k ; u+\frac{\tau v}{2},x_2, \dots ,x_k)| | \nabla_{x_{k+1}} \xi^{(k+1)}_{\ell} (\bx'_k, u-\frac{\tau v}{2})| \right) \, . \end{split}\end{equation}
Through a weighted Schwarz inequality, we find
\[ \begin{split}
\text{I} \leq \; &\int \rd u \rd v \rd x_2 \dots \rd x_k  \rd \bx'_k \int_0^1 \rd \tau \, N V_N (v) |v| \, 
| \nabla_{x_1} J^{(k)} (\bx'_k ; u+\frac{\tau v}{2},x_2, \dots ,x_k)| \\ &\hspace{1.5cm} \times \left( N^{-1/2} \, |\xi^{(k+1)}_{\ell} (u+v/2,x_2, \dots ,x_k,u-v/2)|^2 +  N^{1/2} | \xi^{(k+1)}_{\ell} (\bx'_k, u-\frac{\tau v}{2})|^2 \right)  \\ 
&+\int \rd u \rd v \rd x_2 \dots \rd x_k  \rd \bx'_k \int_0^1 \rd \tau \, N V_N (v) |v| \, 
| J^{(k)} (\bx'_k ; u+\frac{\tau v}{2},x_2, \dots ,x_k)| \\ &\hspace{1.5cm} \times \left( N^{-1/2} \, |\xi^{(k+1)}_{\ell} (u+v/2,x_2, \dots ,x_k,u-v/2)|^2 +  N^{1/2} \, | \nabla_{x_{k+1}} \xi^{(k+1)}_{\ell} (\bx'_k, u-\frac{\tau v}{2})|^2 \right) \, . \end{split}\]
Extracting the observable from the integral (after integrating some of its variables and taking the supremum over the other), and using Sobolev inequalities where needed, we find 
\begin{equation}\label{eq:revI} \begin{split}
\text{I} \leq \; & C_k \,  N^{1/2} \| |v| V_N \|_{3/2} \, \tri J^{(k)} \tri \,  \int \rd u \rd v \rd x_2 \dots \rd x_k |\nabla_v \xi^{(k+1)}_{\ell} (u+v/2, x_2, \dots ,x_k, u-v/2)|^2 \\ &+ C_k \,  N^{3/2} \| |v| V_N \|_1 \,  \tri J^{(k)} \tri \left( \| \xi^{(k)}_{\ell} \|^2 + \int \rd u \rd \bx_k' \, |\nabla_{k+1} \xi^{(k+1)}_{\ell} (\bx'_k, u) |^2 \right) \\ \leq \; & \frac{C_k \,  \tri J^{(k)} \tri}{N^{1/2}} \left\langle W_{N,(1,k+1)}^* \xi_{\ell}^{(k+1)}, \left(-\Delta_1 - \Delta_{k+1} + 1\right) W_{N,(1,k+1)}^* \xi_{\ell}^{(k+1)} \right\rangle \, .
\end{split} 
\end{equation} In the last line we proceeded similarly to (\ref{eq:rev3}) for the two terms with derivatives, using the bounds $\| |v| V_N \|_{3/2} \leq C N^{-1}$ and $\| |v| V_N \|_1 \leq C N^{-2}$.

As for the second term on the r.h.s. of (\ref{eq:rev4}), we can bound it applying a  Schwarz inequality and Lemma \ref{lm:VL1} by
\[ \begin{split}
\text{II} \leq \; &  \int \rd u \rd v \rd x_2 \dots \rd x_k  \rd \bx'_k , N |(W^*_N V_N )(v)| \, |J^{(k)} (\bx'_k ; u,x_2, \dots ,x_k)|  |\xi^{(k+1)}_{\ell} (\bx'_k, u)|^2 \\ &+  \int \rd u \rd v \rd x_2 \dots \rd x_k  \rd \bx'_k \, N |(W^*_N V_N )(v)| \, |J^{(k)} (\bx'_k ; u,x_2, \dots ,x_k)| \\ &\hspace{3cm} \times \left|\left(W_{N,(1,k+1)}^* \xi^{(k+1)}_{\ell}\right) (u+v/2,x_2, \dots ,x_k,u-v/2)\right|^2  \\ \leq \; &  C_k \, N \tri J^{(k)} \tri \, \| W_N^* V_N \|_1 \left\langle W_{N,(1,k+1)}^* \xi^{(k)}_{\ell} , \left[ (\nabla_1 \cdot \nabla_{k+1})^2 -\Delta_1 - \Delta_{k+1} + 1 \right] W_{N,(1,k+1)}^* \xi^{(k)}_{\ell} \right\rangle \, .
\end{split}\]
Since, by Yajima's bounds (Proposition \ref{prop:waveop}, part v)), $\| W_N^* V_N \|_1 \leq C \| V_N \|_1 \leq C N^{-1}$, it follows that
\[ \text{II} \leq C_k \,  \tri J^{(k)} \tri \,  \left\langle W_{N,(1,k+1)}^* \xi^{(k)}_{\ell} , \left[ (\nabla_1 \cdot \nabla_{k+1})^2 -\Delta_1 - \Delta_{k+1} + 1 \right] W_{N,(1,k+1)}^* \xi^{(k)}_{\ell} \right\rangle \, . \]
Inserting this and (\ref{eq:revI}) into the r.h.s. of (\ref{eq:rev4}), it follows from (\ref{eq:comp31}) after resumming over $\ell$ that 
\begin{equation*}\begin{split}
\Big| \tr \, J^{(k)} \, N\, V_N (x_1 - &x_{k+1})  \gamma_{N,s}^{(k+1)}  \Big| \\
\leq \; & C_k \, \tri J^{(k)} \tri \,  \tr \, \left( (\nabla_1 \cdot \nabla_{k+1})^2 -\Delta_1 - \Delta_{k+1} +1 \right) W_{N,(1,k+1)}^*  \gamma_{N,s}^{(k+1)} W_{N,(1,k+1)}  \\
\leq \; & C_k \, \tri J^{(k)} \tri
\end{split} \end{equation*}
for all $s \in \bR$ (in the last line we used Proposition \ref{prop:apri2}).
Since the same bound remain valid if we replace $x_1$ with an arbitrary $x_j$, $j=2, \dots k$ (and also if the potential lies on the right of the marginal density $\gamma_{N,s}^{(k+1)}$), it follows that
\begin{equation}\label{eq:equi-4}
\sum_{j=1}^k  \Big| \tr \; J^{(k)} \, \left[ N V_N (x_j -x_{k+1}), \gamma_{N,s}^{(k+1)} \right] \Big| \leq C_k \, \tri J^{(k)} \tri \,.
\end{equation}

\medskip

{F}rom (\ref{eq:equi-1}), (\ref{eq:equi-2}), (\ref{eq:equi-3}), and (\ref{eq:equi-4}), it follows that \[ \Big| \tr \; J^{(k)} \left( \gamma_{N,t}^{(k)} - \gamma_{N,r}^{(k)} \right) \Big| \leq C_k \, \tri J^{(k)} \tri \, |t-s|  \, . \]  This implies (\ref{eq:equi02}), and thus the equicontinuity of the sequence $\Gamma_{N,t} = \{ \gamma_{N,t}^{(k)} \}_{k=1}^N$ with respect to the metric $\tau_{\text{prod}}$.

\medskip

The proof of the fact that $\gamma_{\infty,t}^{(k)}$ is symmetric w.r.t. permutations, that it is non-negative and such that $\tr \, \gamma_{\infty,t}^{(k)} \leq 1$ can be found in \cite[Theorem 6.1]{ESY}.
\end{proof}

\section{Higher order a-priori estimates on the limit points $\Gamma_{\infty,t}$.}
\label{sec:hoee}
\setcounter{equation}{0}

The goal of this section is to establish strong a-priori bounds for the limit
 points $\Gamma_{\infty,t}$ of the sequence $\Gamma_{N,t} = \{ \gamma^{(k)}_{N,t} \}_{k=1}^N$.
 As we did in \cite{ESY}, we will obtain strong a-priori estimates on
 $\Gamma_{\infty,t}$ by proving higher order energy estimates, which compare
 the expectation of powers of the Hamiltonian $\langle \psi_N, H^k_N \psi_N \rangle$
 with certain Sobolev norms of the $N$-particle wave function $\psi_N$.
 It turns out that the expectation of powers of the Hamiltonian can only control
 the Sobolev norms of $\psi_N$ in appropriate regions of the configuration space.
 To characterize these regions, we introduce the same cutoffs we used in \cite{ESY}.
For a given length scale $\ell >0$ (in our analysis, we will need that $N^{-1/2}
 \ll \ell \ll N^{-1/3}$), we set
\begin{equation}\label{eq:hdef}
h (x): = e^{-\frac{\sqrt{x^2 + \ell^2}}{\ell}}.
\end{equation}
Note that $h \simeq 0$ if $|x| \gg \ell$, and $h \simeq e^{-1}$ if
$|x| \ll \ell$. For $i=1,\dots ,N$ we define the cutoff function
\begin{equation}\label{eq:thetajdef}
\theta_i (\bx) := \exp \left(-\frac{1}{\ell^{\eps}} \sum_{j\neq i} h
(x_i -x_j)\right)
\end{equation}
for some $\eps >0$. Note that  $\theta_i (\bx)$ is exponentially small if there is
 at least one other particle at distance of order $\ell$ from $x_i$, while
$\theta_i(\bx)$ is exponentially close to 1 if there is no other particle
near $x_i$ (on the length scale $\ell$). Next we define
\begin{equation}
\theta_i^{(n)}(\bx) := \theta_i (\bx)^{2^n} = \exp \left( -\frac{2^n}{\ell^{\eps}}
\sum_{j \neq i} h (x_i -x_j) \right) \,
\end{equation}
and their cumulative versions, for $n, k\in \bN$,
\be\label{eq:thetan}
\Theta_k^{(n)} (\bx) := \theta_1^{(n)} (\bx)
\dots \theta_k^{(n)} (\bx) = \exp \left( -\frac{2^n}{\ell^{\eps}}
\sum_{i\leq k}\sum_{j \neq i} h (x_i -x_j) \right) \; . \ee To cover
all cases in one formula, we introduce the notation
$\Theta_k^{(n)}=1$ for any $k\leq 0$, $n\in \bN$.
Some important properties of the function $\Theta_k^{(n)}$, used throughout
 the proof of Proposition \ref{prop:hk}, are collected, for completeness,
in Lemma \ref{lm:theta}.

\bigskip

\begin{proposition}\label{prop:hk} Suppose that $V \geq 0$, with
$|\nabla^{\alpha} V(x)| \leq C$ for all $|\alpha| \leq 2$. Let
$\psi \in L^2_s (\bR^{3N})$ be a function symmetric in all its variables.
Suppose that $\ell\gg N^{-1/2}$ (in the sense that there exists
$\delta >0$ with $N^{1/2} \ell \geq N^{\delta}$). There exists $C_0 >0$
such that for every integer $k\ge 1$ there exists $N_0 = N_0 (k)$ such that
\be
   \langle\psi, (H_N+N)^k\psi\rangle \ge C_0^k N^k \int \rd \bx \;
\Theta_{k-1}^{(k)} (\bx) \,  |\nabla_1\ldots \nabla_k\psi (\bx)|^2
\label{eq:hk}
\ee
for all $N\ge N_0$.
\end{proposition}

\begin{proof} We use induction over $k$. For $k=1$ the statement follows
directly from $V_N\ge0$,
since on the symmetric subspace
\be
    H_N+N \ge  \sum_{i=1}^N \nabla_i^*\nabla_i = N\nabla_j^*\nabla_j
\label{eq:k=1}
\ee
for any fixed $j=1,2, \ldots, N$.
We will present the $k=2$ case in details and then comment on the general case.
 Set $T=H_N+N\ge 0$
for brevity and use the induction hypothesis

\be\label{eq:for}
\begin{split}
   T^2
 &  \ge C_0N T^{1/2} \nabla_1^*\nabla_1 T^{1/2}  \\ &\ge
C_0N T^{1/2} \nabla_1^*\theta_1^4\nabla_1 T^{1/2} \\
 & \ge  \frac{1}{2}C_0N \nabla_1^* T^{1/2}\theta_1^4 T^{1/2}\nabla_1
 -  C_0N [T^{1/2}, \nabla_1]^*\theta_1^4 [ T^{1/2}, \nabla_1 ] \\
& \ge  \frac{1}{4}C_0N \nabla_1^* \theta_1^2 T\theta_1^2\nabla_1
- C_0N \nabla_1^* [T^{1/2},\theta_1^2]^*[ T^{1/2},\theta_1^2]\nabla_1
 -  C_0N [T^{1/2}, \nabla_1]^*\theta_1^4 [ T^{1/2}, \nabla_1 ].
\end{split}
\ee
In the first term we use that $H_N + N \geq \sum_{j=2}^N \nabla^*_j \nabla_j$ to obtain
\be\label{eq:back}
\begin{split}
   \frac{1}{4}C_0 N \nabla_1^* \theta_1^2 T\theta_1^2\nabla_1
&\ge \frac{1}{4}C_0 N (N-1) \nabla_1^* \theta_1^2\nabla_2^*\nabla_2\theta_1^2\nabla_1  \\
& \ge \frac{1}{8}C_0 N^2 \nabla_1^* \nabla_2^*\theta_1^4\nabla_2\nabla_1
- C_0N^2 \nabla_1^*[ \nabla_2, \theta_1^2]^* [ \nabla_2, \theta_1^2]\nabla_1
\end{split}
\ee
for all $N$ large enough. Since $\Theta_1^{(2)} = \theta_1^4$, we would obtain \eqref{eq:hk}
for $k=2$ with $C_0 < 1/8$ once we show that the commutator
terms in \eqref{eq:for} and \eqref{eq:back} are negligible.

The commutator in \eqref{eq:back} on symmetric functions can be estimated by
\be\label{eq:symme}
\begin{split}
   C_0N^2 \nabla_1^*[ \nabla_2, \theta_1^2]^* [ \nabla_2, \theta_1^2]\nabla_1 & =
   \frac{C_0N^2}{N-1}\sum_{j=2}^N  \nabla_1^* (\nabla_j\theta_1^2)^2\nabla_1\\
 &\leq  O(\ell^{-2}N^{-1}) N^2 \nabla_1^* \nabla_1
 \leq O(\ell^{-2}N^{-1})T^2 = o(1) T^2
\end{split}
\ee
where we used (\ref{eq:lmthetaiv}), recalling that $\theta_1^2 = \Theta_1^{(1)}$, and
we also used $T\ge N$ and \eqref{eq:k=1}.

To estimate the two commutators in \eqref{eq:for}, we express
\be
   [T^{1/2}, A] = \frac{1}{\pi} \int_0^\infty \frac{1}{T+s} \,
\left[A,T \right] \, \frac{1}{T+s} \, s^{1/2}\rd s
\label{eq:com}
\ee
for any operator $A$.

To estimate the first commutator term in \eqref{eq:for} we note that,  by Schwarz inequality,
\be
\begin{split}
  [T^{1/2}, \theta_1^2]^*[T^{1/2}, \theta_1^2] \leq & \; C \,
(\log K)\int_0^K \frac{1}{T+s} \, [\theta_1^2,T]^* \,
 \frac{1}{(T+s)^2} \,
 [\theta_1^2,T] \, \frac{1}{T+s} \, \langle s\rangle^2 \, \rd s \; \\
 &+ C \int_K^\infty \frac{1}{T+s} \, [\theta_1^2,T]^* \, \frac{1}{(T+s)^2} \,
 [\theta_1^2,T] \, \frac{1}{T+s} \, \langle s\rangle^{5/2} \, \rd s,
\end{split}
\label{eq:comm}
\ee
where $K=\exp(N^\e)$ for some $\e>0$. Estimating
$(T+s)^{-2}\leq \langle s\rangle ^{-2}$
(using $T\ge N$), we have
\be
\begin{split}
 N \nabla_1^* [T^{1/2},\theta_1^2]^*[ T^{1/2},\theta_1^2]\nabla_1
  \leq & \; c N^{1+\e} \int_0^K \nabla_1^*
\, \frac{1}{T+s} \, [T,\theta_1^2]^*[T,\theta_1^2]
\, \frac{1}{T+s} \, \nabla_1 \, \rd s \\
&+ c N \int_K^\infty \nabla_1^*
\frac{1}{T+s} \, [T,\theta_1^2]^*[T,\theta_1^2]
\, \frac{1}{T+s} \, \nabla_1 \, \langle s\rangle^{1/2} \, \rd s\, ,
\end{split}
\label{eq:comm1}
\ee
and we can estimate
\be
\begin{split}
   [T,\theta_1^2]^*[T,\theta_1^2]  & = \sum_{i,j}
\big(2\nabla_j^* \cdot (\nabla_j\theta_1^2) + (\Delta_j\theta_1^2)\big)
\big(2 (\nabla_i\theta_1^2)\cdot\nabla_i + (\Delta_i\theta_1^2) \big)\\
&\leq c \sum_{i,j} \Big[ \nabla_j^* \cdot (\nabla_j\theta_1^2)(\nabla_i\theta_1^2)\cdot\nabla_i
  +  \nabla^*_i \cdot |\Delta_j\theta_1^2| \nabla_i +
|\nabla_i\theta_1^2| |\Delta_j\theta_1^2| |\nabla_i\theta_1^2|
+ |\Delta_i\theta_1^2||\Delta_j\theta_1^2|\Big]\\
&\leq  c \sum_{i,j} \nabla^*_i \cdot\big( |\nabla_j\theta_1^2|^2 + |\Delta_j\theta_1^2|\big) \nabla_i
 + c\Big(\sum_i |\Delta_i\theta_1^2|\Big)^2 \label{eq:key} \\
&\leq  c \ell^{-2} \sum_i \nabla_i^*\cdot\nabla_i + c\ell^{-4}
 \leq c\ell^{-2} T
\end{split}
\ee
by using (\ref{eq:lmthetaiv}) and (\ref{eq:lmthetavi}) and the fact that $\theta_1^2 =\Theta_1^{(1)}$.
Thus, the first commutator term in \eqref{eq:for}
 is estimated as
\be
\begin{split}
   N \nabla_1^* [T^{1/2},\theta_1^2]^*[ T^{1/2},\theta_1^2]\nabla_1
\leq \; & O(N^{1+\e}\ell^{-2}) \, \int_0^K \nabla_1^*
\, \frac{1}{T+s} \, T \,
\frac{1}{T+s} \, \nabla_1 \rd s \\
&+  O(N\ell^{-2}) \, \int_K^\infty \nabla_1^* \, \frac{1}{T+s} \, T \,
\frac{1}{T+s} \, \nabla_1 \, \langle s\rangle^{1/2} \, \rd s\\
\leq \; &O(N^{1+\e}\ell^{-2}) \, \nabla_1^*\nabla_1
 + O(N\ell^{-2}K^{-1/2}) \nabla_1^* \, T \, \nabla_1\\
\leq \; &O(N^{\e}\ell^{-2})T + o(1) T^2 \\ \leq \; &o(1) T^2
\end{split}
\ee
if we choose $\e > 0$ so small that $\ell^{-2}\ll N^{1-\e}$.
When estimating the term $\nabla_1^* T\nabla_1$ in the last step,
we could afford estimating any commutators, since $K^{-1/2}$ is exponentially
small:
\be\nonumber
\begin{split}
    \nabla_1^*T\nabla_1  & = \frac{1}{N}\sum_j \nabla_j^* T\nabla_j
   =  \frac{1}{N}\sum_j\Big[ (-\Delta_j) T  - \nabla_j^*\cdot \sum_i(\nabla V_N)(x_j-x_i)\Big] \\
& = N^{-1}T^2 - N^{-1} \sum_{ij} \Big[ V_N(x_i-x_j)T -\nabla_j^*\cdot(\nabla V_N)(x_j-x_i)\Big] \\
&\leq 2N^{-1}T^2 + O(N^8)
\end{split}
\ee
using that $|V_N (x)| \leq C N^2$, and $|\nabla V_N (x)|\leq C N^3$, for all $x \in \bR^3$.

Finally, we estimate the second commutator term in \eqref{eq:for} by using again Schwarz
inequality in \eqref{eq:com} but this time we do not split the integration:
\be
\begin{split}
 [T^{1/2}, &\nabla_1]^* \theta_1^4 [ T^{1/2}, \nabla_1 ]
\\  & \leq c \int_0^\infty \frac{1}{T+s} [\nabla_1, T]^*
\frac{1}{T+s}\theta_1^4 \frac{1}{T+s} [\nabla_1, T]\frac{1}{T+s} \langle s\rangle^{5/2}\rd s \\
& \leq cN\sum_{i\neq 1} \int_0^\infty \frac{1}{T+s} (\nabla V_N)(x_1-x_i)
\frac{1}{T+s}\theta_1^4 \frac{1}{T+s}  (\nabla V_N)(x_1-x_i)\frac{1}{T+s} \langle s\rangle^{5/2}\rd s
\end{split}
\label{eq:seccomm}
\ee
where we used  $[\nabla_1, T] =\sum_{i\neq 1} (\nabla V_N)(x_1-x_i)$.

Since $T_0= \sum_j -\Delta_j + N$ is a positivity preserving operator and
$V \geq 0$, $T$ is also positivity preserving and its resolvent kernel satisfies
$$
      \frac{1}{T+s} (\bx ; \by) \leq \frac{1}{ T_0 + s}(\bx ; \by),
$$
and thus
\be \label{eq:TT}
\begin{split}
I:= \Big\| (\nabla V_N)(x_1-x_i)
&\frac{1}{T+s}\, \theta_1^4 \, \frac{1}{T+s}  (\nabla V_N)(x_1-x_i) \Big\| \\
& \leq  \Big\| \left|(\nabla V_N)(x_1-x_i)\right|
\frac{1}{T_0+s} \, e^{-4\ell^{-\e} h(x_1-x_i)} \, \frac{1}{T_0+s} \, \left|(\nabla V_N)(x_1-x_i)\right| \Big\| \, ,
\end{split}
\ee
where we also estimated $\theta_1$ by keeping only one summand in
its definition (\ref{eq:thetajdef}). Introducing the variable $y=x_1 -x_i$,
and observing that
$L^2 (\bR^{3N}, \rd \bx) \simeq L^2 (\bR^3, \rd y ; L^2 (\bR^{3(N-1)} ,
\rd z \rd x_2 \dots \widehat{\rd x_i} \dots \rd x_N))$ (where the hat means
that the variable $x_i$ is omitted), we obtain that
\begin{equation*}\begin{split}
I \leq \; &\sup_{M \geq 0} \Big\| |(\nabla V_N)(y)| \,
\frac{1}{-\Delta_y + M +N+s} e^{-4\ell^{-\e} h(y)} \frac{1}{\Delta_y + M +N+s}
\, | (\nabla V_N)(y)| \Big\| \\ \leq \; & \sup_{M \geq 0} \Big\| |(\nabla V_N)(y)|
\frac{1}{-\Delta_y + M +N+s} e^{-4\ell^{-\e} h(y)} \frac{1}{\Delta_y + M +N+s}
 |(\nabla V_N)(y)| \Big\|_{\text{HS}}
\end{split}
\end{equation*}
where the norms on the last two lines are, respectively, the operator norm and
the Hilbert-Schmidt norm of an operator over $L^2 (\bR^3, \rd y)$. The last equation implies that
\be\label{eq:I}
\begin{split}
I\leq \; & \int \rd y \rd y' e^{-4\ell^{-\e}h(y)} \Big|\frac{1}{\Delta + M +N+s}(y,y')\Big|^2
|\nabla V_N(y')|^2 \\ \leq \; &\int \rd y \rd y' e^{-4\ell^{-\e}h(y)}
\frac{e^{-2\sqrt{N+s}|y-y'|}}{|y-y'|^2}
|\nabla V_N(y')|^2  \\  \le \; & O (e^{-\ell^{-\e}})
\end{split}
\ee
since $h\approx e^{-1}$ on the support of $V_N$. We will use this bound for $s\leq K:=
 \exp(c\ell^{-\e})$ with a sufficiently small $c>0$. From
 \eqref{eq:TT}, we also have the trivial bound
\be
    I \leq \frac{N^6}{\langle s \rangle^2}
\label{eq:trivi}
\ee
that will be used for large $s$. Inserting these estimates into \eqref{eq:seccomm}, we have
$$
[T^{1/2}, \nabla_1]^*\theta_1^4 [ T^{1/2}, \nabla_1 ]  \leq
 O(N^2e^{-\ell^{-\e}} ) \int_0^K \frac{\langle s\rangle^{5/2}\rd s}{(T+s)^2}
+ O(N^8) \int_K^\infty \frac{\langle s\rangle^{1/2}\rd s}{(T+s)^2} = O(e^{-c\ell^{-\e}} ),
$$
i.e. this commutator term is subexponentially small in $N$ and this completes the proof of
\eqref{eq:hk} for $k=2$.

\bigskip

The proof for general $k>2$ follows the same pattern as for $k=2$.
Introduce the notation
$$
   D_k = \nabla_1\nabla_2 \ldots \nabla_k\, .
$$
We recall the summation convention: for any operator $A$, we denote
$$
   D_k^*AD_k := \sum_{\alpha_1=1}^3\ldots \sum_{\alpha_k=1}^3  \nabla_{x_{1,\alpha_1}}^*
   \ldots  \nabla_{x_{k,\alpha_k}}^* A \nabla_{x_{k,\alpha_k}}\ldots\nabla_{x_{1,\alpha_1}}
$$
where $x_j = (x_{j,1}, x_{j,2}, x_{j,3})$ are the three coordinates of $x_j\in \bR^3$.

Using the induction hypothesis,  $\Theta_{k-1}^{(k)} \geq [\Theta_{k}^{(k)}]^2=\Theta_k^{(k+1)}$
and \eqref{eq:k=1} we obtain,  similarly to  \eqref{eq:for} and \eqref{eq:back},
\be
\begin{split}
    T^{k+1} &\ge C_0^{k}N^{k} T^{1/2} D_{k}^* \Theta_{k-1}^{(k)} D_k T^{1/2}
   \\ &\ge C_0^{k}N^{k} T^{1/2} D_{k}^* [\Theta_{k}^{(k)}]^2 D_k T^{1/2}\\
   &\ge \frac{1}{8} C_0^k N^{k+1} D_{k+1}^* \Theta_{k}^{(k+1)} D_{k+1}
   - C_0^kN^k  D_k^* [T^{1/2}, \Theta_{k}^{(k)}]^*[T^{1/2}, \Theta_{k}^{(k)}] D_k \\
   &\;\;\;\; -  C_0^kN^k  [T^{1/2}, D_k]^*\Theta_{k}^{(k+1)} [T^{1/2},  D_k]
  - C_0^kN^{k+1} D_k^* [\nabla_{k+1}, \Theta_{k}^{(k)}]^*[\nabla_{k+1}, \Theta_{k}^{(k)}]D_k ,
\end{split} \label{eq:Tk+1}
\ee
for all $N$ sufficiently large (depending on $k$). The first term gives the
desired result if $C_0<1/8$; in the sequel we show that all three commutator
terms are negligible.

The first commutator in \eqref{eq:Tk+1} is estimated exactly
as the first commutator in \eqref{eq:for}, after replacing
$\theta_1^2 = \Theta_1^{(1)}$ with $\Theta_k^{(k)}$.  The estimates
\eqref{eq:comm} are  \eqref{eq:comm1}
are identical for $k>1$ as well. In the key estimate \eqref{eq:key}, the
only properties we used about $\theta_1^2= \Theta_1^{(1)}$
from Lemma \ref{lm:theta} were those that hold for $\Theta_k^{(k)}$ as well.

The last commutator in \eqref{eq:Tk+1}
can be estimated similarly to \eqref{eq:symme} by using (\ref{eq:lmthetaiv})
\be\label{eq:symmek}
\begin{split}
   C_0^kN^{k+1} D_k^*[ \nabla_{k+1}, \Theta_k^{(k)}]^* [ \nabla_{k+1}, \Theta_k^{(k)}]D_k & =
   \frac{C_0^kN^{k+1}}{N-k}\sum_{j=k+1}^N  D_k^* (\nabla_j\Theta_k^{(k)})^2D_k  \\
 &\leq  O(\ell^{-2}N^{-1}) N^{k+1} D_k^* \Theta_k^{(k-1)}  D_k \\
& \leq O_k(\ell^{-2}N^{-1}) N T^k = o_k(1) T^{k+1}
\end{split}
\ee
by the induction hypothesis and $T\ge N$ (here we use the notation $f = o_k (g)$
if $f/ g \to 0$ as $N \to \infty$ for fixed $k$; analogously for $f=O_k (g)$).

Finally, the estimate of
the second commutator in \eqref{eq:Tk+1} is  similar to that of the second
commutator in \eqref{eq:for}, but more commutators need to be computed.
Similarly to  \eqref{eq:seccomm} and taking the permutation symmetry into account,
we have
\be
\begin{split}
 &[T^{1/2}, D_k]^* \Theta_{k}^{(k+1)} [T^{1/2},  D_k] \\
& \leq C_kN \sum_{i\neq k} \int_0^\infty \frac{1}{T+s} D_{k-1}^*(\nabla V_N)(x_k-x_i)
\frac{1}{T+s}\Theta_{k}^{(k+1)} \frac{1}{T+s}  (\nabla V_N)(x_k-x_i)D_{k-1} \frac{1}{T+s}
 \langle s\rangle^{5/2}\rd s \\
& \quad + C_k \int_0^\infty \frac{1}{T+s} D_{k-2}^*(\nabla^2 V_N)(x_k-x_{k-1})
\frac{1}{T+s}\Theta_{k}^{(k+1)}\\ &\hspace{7cm} \times  \frac{1}{T+s}
 (\nabla^2 V_N)(x_k-x_{k-1})D_{k-2} \frac{1}{T+s}
 \langle s\rangle^{5/2}\rd s
\end{split}
\label{eq:seccommk}
\ee

We will need the following lemma whose proof is postponed.

\begin{lemma}\label{lemma:exp}  Let $\psi \in L^2_s (\bR^{3N})$ be a function
symmetric in all its variables
and let $\delta>0$. Choose a strictly
 increasing sequence of positive constants $\{ c_k \}_{k \geq 1}$.
Then for every integer $k\ge 1$ there exists $N_0 = N_0 (k,\delta)$ such that
\be
\langle\psi, (H_N+N)^k\psi\rangle \ge e^{-c_k N^\delta}
\int  \rd \bx \, |\nabla_1\ldots \nabla_k\psi (\bx)|^2
\label{eq:hk1}
\ee
for all $N\ge N_0$.
\end{lemma}

We demonstrate the estimate of the first term in (\ref{eq:seccommk}),
 the second one is similar. Using
$\Theta_k^{(k+1)} \leq  e^{-\ell^{-\e}h(x_k-x_i)}$, we obtain, similarly
to \eqref{eq:TT}--\eqref{eq:trivi} that
$$
   I: = \Bigg\| (\nabla V_N)(x_k-x_i)
\frac{1}{T+s}\Theta_{k}^{(k+1)} \frac{1}{T+s}  (\nabla V_N)(x_k-x_i)\Bigg\| \leq O(e^{-\ell^{-\e}})
$$
and also
$$
    I\leq \frac{N^6}{\langle s\rangle^2}.
$$
Let  $K:=
 \exp(c\ell^{-\e})$ with a sufficiently small $c>0$.
Choosing a sufficiently small $\delta$, so that $N^\delta \ll \ell^{-\e}$, by
using \eqref{eq:hk1},
 we have,
\be\label{eq:nablaVterm}
\begin{split}
\int_0^\infty \frac{1}{T+s}  & D_{k-1}^*(\nabla V_N)(x_k-x_i)
\frac{1}{T+s}\Theta_{k}^{(k+1)} \frac{1}{T+s}  (\nabla V_N)(x_k-x_i)D_{k-1} \frac{1}{T+s}
 \langle s\rangle^{5/2}\rd s \\
&\leq O(e^{-\ell^{-\e} +c_{k-1}N^\delta})
 \int_0^K \frac{T^{k-1}}{(T+s)^2}\langle s\rangle^{5/2}\rd s
  + O(e^{c_{k-1}N^\delta})\int_K^\infty \frac{T^{k-1}}{(T+s)^2}\frac{N^6}{\langle s\rangle^2}
\langle s\rangle^{5/2}\rd s \\
&\leq   O(e^{-c'\ell^{-\e}})T^{k-1} \leq  o(1) T^{k+1}.
\end{split}
\ee
This completes the proof of Proposition \ref{prop:hk}.
\end{proof}

\begin{proof}[Proof of Lemma \ref{lemma:exp}]
We proceed by a step two induction on $k$; for $k=1$
the claim follows from \eqref{eq:k=1}. We now consider the $k=2$ case.
Similarly to \eqref{eq:h1}, but keeping also the $h_j^2$ terms in the
expansion of $H_N^2$, we find
\be\label{eq:lm62}
\begin{split}
  T^2 & \ge N (N-1) \Big( -\Delta_1 + \frac{1}{2} V_N(x_1-x_2)\Big)
  \Big( -\Delta_2 + \frac{1}{2} V_N(x_1-x_2)\Big)  + N
\Big( -\Delta_1 + \frac{1}{2} V_N(x_1-x_2)\Big)^2 \\
 & \ge (N^2/2) \Big( D_2^*D_2 - 2 \nabla_1^*\nabla_1 -
 2 \nabla_2^*\nabla_2 - 4\| \nabla V_N\|_\infty^2
\Big) + N\Big( (\nabla_1^*\nabla_1)^2  -  2 \nabla_1^*\nabla_1- 4\| \nabla V_N\|_\infty^2
\Big)\\
&\ge (N^2/2) D_2^*D_2 + N (\nabla_1^*\nabla_1)^2  - CN^2 \nabla_1^*\nabla_1 - CN^8 \\
&\ge  (N^2/2) D_2^*D_2 - O(N^8).
\end{split}
\ee
Combining this bound with $T^2 \geq N^2$, it follows that
\be
    T^2\ge cN^{-4}D_2^*D_2
\label{eq:T}
\ee
for a sufficiently small positive $c$.

Now we show how to go from $k$ to $k+2$.
By the induction hypothesis, we have
\be
   T^{k+2} \ge e^{-c_kN^\delta}TD_k^* D_k T \ge
   e^{-c_kN^\delta}\Big( \frac{1}{2} D_k^* T^2 D_k - 2 [D_k, T]^*[D_k, T]\Big).
\label{eq:Tk+2}
\ee
In the first term we can use \eqref{eq:T} in the form
$T^2\ge cN^{-4} \nabla_{k+1}^*\nabla_{k+2}^*\nabla_{k+2}\nabla_{k+1}$,
which holds for all $N$ large enough (because of the factors $D_k$,
we only have symmetry on the last $N-k$ variables; this means that
instead of (\ref{eq:lm62}), we are going to obtain
$T^2 \geq (N-k) (N-k-1) \Delta_{k+1} \Delta_{k+2}
\geq (N^2/2) \Delta_{k+1} \Delta_{k+2}$ for all $N$ large enough).

The commutator term, after several Schwarz inequalities, can be estimated as
\be\label{eq:commm}
\begin{split}
    [D_k, T]^*[D_k, T] & \leq C_k \Big( N^2  D_{k-1}^* \|\nabla V_N \|^2 D_{k-1} +
    D_{k-2}^* \|\nabla^2 V_N \|^2 D_{k-2}\Big) \\
  &\leq C_k N^8 \Big(  D_{k-1}^*  D_{k-1} +  D_{k-2}^*  D_{k-2}\Big)
 \\ &\leq C_k N^8 \, e^{c_{k-1} N^\delta}T^{k-1}
 \leq C_k N^8 \, e^{c_{k-1} N^\delta}T^{k+2} \, ,
\end{split}
\ee
where we used the induction hypothesis for $k-1$ and $k-2$ and, by convention, $D_m=1$
for $m\leq 0$. Inserting this estimate into \eqref{eq:Tk+2}, we obtain
$$
    T^{k+2} \ge cN^{-4} e^{-c_kN^\delta} D_{k+2}^*D_{k+2} -
C_k N^8 e^{-(c_k-c_{k-1})N^\delta} T^{k+2} \; .
$$
Since $c_k$ is strictly increasing, we obtain \eqref{eq:hk1} for $k+2$.

Actually the proof shows that a sufficiently large  $k$-dependent negative power,
$N^{-\beta_k}$,
would suffice on the r.h.s. of \eqref{eq:hk1} instead of the subexponentially small prefactor.
\end{proof}

\bigskip

The higher order energy estimates proved in Proposition \ref{prop:hk} are used
to show the following strong a-priori estimates on the limit points
$\Gamma_{\infty,t} = \{ \gamma^{(k)}_{\infty,t} \}_{k \geq 1}$ of the sequence $\Gamma_{N,t}$.

\begin{theorem}\label{thm:aprik}
Suppose that the assumptions of Theorem \ref{thm:main2} are satisfied
and fix $T>0$. Assume moreover that $\Gamma_{\infty,t}^{(k)} =
\{ \gamma^{(k)}_{\infty,t} \}_{k \geq 1} \in \bigoplus_{k \geq 1} C([0,T], \cL^1_k)$
is a limit point of the sequence $\Gamma_{N,t} = \{ \gamma^{(k)}_{N,t} \}_{k=1}^N$
 with respect to the topology $\tau_{\text{prod}}$. Then
\begin{equation}\label{eq:aprik}
\tr \; (1-\Delta_1) \dots (1-\Delta_k) \gamma^{(k)}_{\infty,t} \leq C^k
\end{equation}
for all $k \geq 1$ and $t \in [0,T]$.
\end{theorem}

\begin{proof}
Theorem \ref{thm:aprik} follows from the higher order energy estimates of
Theorem \ref{prop:hk};
the proof of this fact can be found in \cite[Proposition 6.3]{ESY}.
\end{proof}

\section{Convergence to the infinite hierarchy.}
\setcounter{equation}{0}

In order to prove Theorem \ref{thm:main2}, we need to prove the convergence of
the BBGKY hierarchy towards a hierarchy of infinitely many equations.
In the argument, we will make use of the apriori bounds from Propositions \ref{prop:apri2}
 and Theorem \ref{thm:aprik} for $k=2$.

\begin{theorem}\label{thm:conv}
Suppose that the assumptions of Theorem \ref{thm:main2} are satisfied and fix $T >0$.
  Suppose that $\Gamma_{\infty,t} = \{ \gamma^{(k)}_{\infty,t} \}_{k \geq 1} \in
\bigoplus_{k \geq 1} C([0,T] , \cL_k^1)$ is a limit point of $\Gamma_{N,t} =
\{ \gamma_{N,t}^{(k)} \}_{k =1}^N$ with respect to the topology $\tau_{\text{prod}}$.
 Then $\Gamma_{\infty,t}$ is a solution to the infinite hierarchy
\begin{equation}\label{eq:conv}
\gamma^{(k)}_{\infty,t} = \cU^{(k)} (t) \gamma^{(k)}_{\infty,0} -
8 \pi a_0 i \sum_{j=1}^k \int_0^t \rd s \, \cU^{(k)} (t-s) \tr_{k+1}
\left[ \delta (x_j - x_{k+1}), \gamma_{\infty,s}^{(k+1)} \right]
\end{equation}
with initial data $\gamma_{\infty,0}^{(k)} = |\ph \rangle \langle \ph|^{\otimes k}$.
Here $\cU^{(k)} (t)$ denotes the free evolution of $k$ particles defined in~(\ref{eq:Uk}).
\end{theorem}

\begin{proof}
Fix $k \geq 1$. Passing to an appropriate subsequence, we can assume that,
 for every $J^{(k)} \in \cK_k$,
\begin{equation}\label{eq:conv-0}
\sup_{t \in [0,T]} \, \tr \; J^{(k)} \, \left( \gamma_{N,t}^{(k)} -
\gamma_{\infty,t}^{(k)} \right) \to 0 \qquad \text{as } N \to \infty\,.
\end{equation}
We will prove (\ref{eq:conv}) by testing the limit point against a certain
 class of observables, that is dense in $\cK_k$. To characterize the class of
observables we are going to consider, we define, for an arbitrary integer $k \geq 1$,
$$
\Omega_k : =  \prod_{j=1}^k \left( \la x_j \ra + \la i \nabla_j \ra \right)   \; .
$$
We will consider $J^{(k)} \in \cK_k$ such that
\begin{equation}\label{eq:assJ} \Big\| \Omega_k^7
J^{(k)}\Omega_k^7 \Big\|_{\text{HS}} < \infty ,
\end{equation}
where  $\| A \|_{\text{HS}}$ denotes the Hilbert-Schmidt norm of the
operator $A$. Note that the set of observables $J^{(k)}$ satisfying the
condition (\ref{eq:assJ}) is a dense subset of $\cK_k$. Moreover, using
the fact that $e^{i\Delta_j t} \la x_j \ra e^{-i\Delta_j t} = \la x_j - i t \nabla_j \ra$,
 it follows that
\begin{equation}\label{eq:UkHS}
\left\| \Omega_k^7 \, \cU^{(k)} (t) J^{(k)} \Omega_k^7 \right\|_{\text{HS}}
 \leq C \, (1 + |t|)^7 \left\| \Omega_k^7 \, J^{(k)} \Omega_k^7 \right\|_{\text{HS}} \,.
\end{equation}
Note also that, with the norm $\tri J^{(k)} \tri$ defined in (\ref{eq:tri}), we have
\begin{equation}\label{eq:triJj} \tri J^{(k)} \tri \leq C_k
\left\| \Omega_k^7 \, J^{(k)} \Omega_k^7 \right\|_{\text{HS}} \, \end{equation}
for a constant $C_k$ only depending on $k$ (see \cite{ESY}, Eq. (7.8)).
Combining (\ref{eq:UkHS}) with (\ref{eq:triJj}),
we also have
\begin{equation}\label{eq:triJjt}
\tri \cU^{(k)} (t) J^{(k)} \tri \leq C_k \, (1 + |t|)^7 \,
\left\| \Omega_k^7 \, J^{(k)} \Omega_k^7 \right\|_{\text{HS}}.
\end{equation}

\medskip

In order to prove Theorem \ref{thm:conv} it is enough to show that, for every
$J^{(k)} \in \cK_k$ satisfying (\ref{eq:assJ}),
\begin{equation}\label{eq:conv-1}
\tr \, J^{(k)}  \gamma_{\infty,0}^{(k)} = \tr \, J^{(k)} |\ph
\rangle \langle \ph|^{\otimes k}
\end{equation}
and
\begin{equation}\label{eq:conv-2}
\begin{split}
\tr \; J^{(k)}  \gamma_{\infty,t}^{(k)} = \tr \; J^{(k)} \cU^{(k)}
(t)  \gamma_{\infty,0}^{(k)} -8\pi a_0 i \sum_{j=1}^k \int_0^t \rd s
\tr \, J^{(k)} \cU^{(k)} (t-s) \left[ \delta (x_j -x_{k+1}),
\gamma^{(k+1)}_{\infty,s} \right]\,
\end{split}
\end{equation}
for all $t \in [0,T]$.

\medskip

The relation (\ref{eq:conv-1}) follows from the assumption
(\ref{eq:init}) and from (\ref{eq:conv-0}).

\medskip

In order to prove (\ref{eq:conv-2}), we fix $t \in [0,T]$, we
rewrite the BBGKY hierarchy (\ref{eq:BBGKY}) in integral form and we
test it against the observable $J^{(k)}$. We obtain
\begin{equation}\label{eq:conv-4}
\begin{split}
\tr \; J^{(k)} \, \gamma_{N,t}^{(k)} = \; & \tr \; J^{(k)} \,
\cU^{(k)} (t) \gamma_{N,0}^{(k)} - i \sum_{i<j}^k \int_0^t  \rd s \,
\tr \; J^{(k)} \, \cU^{(k)}
(t-s) [ V_N (x_i -x_j), \gamma_{N,s}^{(k)} ] \\
& -i (N-k) \sum_{j=1}^k \int_0^t \rd s \, \tr J^{(k)} \cU^{(k)}
(t-s) [ V_N (x_j -x_{k+1}) , \gamma_{N,s}^{(k+1)} ] \,.
\end{split}
\end{equation}
{F}rom (\ref{eq:conv-0}) it follows immediately that
\begin{equation}\label{eq:conv-4-1}
\tr \; J^{(k)} \, \gamma_{N,t}^{(k)} \to \tr \; J^{(k)}
\gamma_{\infty,t}^{(k)}
\end{equation}
and also that, as $N \to \infty$,
\begin{multline}\label{eq:conv-4-2}
\tr \; J^{(k)} \,\cU^{(k)} (t)  \gamma_{N,0}^{(k)} =  \tr \;
\left(\cU^{(k)} (-t) J^{(k)}\right) \,  \gamma_{N,0}^{(k)} \\ \to
\tr\;\left(\cU^{(k)} (-t) J^{(k)}\right)  \gamma_{\infty,0}^{(k)}
=\tr \; J^{(k)} \,\cU^{(k)} (t)  \gamma_{\infty,0}^{(k)}\,.
\end{multline}
Here we used that, if $J^{(k)} \in \cK_k$, then
also $\cU^{(k)} (-t) J^{(k)} \in \cK_k$.

\medskip

Next we consider the second term on the r.h.s. of (\ref{eq:conv-4})
and we prove that it converges to zero, as $N \to \infty$. To this end, we note that, setting $J_{t}^{(k)} = \cU^{(k)} (t) J^{(k)}$, we have  \[
\sum_{i<j}^k \, \tr \; J^{(k)} \, \cU^{(k)}
(t-s) [ V_N (x_i -x_j), \gamma_{N,s}^{(k)} ] = \sum_{i<j}^k \,
\tr \;  J_{s-t}^{(k)} \, [ V_N (x_i -x_j), \gamma_{N,s}^{(k)} ]
\] and therefore, from (\ref{eq:equi-3}) and (\ref{eq:triJjt}), we obtain that
\[ \sum_{i<j}^k \, \Big| \tr \; J^{(k)} \, \cU^{(k)}
(t-s) [ V_N (x_i -x_j), \gamma_{N,s}^{(k)} ] \Big| \leq \frac{C_k \, \tri
J_{s-t}^{(k)} \tri}{N} \leq \frac{C_k (1 +T^7) \| \Omega_k^7 J^{(k)}
 \Omega_k^7 \|_{\text{HS}}}{N} \]
for all $0 \leq s \leq t \leq T$. This implies that for all $k \in \bN$,
for all $t \in [0,T]$ and for all $J^{(k)}$ such that (\ref{eq:assJ}) is true,
\begin{equation}\label{eq:conv-5ba}
\sum_{i<j}^k \int_0^t  \rd s \,
\tr \; J^{(k)} \, \cU^{(k)}
(t-s) \left[ V_N (x_i -x_j), \gamma_{N,s}^{(k)} \right] \to 0
\end{equation}
as $N \to \infty$.

\medskip

Finally we consider the last term on the r.h.s. of
(\ref{eq:conv-4}). First of all, we observe that, for every $k \in \bN$,
$t\in [0,T]$ and $J^{(k)} \in \cK_k$ such that (\ref{eq:assJ}) is satisfied, we have
\begin{equation}\label{eq:conv-5b}
k \sum_{j=1}^k \int_0^t \rd s \, \tr \, J^{(k)} \cU^{(k)} (t-s) \left[
V_N (x_j -x_{k+1}) , \gamma_{N,s}^{(k+1)}\right] \to 0
\end{equation}
as $N \to \infty$. This follows (similarly to (\ref{eq:conv-5ba}))
from (\ref{eq:equi-4}), and from the bound (\ref{eq:triJjt}).

\medskip

It remains to show that, for every fixed $k \in \bN$, $t \in [0,T]$,
and for every $J^{(k)} \in\cK_k$ with (\ref{eq:assJ})
\begin{equation}\label{eq:conv-6}
\begin{split}
N \sum_{j=1}^k \int_0^t \rd s \, &\tr \, J^{(k)} \cU^{(k)} (t-s) [
V_N (x_j -x_{k+1}) , \gamma_{N,s}^{(k+1)} ] \\ &\to 8\pi a_0 \sum_{j=1}^k
\int_0^t \rd s \, \tr \; J^{(k)} \cU^{(k)} (t-s) [ \delta (x_j - x_{k+1}) , \gamma_{\infty,s}^{(k+1)} ]
\end{split}
\end{equation}
as $N \to \infty$. To prove (\ref{eq:conv-6}), we fix $s \in [0,t]$, and we
 consider, for example, the contribution with $j=1$. We write
\begin{equation}\label{eq:715}
\begin{split} \tr \; J^{(k)} \cU^{(k)} (t-s) N &V_N (x_1 -x_{k+1})
 \gamma^{(k+1)}_{N,s} \\ &= \tr \; J_{s-t}^{(k)}
 \, N V_N (x_1 -x_{k+1}) W_{N,(1,k+1)}  W_{N,(1,k+1)}^* \gamma^{(k+1)}_{N,s} ,
\end{split}\end{equation}
where $W_{N,(1,k+1)}$ denotes the wave operator associated with the
Hamiltonian $-\Delta + (1/2) V_N$ acting only on the variable $x_{k+1} - x_1$
 (as defined in (\ref{eq:Wij})). Therefore, if we choose a probability density
 $h \in L^1 (\bR^3)$, with $h \geq 0$, $\int \rd x h(x) = 1$, and we denote
$h_{\alpha} (x) = \alpha^{-3} h (x/\alpha)$ for all $\alpha >0$, we have
\begin{equation}\label{eq:NVN-term}
\begin{split}
\Big| \tr \; J^{(k)} &\cU^{(k)} (t-s) N V_N (x_1 -x_{k+1}) \gamma^{(k+1)}_{N,s} -
8 \pi a_0  \tr \; J^{(k)} \cU^{(k)} (t-s) \delta (x_1 -x_{k+1}) \gamma^{(k+1)}_{\infty,s}
\Big|    \\ \leq \; & \Big| \tr \;  J_{s-t}^{(k)} \,
\left[ N V_N (x_1 -x_{k+1}) W_{N,(1,k+1)} - 8\pi a_0 \, \delta (x_1 - x_{k+1}) \right]
W^*_{N,(1,k+1)} \gamma^{(k+1)}_{N,s} \Big| \\&+ 8 \pi a_0 \, \Big| \tr \;
 J_{s-t}^{(k)} \left[ \delta (x_1 -x_{k+1}) - \,
 h_{\alpha} (x_1 - x_{k+1}) \right] W^*_{N,(1,k+1)} \gamma^{(k+1)}_{N,s} \Big|\\
&+ 8 \pi a_0  \, \Big| \tr\; J_{s-t}^{(k)} \,
 h_{\alpha} (x_1 - x_{k+1}) \, (W_{N,(1,k+1)}^* -1) \gamma^{(k+1)}_{N,s}
\Big| \\ &+ 8 \pi a_0  \, \Big| \tr\; J_{s-t}^{(k)} \,
 h_{\alpha} (x_1 - x_{k+1}) \, \left( \gamma^{(k+1)}_{N,s} -
\gamma^{(k+1)}_{\infty,s} \right) \Big| \\ &+ 8 \pi a_0  \, \Big| \tr\;
J_{s-t}^{(k)} \; \left[ h_{\alpha} (x_1 - x_{k+1})
 -\delta (x_1 -x_{k+1}) \right] \, \gamma^{(k+1)}_{\infty,s} \Big|.
\end{split}
\end{equation}
Here we insert the wave operator $W_{N, (1,k+1)}$, because we only have a-priori bounds on the quantity $W^*_{N,(1,k+1)} \gamma_{N,s}^{(k+1)}$ . Then we replace $NV_N (x_1 -x_{k+1}) W_{N,(1,k+1)}$ by $8 \pi a_0 \delta (x_1 -x_{k+1})$. Afterwards, in order to remove the inverse wave operator $W^*_{N, (1,k+1)}$ and to take the limit $\gamma_{N,s}^{(k+1)} \to \gamma_{\infty,s}^{(k+1)}$, we need to replace the delta-function by the bounded potential $h_{\alpha}$ independent of $N$. At the end, $h_{\alpha}$ is changed back to the $\delta$-function.

\medskip

In Lemma \ref{lm:term1} and Lemma \ref{lm:term2} we prove that, for every
$k \in \bN$, for every $0 \leq s \leq t \leq T$, and for every $J^{(k)}
\in \cK_k$ with (\ref{eq:assJ}),
\begin{equation}\label{eq:tm1to0} \Big| \tr \;   J_{s-t}^{(k)} \,
 \left[ N V_N (x_1 -x_{k+1}) W_{N,(1,k+1)} - 8\pi a_0 \, \delta (x_1 - x_{k+1}) \right]
 W^*_{N,(1,k+1)} \gamma^{(k+1)}_{N,s} \Big| \to 0 \end{equation} as $N \to \infty$
 and that \begin{equation}\label{eq:tm2to0} \Big| \tr \;
 J_{s-t}^{(k)} \, \left[ \delta (x_1 -x_{k+1}) - \, h_{\alpha} (x_1 - x_{k+1}) \right]
 W^*_{N,(1,k+1)} \gamma^{(k+1)}_{N,s} \Big| \to 0 \end{equation} as $\alpha \to 0$,
 uniformly in $N$.

\medskip

As for the third term on the r.h.s. of (\ref{eq:NVN-term}) we remark that,
for fixed $k \in \bN$, $s \in [0,T]$, $J^{(k)} \in \cK_k$ and $\alpha >0$,
\begin{equation}\label{eq:WNto1}
\Big|  \tr \; J_{s-t}^{(k)} \, h_{\alpha} (x_1 -x_{k+1})
(W_{N,(1,k+1)}^* -1 ) \gamma^{(k+1)}_{N,s} \Big| \to 0
\end{equation}
as $N \to \infty$. In fact, for the bounded operator $A=
J_{s-t}^{(k)}\, h_{\alpha} (x_1 -x_{k+1}) $,
we can use the spectral decomposition $\gamma^{(k+1)}_{N,s} =
 \sum_j \lambda_j |\xi^{(k+1)}_j \rangle \langle \xi^{(k+1)}_j|$
with $\sum_j \lambda_j =1$, $\lambda_j >0$, $\| \xi^{(k+1)}_j \|=1$,
and estimate
\begin{equation}
\begin{split}
   \Big |\tr \; A  \, (W_{N,(1,k+1)}^* -1 ) \, \gamma^{(k+1)}_{N,s} \Big|
 & \leq \| A \| \sum_j \lambda_j \|
   (W_{N,(1,k+1)}^* -1 ) \xi^{(k+1)}_j\|^2 \cr
   &\leq C\|A \| N^{-1/3} \; \tr (1-\Delta_1 - \Delta_{k+1})\gamma^{(k+1)}_{N,s} \cr
  &\leq  C\|A \|N^{2/3}\langle \psi_{N, s}, (H_N + N)\psi_{N, s}\rangle
\leq C\| A\|N^{-1/3}
\end{split}
\end{equation}
by the energy conservation and \eqref{eq:assH1}. From the first to
the second line we used
Lemma \ref{lm:WNto1}. Since the operator $A$ is bounded for any fixed
 $J^{(k)}$ and $\alpha>0$,
we obtain \eqref{eq:WNto1}.

\medskip

To control the fourth term on the r.h.s. of (\ref{eq:NVN-term}) we observe that, for arbitrary $\delta >0$,
\begin{equation}
\begin{split}
\tr\; J_{s-t}^{(k)} \, h_{\alpha} &(x_1 - x_{k+1}) \,
\left( \gamma^{(k+1)}_{N,s} - \gamma^{(k+1)}_{\infty,s} \right) \\ = \;
&\tr \; J_{s-t}^{(k)} \, h_{\alpha} (x_1 -x_{k+1})
\frac{1}{1+ \delta (1-\Delta_{k+1})^{1/2}} \left( \gamma^{(k+1)}_{N,s} -
\gamma_{\infty,s}^{(k+1)}\right) \\
&+ \tr \; J_{s-t}^{(k)} \, h_{\alpha} (x_1 -x_{k+1})
\left(1-\frac{1}{1+ \delta ( 1-\Delta_{k+1})^{1/2}}\right) \left(\gamma^{(k+1)}_{N,s}
-\gamma^{(k+1)}_{\infty,s}\right)\,.
\end{split}
\end{equation}
The first term on the r.h.s. of the last equation converges to zero, as $N \to \infty$,
for every fixed $\delta,\alpha >0$; this follows from the assumption (\ref{eq:conv-0})
and from the observation that $J_{s-t}^{(k)} h_{\alpha} (x_1 - x_{k+1})
(1+ \delta (1-\Delta_{k+1}))^{-1}$ is a compact operator on $L^2 (\bR^{3(k+1)})$.
As for the second term, we notice that it can be bounded by
\begin{equation}
\begin{split}
\Big| \tr \; J_{s-t}^{(k)} \, h_{\alpha} (x_1 -x_{k+1}) &
\left(1-\frac{1}{1+ \delta \left( 1-\Delta_{k+1})^{1/2}\right)}\right)
\left(\gamma^{(k+1)}_{N,s}-\gamma^{(k+1)}_{\infty,s}\right) \Big| \\ & \leq \delta
 \| J^{(k)} \| \, \| h_{\alpha} \|_{\infty} \tr \Big| (1-\Delta_{k+1})^{1/2}
\left( \gamma^{(k+1)}_{N,s}-\gamma^{(k+1)}_{\infty,s}\right) \Big| \\ &
\leq \delta \alpha^{-3} \| J^{(k)} \| \| h \|_{\infty} \left( \tr \;
(1-\Delta_{k+1}) \gamma^{(k+1)}_{N,s} + \tr \; (1- \Delta_{k+1})
\gamma_{\infty,s}^{(k+1)} \right) \\ & \leq C \delta \alpha^{-3}
\end{split}\end{equation}
uniformly in $N$. Choosing, for example, $\delta = \alpha^4$, it follows that
\begin{equation}\label{eq:eta}
\Big| \tr\; J_{s-t}^{(k)} \, h_{\alpha} (x_1 - x_{k+1}) \,
\left( \gamma^{(k+1)}_{N,s} - \gamma^{(k+1)}_{\infty,s} \right) \Big| \leq \eta (\alpha,N) + C \alpha
\end{equation}
where $\eta (\alpha,N) \to 0$ as $N \to \infty$, for every fixed $\alpha >0$,
and where the constant $C$ only depends on $J^{(k)}$.

\medskip

Finally, using Lemma \ref{lm:sobsob} and Theorem \ref{thm:aprik}, the last term
on the r.h.s. of (\ref{eq:NVN-term}) can be controlled by
\begin{equation}\label{eq:term5}
\begin{split}
\Big| \tr\; J_{s-t}^{(k)} \; \left[ h_{\alpha}
 (x_1 - x_{k+1}) -\delta (x_1 -x_{k+1}) \right] \, \gamma^{(k+1)}_{\infty,s}
\Big| \leq \; &C \alpha^{1/2} \, \tri J_{s-t}^{(k)} \tri \, \tr \;
(1-\Delta_1) (1-\Delta_{k+1}) \gamma^{(k+1)}_{\infty,s} \\ \leq \; &C (k,T,J^{(k)}) \, \alpha^{1/2}\,.
\end{split}
\end{equation}

\medskip

{F}rom (\ref{eq:NVN-term}), (\ref{eq:tm1to0}), (\ref{eq:tm2to0}), (\ref{eq:WNto1}),
(\ref{eq:eta}), and (\ref{eq:term5}) it follows that, for every $k \geq 1$, $0
\leq s \leq t \leq T$, and $J^{(k)} \in \cK_k$ with (\ref{eq:assJ}),
\begin{equation}\label{eq:tmNto0} \Big| \tr \; J^{(k)} \cU^{(k)} (t-s) N V_N (x_1 -x_{k+1})
 \gamma^{(k+1)}_{N,s} - 8 \pi a_0  \tr \; J^{(k)} \cU^{(k)} (t-s) \delta (x_1 -x_{k+1})
 \gamma^{(k+1)}_{\infty,s} \Big| \to 0 \end{equation} as $N \to \infty$.
Similarly to (\ref{eq:tmNto0}), we can also prove that
\begin{equation}\label{eq:tmNto0b} \Big| \tr \; J^{(k)} \cU^{(k)} (t-s)
\gamma^{(k+1)}_{N,s} N V_N (x_1 -x_{k+1}) - 8 \pi a_0  \tr \; J^{(k)} \cU^{(k)}
(t-s) \gamma^{(k+1)}_{\infty,s} \delta (x_1 -x_{k+1}) \Big| \to 0 \end{equation}
as $N \to \infty$. Since (\ref{eq:tmNto0}) and (\ref{eq:tmNto0b}) remain valid
if we replace $x_1$ by any $x_j$, $j =2, \dots ,k$ in the potentials, it follows
that, for every $k \geq 1$, $0\leq s \leq t \leq T$, and $J^{(k)} \in \cK_k$ with (\ref{eq:assJ}),
\begin{equation}\label{eq:tmNto0c}
\begin{split}
\Big| \sum_{j=1}^k  \Big( \tr \; J^{(k)} &\cU^{(k)} (t-s) \left[ N V_N (x_j -x_{k+1}),
\gamma^{(k+1)}_{N,s}\right]  \\ &- 8 \pi a_0  \tr \; J^{(k)} \cU^{(k)} (t-s)
\left[\delta (x_j -x_{k+1}), \gamma^{(k+1)}_{\infty,s}\right] \Big) \Big| \to 0
\end{split}\end{equation} as $N \to \infty$. {F}rom (\ref{eq:equi-4}) (with $J^{(k)}$
replaced by $\cU^{(k)} (s-t) J^{(k)}$, using the fact that $\tri \cU^{(k)} (s-t)
J^{(k)} \tri \leq C_T$ for all $0 \leq s \leq t \leq T$), and from an estimate
similar to (\ref{eq:equi-4}) but with $\gamma_{N,s}^{(k+1)}$ replaced by
$\gamma_{\infty,s}^{(k+1)}$ and $NV_N (x_j - x_{k+1})$ replaced by
$\delta (x_j - x_{k+1})$, we can now apply the dominated convergence theorem to conclude (\ref{eq:conv-6}).
\end{proof}

The following lemmas are important ingredients in the proof of Theorem \ref{thm:conv}.

\begin{lemma}\label{lm:term1} Under the same assumptions of Theorem
\ref{thm:conv}, and using the notation $J_t^{(k)} = \cU^{(k)} (t) J^{(k)}$,
we have, for every $k \geq 1$, $\ell =1, \dots, k$, $0
\leq s \leq t \leq T$, $J^{(k)} \in \cK_k$ such that (\ref{eq:assJ}) is satisfied,
\begin{equation*}\Big| \tr \;  J_{s-t}^{(k)} \,
\left[ N V_N (x_{\ell} -x_{k+1}) W_{N,(\ell,k+1)} - 8\pi a_0 \,
\delta (x_{\ell} - x_{k+1}) \right] W^*_{N,(\ell,k+1)} \gamma^{(k+1)}_{N,s} \Big| \to 0 \end{equation*}
as $N \to \infty$.
\end{lemma}

\begin{proof}
We fix $\ell =1$. Decomposing $\gamma_{N,s}^{(k+1)} = \sum_j \lambda_j \, |\xi^{(k+1)}_j
 \rangle \langle \xi^{(k+1)}_j |$, and introducing the variables $u= (x_1 + x_{k+1})/2$
and $v= x_1 -x_{k+1}$,  we find
\begin{equation}
\begin{split}
\tr \;  J_{s-t}^{(k)} \, N V_N (x_1 -x_{k+1}) \gamma^{(k+1)}_{N,s}  =\; & \sum_j \lambda_j
\int \rd u \rd v \rd x_2 \dots \rd x_k \rd \bx'_k
J_{s-t}^{(k)} (\bx'_k; u+v/2, x_2 , \dots ,x_k) N V_N (v) 
 \\ &\hspace{.2cm} \times  \,
\xi^{(k+1)}_j (u+v/2, x_2 , \dots ,x_k , u-v/2) \, \overline{\xi}^{(k+1)}_j (\bx'_k, u-v/2).
\end{split}\end{equation}
The potential $V_N (v)$ forces $v$ to be of order $1/N$. Using this fact, we are going to remove the $v$-dependence from the observable and from the wave function $\overline{\xi}_j^{(k+1)}$. After removing this $v$-dependence, we introduce the wave operator using its $L^2$-unitarity. We find
\begin{equation}\label{eq:termI1}
\begin{split}
\tr \;  &J_{s-t}^{(k)} \,  N V_N (x_1 -x_{k+1}) \gamma^{(k+1)}_{N,s}
\\ =\; &\sum_j \lambda_j  \int \rd u \rd v \rd x_2 .. \rd x_k \rd \bx'_k
\, N V_N (v) \,  \xi^{(k+1)}_j (u+v/2, x_2 , .. ,x_k , u-v/2)
\\ &\hspace{.2cm} \times \left[ J_{s-t}^{(k)} (\bx'_k; u+v/2, x_2 , \dots ,x_k)
\overline{\xi}^{(k+1)}_j (\bx'_k, u-v/2)  -
J_{s-t}^{(k)}  (\bx'_k; u, x_2 , \dots ,x_k)
\overline{\xi}^{(k+1)}_j (\bx'_k, u) \right] \\ &+\sum_j \lambda_j  \int \rd u \rd v 
\rd x_2 .. \rd x_k \rd \bx'_k \, N (W_N^* V_N) (v)  \, J_{s-t}^{(k)} (\bx'_k; u, x_2 , .. ,x_k) \,
 \overline{\xi}^{(k+1)}_j (\bx'_k, u) \\ &\hspace{.2cm} \times
\left[ (W_{N,(1,k+1)}^* \xi^{(k+1)}_j) (u+v/2, x_2 , .. ,x_k , u-v/2) -
 (W_{N,(1,k+1)}^* \xi^{(k+1)}_j) (u, x_2 , .. ,x_k , u) \right] \\ &+\sum_j \lambda_j
 \left( \int \rd v \, (W_N^* V_N) (v) \right)
\int \rd u \rd x_2 \dots \rd x_k \rd \bx'_k \, J_{s-t}^{(k)} (\bx'_k; u, x_2 , \dots ,x_k) \\ &\hspace{.2cm} \times
 \overline{\xi}^{(k+1)}_j (\bx'_k, u) (W_{N,(1,k+1)}^* \xi^{(k+1)}_j) (u, x_2 , \dots ,x_k , u).
\end{split}
\end{equation}
{F}rom Lemma \ref{lm:VW=a_0}, we know that
\[ \int \rd v \, N (W_N^* V_N) (v) = 8 \pi a_0 \, .\] Therefore, from (\ref{eq:termI1}),
we obtain that
\begin{equation}\label{eq:Ij+IIj}
\begin{split}
\Big| \tr \;  &J^{(k)}_{s-t} \, \left[ N V_N (x_1 -x_{k+1})
W_{N,(1,k+1)} - 8\pi a_0 \, \delta (x_1 - x_{k+1}) \right] W^*_{N,(1,k+1)} \gamma^{(k+1)}_{N,s} \Big| \\
\leq \; & \sum_j \lambda_j \, \int \rd u \rd v \rd x_2 \dots \rd x_k \rd \bx'_k \, N V_N (v) \,  | \xi^{(k+1)}_j (u+v/2, x_2 , \dots ,x_k , u-v/2)| \\ &\hspace{.3cm}
\times  \Big| J_{s-t}^{(k)} (\bx'_k; u+v/2, x_2 , \dots ,x_k)
\overline{\xi}^{(k+1)}_j (\bx'_k, u-v/2) 
 -J^{(k)}_{s-t} (\bx'_k; u, x_2 , \dots ,x_k) \overline{\xi}^{(k+1)}_j (\bx'_k, u) \Big| \\ &+ \sum_j \lambda_j
\Big| \int \rd u \rd v \rd x_2 \dots \rd x_k \rd \bx'_k \,  N (W_N^* V_N) (v)  \, J_{s-t}^{(k)} (\bx'_k; u, x_2 , \dots ,x_k)
\overline{\xi}^{(k+1)}_j (\bx'_k, u) \\ &\hspace{.3cm} \times 
\left[ (W_{N,(1,k+1)}^* \xi^{(k+1)}_j) (u+v/2, x_2 , \dots ,x_k , u-v/2)  - (W_{N,(1,k+1)}^* \xi^{(k+1)}_j) (u, x_2 , \dots ,x_k , u) \right] \Big| \\
= : \; &\sum_j \lambda_j \left(\text{I}_j + \text{II}_j \right) \, .
\end{split}
\end{equation}
The terms $\text{I}_j$ can be bounded exactly like the term $\text{I}$ on the r.h.s. of (\ref{eq:rev4}), after replacing $J^{(k)}$ by $J^{(k)}_{s-t}$. Following the steps (\ref{eq:rev5})-(\ref{eq:revI}), we find that
\begin{equation}\label{eq:Ijto0}
\sum_j \lambda_j \, \text{I}_j \leq \frac{C_k \, \tri J^{(k)}_{s-t} \tri}{N^{1/2}} \, \tr \left( -\Delta_1 -\Delta_{k+1} +1 \right) W_{N, (1,k+1)}^* \gamma_{N,s}^{(k+1)} W_{N, (1,k+1)} 
\end{equation}
which converges to zero as $N \to \infty$ for all $0 \leq s \leq t \leq T$, and all observables $J^{(k)}$ satisfying (\ref{eq:assJ}) (here we used the a-priori estimate 
given in Proposition \ref{prop:apri2} and the observation (\ref{eq:triJjt})).

\bigskip

Next, we consider the second term on the r.h.s. of (\ref{eq:Ij+IIj}):
\begin{equation}\label{eq:IIJ1}
\begin{split}
\text{II}_j
= \; & \Big| \int \rd u \rd v \rd x_2 \dots \rd x_k \rd \bx'_k \,
J^{(k)}_{s-t} (\bx'_k; u, x_2 , \dots ,x_k)
\overline{\xi}^{(k+1)}_j (\bx'_k, u)  \\ &\hspace{1cm} \times
\left( N^3 (W^* V)(Nv) - \delta (v) \right) (W_{N,(1,k+1)}^* \xi^{(k+1)}_j)
 (u+v/2, x_2 , \dots ,x_k , u-v/2)\Big| \,.
\end{split}
\end{equation}
To control this contribution, we first insert a cutoff $\chi (v)$; this
 will allow us to apply Lemma \ref{lm:VL12} to bound the integral over $u$ and $v$.
To this end, we choose a function $\chi \in C^{\infty}_0 (\bR^3)$ such that
$0 \leq \chi (x) \leq 1$, $\chi (x) = 1$ for $|x| \leq 1$ and $\chi (x) =0$ for $|x| \geq 2$,
 and we put $\bar{\chi} = 1-\chi$. Using $\chi$, we decompose the r.h.s. of (\ref{eq:IIJ1}) in two parts
\begin{equation}\label{eq:IIJ2}
\begin{split}
\text{II}_j \leq \; &\Big|\int \rd u \rd v \rd x_2 \dots \rd x_k \rd \bx'_k \,
J_{s-t}^{(k)} (\bx'_k; u, x_2 , \dots ,x_k) \chi ( v )
 \overline{\xi}^{(k+1)}_j (\bx'_k, u) \\ &\hspace{1cm} \times \left[ N^3 (W^* V)(Nv)
- \delta (v) \right] (W_{N,(1,k+1)}^* \xi^{(k+1)}_j ) (u+v/2, x_2 , \dots ,x_k , u-v/2) \Big| \\
&+ \Big| \int \rd u \rd v \rd x_2 \dots \rd x_k \rd \bx'_k \, J_{s-t}^{(k)} (\bx'_k; u, x_2 , \dots ,x_k) \bar{\chi} (v)
\overline{\xi}^{(k+1)}_j (\bx'_k, u) \\ &\hspace{1cm} \times
N^3 (W^* V)(Nv) (W_{N,(1,k+1)}^* \xi^{(k+1)}_j) (u+v/2, x_2 , \dots ,x_k , u-v/2)\Big| \\
=: \; & \text{A}_j + \text{B}_j\, .
\end{split}
\end{equation}
The term $\text{B}_j$ can be bounded by
\begin{equation*}
\begin{split}
\text{B}_j \leq \; &\int \rd u \rd v \rd x_2 \dots \rd x_k \rd \bx'_k \,
\Big| J_{s-t}^{(k)} (\bx'_k; u, x_2 , \dots ,x_k)
\Big| \,  \bar{\chi} (v) N^3 (W^* V)(Nv) \\  &\hspace{.5cm} \times
\left( |\xi^{(k+1)}_j (\bx'_k, u)|^2  +
\Big|(W_{N,(1,k+1)}^* \xi^{(k+1)}_j) (u+v/2, x_2 , \dots ,x_k , u-v/2)\Big|^2 \right)
\\ \leq \; & \, \| \xi^{(k+1)}_j \|^2 \, \left( \sup_{u,\bx'_k}
\int \rd x_2 \dots \rd x_k \; \Big|
 J_{s-t}^{(k)} (\bx'_k; u, x_2 , \dots ,x_k) \Big| \right) \;
\int_{|v| \geq N} |(W^* V) (v)| \rd v\\
&+ \left( \sup_{\bx_k} \int \rd \bx'_k \,
\Big| J_{s-t}^{(k)} (\bx'_k; \bx_k)\Big| \right)  \\
& \hspace{.5cm} \times \int \rd x_1 \dots \rd x_{k+1}
\bar{\chi} (x_1 -x_{k+1}) N^3 (W^* V)(N (x_1 -x_{k+1}))
\Big|(W_{N,(1,k+1)}^* \xi^{(k+1)}_j) (\bx_k, x_{k+1})\Big|^2.
\end{split}
\end{equation*}
{F}rom Lemma \ref{lm:VL1}, we obtain
\begin{equation}\label{eq:Fjto0}
\begin{split}
\sum_j \lambda_j \, \text{B}_j \leq \; &C (k,T, J^{(k)}) \,
\left( \int_{|v| \geq N} \rd v \; |(W^* V) (v)| \right) \,  \\
&\hspace{1cm} \times \left( \tr \; \left( (\nabla_1 \cdot \nabla_2)^2 -
 \Delta_1 -\Delta_2 + 1 \right) W_{N,(1,k+1)} \gamma_{N,s}^{(k+1)} W_{N,(1,k+1)}^* \right) \to 0
\end{split}
\end{equation}
as $N \to \infty$. Here we used Proposition \ref{prop:apri2} and the fact that,
since $W^* V \in L^1 (\bR^3)$,
 \[ \int_{|x| >N} |W^* V (x)| \rd x \to 0 \qquad \text{ as} \quad N \to \infty. \]
As for the term $\text{A}_j$ on the r.h.s. of (\ref{eq:IIJ2}), Lemma \ref{lm:VL12}
implies that there exists a sequence $\delta_{N} \to 0$ as $N \to \infty$
 ($\delta_N$ corresponds to the sequence $\beta_{1/N}$ defined in Lemma \ref{lm:VL12},
 with $V$ replaced by $W^*V$) such that
\begin{equation}
\begin{split}
\text{A}_j \leq \; &\delta_{N}  \int \rd x_2 \dots \rd x_k \rd \bx'_k \, \\ & \hspace{.5cm}
 \times \Big( \int \rd u \rd v \;   \Big| \left( (\Delta_u - \Delta_v)^2 -\Delta_u - \Delta_v + 1 \right)^{1/2} \\
&\hspace{5cm} \times \left( W_{N,(1,k+1)}^* \xi^{(k+1)}_j\right) (u+v/2, x_2, \dots , x_k, u-v/2)
\Big|^{2} \Big)^{1/2} \\ &\hspace{.5cm} \times \left( \int
\rd u \rd v \, \Big| (1-\Delta_u+\Delta_v^2)^{1/2} \chi (v) \, J^{(k)}_{s-t}
(\bx'_k; u, x_2 , \dots ,x_k) \overline{\xi}^{(k+1)}_j (\bx'_k, u) \Big|^2 \right)^{1/2}
\\ \leq \; & \delta_{N}  \| \chi \|_{H^2} \int \rd x_2 \dots \rd x_k \rd \bx'_k \,
\left( \int
\rd u \Big| (1-\Delta_u)^{1/2}  \overline{\xi}^{(k+1)}_j (\bx'_k, u) \Big|^2 \right)^{1/2} \\
&\hspace{.5cm} \times \sup_u \left[ \Big| J^{(k)}_{s-t} (\bx'_k; u, x_2 , \dots ,x_k)\Big| + \Big| \nabla_u \,
 J_{s-t}^{(k)} (\bx'_k; u, x_2 , \dots ,x_k)\Big| \right] \\ & \hspace{.5cm} \times
\Big( \int \rd u \rd v \;   \Big| \left( (\Delta_u - \Delta_v)^2 -\Delta_u - \Delta_v + 1 \right)^{1/2} \\
&\hspace{5cm} \times \left( W_{N,(1,k+1)}^* \xi^{(k+1)}_j\right)
 (u+v/2, x_2, \dots , x_k, u-v/2) \Big|^{2} \Big)^{1/2}.
\end{split}
\end{equation}
With a Schwarz inequality, we find
\begin{equation}
\begin{split}
\text{A}_j \leq &\; \delta_{N} \| \chi \|_{H^2} \int \rd x_2 \dots \rd x_k \rd \bx'_k \, \\
 &\hspace{.05cm} \times \sup_u \left[ \Big| J_{s-t}^{(k)} (\bx'_k; u, x_2 , \dots ,x_k)\Big|
 + \Big| \nabla_u \, J^{(k)}_{s-t} (\bx'_k; u, x_2 , \dots ,x_k)\Big| \right] \\
&\hspace{.05cm} \times \left(\int
\rd u \Big| (1-\Delta_u)^{1/2}  \overline{\xi}^{(k+1)}_j (\bx'_k, u) \Big|^2  \right. \\
&\hspace{.1cm} \left.+  \int \rd u \rd v \Big| \left( (\Delta_u - \Delta_v)^2 -\Delta_u - \Delta_v + 1 \right)^{1/2}
\left( W_{N,(1,k+1)}^* \xi^{(k+1)}_j\right) (u+v/2, x_2,.., x_k, u-v/2) \Big|^{2} \right) \\ \leq \; &
C (k,T,J^{(k)}) \, \delta_{N} \, \left( \langle \xi_j^{(k+1)}, (1-\Delta_u) \xi^{(k+1)}_j \rangle \right. \\
&\left. \hspace{2cm} + \langle W_{N,(1,k+1)}^* \xi_j^{(k+1)},
\left( (\nabla_1 \cdot \nabla_{k+1})^2 -\Delta_1 -\Delta_{k+1} + 1 \right) W_{N,(1,k+1)}^* \xi_j^{(k+1)}
\rangle \right)\,.
\end{split}
\end{equation}
{F}rom Lemma \ref{lm:VL12} and Proposition \ref{prop:apri2}, we find
\begin{equation}\label{eq:Ejto0}
\sum_j \lambda_j \, \text{A}_j \leq C \delta_{N} \to 0 \qquad \text{as } N \to \infty
\end{equation}
and this, with (\ref{eq:Fjto0}), implies that
\begin{equation*} \sum_j \lambda_j \text{II}_j \to 0 \qquad \text{as } N \to \infty \, .
\end{equation*}
Together with (\ref{eq:Ijto0}) and (\ref{eq:Ij+IIj}), this concludes the proof of the lemma.
\end{proof}

\begin{lemma}\label{lm:term2}
Under the same conditions as in Theorem \ref{thm:conv}, we have, for every $k \geq 1$,
$\ell=1,\dots,k$, $0\leq s \leq T$, and $J^{(k)} \in \cK_k$ satisfying (\ref{eq:assJ}),
\begin{equation*} \Big| \tr \;  J_{s-t}^{(k)} \,
\left[ \delta (x_{\ell} -x_{k+1}) - \, h_{\alpha} (x_{\ell} - x_{k+1}) \right] W^*_{N,(\ell,k+1)}
 \gamma^{(k+1)}_{N,s} \Big| \to 0 \end{equation*} as $\alpha \to 0$, uniformly in $N$. Here we use notation $J^{(k)}_t = \cU^{(k)} (t) J^{(k)}$. 
\end{lemma}
\begin{proof}
We fix $\ell =1$. Using the decomposition $\gamma_{N,s}^{(k+1)} =
\sum_j \lambda_j |\xi^{(k+1)}_j \rangle \langle \xi^{(k+1)}_j |$, we find that
\begin{equation*}
\begin{split}
\Big| \tr \;  &J_{s-t}^{(k)} \, \left[ \delta (x_1 - x_{k+1}) -
 \, h_{\alpha} (x_1 - x_{k+1}) \right] W^*_{N,(1,k+1)} \gamma^{(k+1)}_{N,s} \Big| \\
\leq \; &\sum_j \lambda_j \, \Big|
\int \rd u \rd v \rd x_2 \dots \rd x_k \rd \bx'_k \, J_{s-t}^{(k)}
(\bx'_k; u+v/2, x_2 , \dots ,x_k) \,  \left[ \delta (v) -h_{\alpha} (v)  \right]\\ &\hspace{.5cm} \times
 \; \overline{\xi}^{(k+1)}_j (\bx'_k, u-v/2) (W_{N,(1,k+1)}^* \xi^{(k+1)}_j)
 (u+v/2, x_2 , \dots ,x_k , u-v/2) \Big| \\
\leq \; &\sum_j \lambda_j \, \Big|
\int \rd u \rd v \rd x_2 \dots \rd x_k \rd \bx'_k \, h_{\alpha} (v) \\ &\hspace{.5cm}\times \left[
J_{s-t}^{(k)} (\bx'_k; u+v/2, x_2 , \dots ,x_k)
\overline{\xi}^{(k+1)}_j (\bx'_k, u-v/2) \right. \\ &\hspace{4cm} \times
(W_{N,(1,k+1)}^* \xi^{(k+1)}_j) (u+v/2, x_2 , \dots ,x_k , u-v/2) \\ &\hspace{1cm} -
\left. J_{s-t}^{(k)} (\bx'_k; u, x_2 , \dots ,x_k)
\overline{\xi}^{(k+1)}_j (\bx'_k, u) (W_{N,(1,k+1)}^* \xi^{(k+1)}_j) (u, x_2 , \dots ,x_k , u) \right] \Big| \, .
\end{split}
\end{equation*}
Similarly to (\ref{eq:termI1}), we first replace $v$ by $0$ in
$J_{s-t}^{(k)} (\bx'_k; u+v/2, x_2 , \dots ,x_k)
\overline{\xi}^{(k+1)}_j (\bx'_k, u-v/2)$ and then in $(W_{N,(1,k+1)}^* \xi^{(k+1)}_j)
 (u+v/2, x_2 , \dots ,x_k , u-v/2)$. We obtain
\begin{equation}\label{eq:alpha1}
\begin{split}
\Big| \tr \;  &J_{s-t}^{(k)} \, \left[ \delta (x_1 - x_{k+1}) -
  \, h_{\alpha} (x_1 - x_{k+1}) \right] W^*_{N,(1,k+1)} \gamma^{(k+1)}_{N,s} \Big| \\
\leq \; & \sum_j \lambda_j \, \Big|
\int \rd u \rd v  \rd x_2 \dots \rd x_k \rd \bx'_k \, h_{\alpha} (v) \,
(W_{N,(1,k+1)}^* \xi^{(k+1)}_j) (u+v/2, x_2 , \dots ,x_k , u-v/2) \, \\
&\hspace{.5cm} \times \left[ J_{s-t}^{(k)}
(\bx'_k; u+v/2, x_2 , \dots ,x_k) \overline{\xi}^{(k+1)}_j (\bx'_k, u-v/2) \right. \\
&\hspace{4cm}\left. - J_{s-t}^{(k)} (\bx'_k; u, x_2 , \dots ,x_k)
\overline{\xi}^{(k+1)}_j (\bx'_k, u) \right] \Big|
\\ & + \sum_j \lambda_j \, \Big|
\int \rd u \rd v \rd x_2 \dots \rd x_k \rd \bx'_k \, h_{\alpha} (v) \, J_{s-t}^{(k)}
(\bx'_k; u, x_2 , \dots ,x_k)\overline{\xi}^{(k+1)}_j (\bx'_k, u) \\ &\hspace{.5cm} \times
\left[(W_{N,(1,k+1)}^* \xi^{(k+1)}_j) (u+v/2, x_2 , \dots ,x_k , u-v/2) -
(W_{N,(1,k+1)}^* \xi^{(k+1)}_j) (u, x_2 , \dots ,x_k , u) \right] \Big| \\
=: \; & \sum_j \lambda_j \; \left( \text{III}_j + \text{IV}_j \right).
\end{split}
\end{equation}
To bound the first term, we expand the difference in an integral
\begin{equation}
\begin{split}
\Big[ J_{s-t}^{(k)} (\bx'_k; u+v/2, &x_2 , \dots ,x_k)
\overline{\xi}^{(k+1)}_j (\bx'_k, u-v/2) 
- J_{s-t}^{(k)} (\bx'_k; u, x_2 , \dots ,x_k)
\overline{\xi}^{(k+1)}_j (\bx'_k, u) \Big] \\ &\hspace{1cm} =
\int_0^{1/2} \rd r \, v \cdot \nabla_1 J_{s-t}^{(k)}
 (\bx'_k; u+rv, x_2 , \dots ,x_k) \overline{\xi}^{(k+1)}_j (\bx'_k, u-rv) \\
&\hspace{1.5cm} - \int_0^{1/2} \rd r \, J_{s-t}^{(k)}
(\bx'_k; u+rv, x_2 , \dots ,x_k)  v \cdot \nabla_{k+1} \overline{\xi}^{(k+1)}_j (\bx'_k, u-rv)
\end{split}
\end{equation}
and we obtain that
\begin{equation}
\begin{split}
\text{III}_j \leq \; &
\int \rd u \rd v  \rd x_2 \dots \rd x_k \rd \bx'_k \, \int_0^{1/2} \rd r \,
 h_{\alpha} (v) |v|  \Big|(W_N^* \xi^{(k+1)}_j) (u+v/2, x_2 , \dots ,x_k , u-v/2)  \Big| \\
 &\hspace{.5cm} \times  \left( \Big| \nabla_1 J_{s-t}^{(k)}
 (\bx'_k; u+rv, x_2 , \dots ,x_k)\Big|  | \xi^{(k+1)}_j (\bx'_k, u-rv)|  \right. \\
&\hspace{2cm} \left. + \Big|  J_{s-t}^{(k)}
 (\bx'_k; u+rv, x_2 , \dots ,x_k)\Big|  \Big| \nabla_{k+1} \xi^{(k+1)}_j (\bx'_k, u-rv) \Big| \right)
\end{split}
\end{equation}
which implies that
\begin{equation*}
\begin{split}
\sum_j \lambda_j \; \text{III}_j  \leq \; & \alpha \; C (k,T,J^{(k)}) \,
 \left( \tr \; (1-\Delta_{k+1}) \gamma^{(k+1)}_{N,s} + \tr\;
\left( (\nabla_1 \cdot \nabla_{k+1})^2 - \Delta_1 -\Delta_{k+1} + 1\right) \gamma^{(k+1)}_{N,s} \right) \\
\leq \; &C \; \alpha.
\end{split}
\end{equation*}
The terms $\text{IV}_j$ can be estimated similarly to the terms $\text{II}_j$
 considered in (\ref{eq:IIJ1}); in particular, analogously to (\ref{eq:Fjto0}) and (\ref{eq:Ejto0}), we also find
\begin{equation*}
\begin{split}
\sum_j \lambda_j \; \text{IV}_j \leq \; & C (k,T,J^{(k)}) \, \beta_{\alpha} \,
\left( \tr \; (1-\Delta_{k+1}) \gamma^{(k+1)}_{N,s} + \tr \;
\left( (\nabla_1 \cdot \nabla_{k+1})^2 - \Delta_1 -\Delta_{k+1} + 1\right)
\gamma^{(k+1)}_{N,s} \right) \\ \leq \; &C \beta_{\alpha}
\end{split}
\end{equation*}
where $\beta_{\alpha} \to 0$ as $ \alpha \to 0$ uniformly in $N$ (the
sequence $\beta_{\alpha}$ comes from Lemma \ref{lm:VL12}, with $V$ replaced by $h$).
This concludes the proof of the lemma.
\end{proof}

\bigskip

\begin{lemma}\label{lm:VW=a_0}
Suppose that $V \geq 0$, with $V (x) \leq C \langle x \rangle^{-\sigma}$ for some
$\sigma >5$ (this implies, in particular that $V \in L^1 (\bR^3) \cap L^2 (\bR^3)$
and thus that $V$ is in the Rollnik class of potentials). Let $W$ denote the wave operator
(as defined in Proposition \ref{prop:waveop}) associated with the Hamiltonian $\fh= -\Delta + (1/2) V (x)$. Then
\[ \int \rd x \; (W^* V) (x) = 8 \pi a_0 \] where $a_0$ is the scattering length of the potential $V$.
\end{lemma}

\begin{proof}
First of all, we observe that, under the assumption that $V \geq 0$ and $V (x) \leq C
 \langle x \rangle^{-\sigma}$, for some $\sigma >5$, the operator $\fh = -\Delta + (1/2) V$
 cannot have a zero energy resonance (recall that a zero-energy resonance of $\fh$ is a
 solution $\ph$ of $(-\Delta + (1/2) V) \ph = 0$ such that $|\ph (x)| \leq C/|x|$ for all
$x \in \bR^3$); this can be proven using the maximum principle. We will make use of this
observation in the proof of this lemma.

\medskip

Next, we note that, since $W^*$ maps $L^1 (\bR^3)$ into $L^1 (\bR^3)$ (see
Proposition \ref{prop:waveop}), we have that $(W^* V) \in L^1 (\bR^3)$ and thus
\begin{equation}\label{eq:WV0}
\begin{split}
\int \rd x \; (W^* V) (x) &= \lim_{\eps \to 0} \int \rd x \; (W^* V) (x) \; \chi_{\eps} (x)
= \lim_{\eps \to 0} \int \rd x \, V(x) (W \chi_{\eps}) (x)
\end{split}
\end{equation}
with \[ \chi_{\eps} (x) = \frac{1}{1+\eps x^2} \,. \]

\medskip

We expand $W \chi_{\eps}$ in terms of solutions $\ph (x,k)$ of the Lippman-Schwinger equation
\begin{equation}\label{eq:LSE} \ph (x,k) = e^{i k \cdot x} - \frac{1}{8\pi}
\int \rd y \frac{e^{i|k||x-y|}}{|x-y|} \, V(y) \ph (y,k) \, .\end{equation}
It follows from \cite[Theorem XI.41, a)]{RS3} that Eq. (\ref{eq:LSE}) has a unique
solution $\ph (x,k)$, such that $\ph (x,k) V^{1/2} (x) \in L^2 (\bR^3)$, for all
$k \in \bR^3$ such that $k^2 \not \in \cE$, for an exceptional set $\cE$ with
 Lebesgue measure zero. The set $\cE$ consists of all values of $k^2$ for which
zero is an eigenvalue of the operator
\begin{equation}\label{eq:Mk} M_{|k|} = 1 + \frac{1}{2} V^{1/2} \frac{1}{-\Delta -k^2} V^{1/2} \, .
 \end{equation}
{F}rom the observation that the operator $\fh= -\Delta + (1/2) V$ does not have a
 zero energy resonance, it follows immediately that $0 \not \in \cE$; in fact, if
 $M_{0} \psi = 0$ for some $\psi \in L^2 (\bR^3)$, then \[ \psi (x) = -\frac{1}{2} V^{1/2} (x)
\int \rd y \, \frac{1}{|x-y|} V^{1/2} (y) \psi (y) \, , \] which implies that
$\psi (x)/ V^{1/2} (x) \leq C / |x|$ for $|x| \gg 1$ and thus that
$\ph (x) := \psi (x) / V^{1/2} (x)$ is a zero-energy resonance solution of
 $\left(-\Delta + (1/2) V \right) \ph = 0$. Since $M_0$ is a non-negative
Fredholm operator with no eigenvalue at zero, it follows that there exists $\lambda > 0$
 with $\sigma (M_0) \subset (\lambda,\infty)$ (here $\sigma (M_0)$ indicates the spectrum
 of $M_0$). Since moreover $M_{|k|} - M_0$ is a compact operator with kernel
\begin{equation}
\left(M_{|k|} - M_0\right) (x;y) = \frac{1}{2} V^{1/2} (x) \frac{e^{i|k||x-y|} - 1}{|x-y|} V^{1/2} (y)
\end{equation}
we obtain that
\begin{equation}
\begin{split}
\| M_{|k|} - M_0 \|^2_{\text{HS}} = \; &\frac{1}{4} \int \rd x \rd y \, V(x) V(y) \,
\frac{\left|e^{i|k||x-y|} - 1\right|^2}{|x-y|^2}  \leq \;\frac{|k|^2 \| V \|_{L^1}^2}{4}
\end{split}
\end{equation}
and thus that there exists $\kappa >0$ such that $\sigma (M_{|k|}) \subset (\lambda/2, \infty)$
 for all $|k| \leq \kappa$. In particular it follows that \begin{equation}\label{eq:Mk-1}
 \| M_{|k|}^{-1} \| \leq 2/\lambda \qquad \text{for all $k \in \bR^3$ with $|k| \leq \kappa$.}
\end{equation}

\medskip

{F}rom \cite[Theorem XI.41, e)]{RS3} we also find
\begin{equation}\label{eq:RS1}
\begin{split}
(W \chi_{\eps}) (x) = \text{L.I.M.} \, (2\pi)^{-3/2} \int \rd k\; \ph (x,k) \,
\widehat{\chi}_{\eps} (k) = \text{L.I.M.} \int \rd k\; \ph (x,k) \frac{e^{-|k|/\sqrt{\eps}}}{4\pi |k|\eps} \, ,
\end{split}\end{equation}
where L.I.M. denotes the $L^2$-limit as $M \to \infty$ and $\delta \to 0$ of the
integral over $\{ k \in \bR^3: |k| \leq M \text{ and } \text{dist } (k^2, \cE) > \delta \}$.
 Inserting (\ref{eq:RS1}) in the r.h.s. of (\ref{eq:WV0}), we find
(recalling that $\kappa >0$ is chosen such that (\ref{eq:Mk-1}) holds true)
\begin{equation}\label{eq:WV1}
\begin{split}
\int \rd x \; (W^* V) (x) \; \chi_{\eps} (x) = \; & \int_{|k| > \kappa} \rd x \rd k \, V(x)
\ph (x,k) \frac{e^{-|k|/\sqrt{\eps}}}{4\pi |k|\eps}+ \int_{|k| \leq \kappa} \rd x \rd k \,
V(x) \ph (x,k) \frac{e^{-|k|/\sqrt{\eps}}}{4\pi |k|\eps}\,.
\end{split}
\end{equation}
The first term on the r.h.s. of (\ref{eq:WV1}) can be controlled by
\begin{equation}\label{eq:WV2}
\begin{split}
\left| \int_{|k| > \kappa} \rd x \rd k \, V(x) \ph (x,k) \frac{e^{-|k|/\sqrt{\eps}}}{4\pi |k|\eps}
 \right| & = \left| \int_{|k| > \kappa} \rd k \, V^{\sharp} (k) \frac{e^{-|k|/\sqrt{\eps}}}{4\pi |k|\eps}
 \right| \\ &\leq C \, \| V^{\sharp} \|_{L^2} \left( \int_{|k| \geq \kappa} \rd k \,
\frac{e^{-2|k|/\sqrt{\eps}}}{|k|^2 \eps^2} \right)^{1/2} \\ &\leq C
\| V \|_{L^2} \frac{e^{-\kappa /(2 \sqrt{\eps})}}{\eps^{3/4}} \to 0
\end{split}
\end{equation}
as $\eps \to 0$. Here we introduced the function \[ V^{\sharp} (k) =
 \text{l.i.m.} (2\pi)^{-3/2} \int \rd x \, V(x) \ph (x,k) \] where l.i.m.
denotes the $L^2$-limit of the integral over $|x| \leq M$ as $M \to \infty$;
the existence of $V^{\sharp}$ for $V \in L^2 (\bR^3)$ and the fact that
$\| V^{\sharp}\|_{L^2} \leq \| V \|_{L^2}$ (actually, in our case,
$\| V^{\sharp}\|_{L^2} = \| V \|_{L^2}$) are proven in \cite[Theorem IX.41]{RS3}.
As for the second term on the r.h.s. of (\ref{eq:WV1}), we have
\begin{equation}\label{eq:WV3}
\begin{split}
 \int_{|k| \leq \kappa} \rd k \rd x \, V(x) \ph (x,k) \frac{e^{-|k|/\sqrt{\eps}}}{4\pi |k|\eps}
= \; & \int_{|k| \leq \kappa} \rd k \rd x \, V(x) \ph (x,0) \frac{e^{-|k|/\sqrt{\eps}}}{4 \pi |k|\eps} \\
 &+ \int_{|k| \leq \kappa} \rd k \rd x \, V(x) \left( \ph (x,k) -\ph (x,0) \right)
\frac{e^{-|k|/\sqrt{\eps}}}{4\pi |k|\eps} \\
= \; & \left(1 - (1+ \kappa \eps^{-1/2}) e^{-\kappa \eps^{-1/2}} \right)
\int \rd x \, V(x) \ph (x,0) \\ &+
\int_{|k| \leq \kappa} \rd k \rd x \, V(x) \left( \ph (x,k) -\ph (x,0) \right)
 \frac{e^{-|k|/\sqrt{\eps}}}{4\pi |k|\eps}\, .
\end{split}
\end{equation}
Using that $\ph (x,0)$ is the solution of the zero energy scattering equation
\[ \left( -\Delta + (1/2) V(x) \right) \ph (x,0) = 0 \] with the boundary
condition $\ph (x,0) \to 1$ as $|x| \to \infty$, it follows that (see (\ref{eq:a0}))
\begin{equation}\label{eq:WV4}
\int \rd x \, V(x) \ph (x,0) = 8 \pi a_0 \, . \end{equation}
To bound the second term on the r.h.s. of (\ref{eq:WV3}), we define
\[ \psi_k (x) = V^{1/2} (x) \ph (x,k) \] and we observe that, from the
 Lippman-Schwinger equation (\ref{eq:LSE}),
\begin{equation}
\begin{split}
\psi_k (x) - \psi_0 (x) = \; & V^{1/2} (x) \left( e^{ik \cdot x} - 1 \right)
-\frac{1}{8\pi} \int \rd y V^{1/2} (x) \frac{e^{i|k||x-y|}}{|x-y|} V^{1/2} (y)
\left( \psi_k (y) - \psi_0 (y) \right) \\ &-\frac{1}{8\pi} \int \rd y V^{1/2} (x)
\frac{e^{i|k||x-y|}-1}{|x-y|} V^{1/2} (y) \psi_0 (y) \, , \end{split}
\end{equation}
which implies that (with $M_{|k|}$ defined in (\ref{eq:Mk}))
\begin{equation}
\begin{split}
\left[ M_{|k|} \left( \psi_k - \psi_0 \right) \right] (x) = \; & V^{1/2} (x)
\left( e^{ik\cdot x} - 1 \right) - \frac{1}{4\pi} \int \rd y V^{1/2} (x)
\frac{e^{i|k||x-y|}-1}{|x-y|} V^{1/2} (y) \psi_0 (y)\, .
\end{split}
\end{equation}
By (\ref{eq:Mk-1}), we have
\begin{equation}
\begin{split}
\Big\| \psi_k - \psi_0 \Big\|_{L^2} \leq \; &C \left( \Big\| V^{1/2} (x)
\left(e^{i k \cdot x} -1\right) \Big\|_{L^2} + \| V^{1/2} \|_{L^2} |k|
\int V^{1/2} (y) |\psi_0 (y)| \right) \\
\leq \; & C |k| \left( \| |x|^2 V \|^{1/2}_{L^1} + \| V \|_{L^1}^{3/2} \right) \\ \leq \; & C |k| \, .
\end{split}
\end{equation}
Therefore, the second term on the r.h.s. of (\ref{eq:WV3}) can be bounded by
\begin{equation}\begin{split}
\Big| \int_{|k| \leq \kappa} \rd k \rd x \, V(x) \left( \ph (x,k) -\ph (x,0) \right)
 \frac{e^{-|k|/\sqrt{\eps}}}{4\pi |k|\eps} \Big| \leq \; &\int_{|k| \leq \kappa} \rd k \rd x \,
 V^{1/2} (x) \left| \psi_k (x) -\psi_0 (x) \right|  \frac{e^{-|k|/\sqrt{\eps}}}{4\pi |k|\eps} \\
\leq \; &\| V \|_{L^1}^{1/2}  \int_{|k| \leq \kappa} \rd k  \, \| \psi_k -\psi_0 \|_{L^2}
\frac{e^{-|k|/\sqrt{\eps}}}{4\pi|k|\eps} \\ \leq \; & C \| V \|_{L^1}^{1/2}
 \int_{|k| \leq \kappa} \rd k \frac{e^{-|k|/\sqrt{\eps}}}{\eps} \\ \leq \; & C \eps^{1/2}
\end{split}
\end{equation}
and thus it converges to zero as $\eps \to 0$. The last equation, together with (\ref{eq:WV0}),
(\ref{eq:WV1}), (\ref{eq:WV2}), (\ref{eq:WV3}), and (\ref{eq:WV4}), concludes the proof of the lemma.
\end{proof}

\begin{lemma}\label{lm:WNto1}
Suppose that $V \geq 0$ and $V(x) \leq C \langle x \rangle^{-\sigma}$, for some $\sigma \geq 5$.
Then, for every $g \in L^2 (\bR^3, \rd x)$, we have
\[ \Big\| \left(W_{N} - 1 \right) g \Big\| \leq CN^{-1/6} \| g\|_{H^1} \, .\]
\end{lemma}
\begin{proof}
Let $\frak{h}_N = -\Delta + (1/2) V_N (x)$. Since
\[ W_N = s-\lim_{t \to \infty} e^{i \frak{h}_N t} e^{i\Delta t} \] it is enough to prove that
\begin{equation}\label{eq:equiv}
\sup_{t \in \, \bR} \left\| \left(e^{-i \frak{h}_N t} - e^{i\Delta t} \right)
 g \right\|\leq CN^{-1/6} \| g\|_{H^1}\,.
\end{equation}
Note that
\begin{equation*}
\begin{split}
\frac{\rd}{\rd t} \;  \left\| \left(e^{-i \frak{h}_N t} -
e^{i\Delta t} \right) g \right\|^2 = 2 \text{Im } \langle e^{-i \fh_N t} g, V_N (x) e^{i\Delta t} g \rangle
\end{split}
\end{equation*}
which implies that
\begin{equation}\label{eq:ints}
\left\| \left(e^{-i \fh_N t} - e^{i\Delta t} \right) g \right\|^2
\leq 2 \int_0^t \rd s \; \left|\langle e^{-i \fh_N s} g, V_N (x) e^{i\Delta s} g \rangle\right|\,.
\end{equation}
Next we observe that
\begin{equation}\label{eq:es1} \left| \langle e^{-i \fh_N s} g, V_N (x) e^{i\Delta s} g \rangle \right|
\leq \| e^{-i \fh_N s} g \|_{\infty} \, \| e^{i\Delta s} g  \|_{\infty} \, \| V_N \|_1
\leq \frac{\| V \|_1 \, \| g \|^2_{1} \, \| W \|_{\infty \to \infty}\, \| W^* \|_{1 \to 1}}{Ns^3}\, ,
\end{equation}
where we used the fact that \[ \| W_N \|_{p \to p} = \| W \|_{p \to p} \]
 for every $N$ and $1 \leq p \leq \infty$. For small $s$ we need a different
estimate of the integrand
on the r.h.s. of (\ref{eq:ints}). To this end we remark that
\begin{equation}\label{eq:es2}
\begin{split}
\left|\langle e^{-i \fh_N s} g, V_N (x) e^{i\Delta s} g \rangle\right| \leq \; &
\langle e^{i \fh_N s} g, V_N (x) e^{i \fh_N s} g \rangle^{1/2} \,
\langle e^{i \Delta s} g, V_N (x) e^{i\Delta s} g \rangle \\ \leq \; &
C \, \| V_N \|_{3/2} \| \nabla e^{i \fh_N s} g \| \| \nabla e^{i\Delta s} g \| \\ \leq \; &
 C\, \| V \|_{3/2} (1 + \| V \|_{3/2} )^{1/2} \, \| g \|^2_{H^1}\, , \end{split}
\end{equation}
where we used that $\| V_N \|_{3/2} = \| V \|_{3/2}$ and we estimated
\begin{equation*}
\begin{split}
\|  \nabla e^{i \fh_N s} g \|^2 = \langle e^{i \fh_N s} g, -\Delta e^{i \fh_N s} g \rangle
\leq \langle g, \fh_N g \rangle  \leq (1 + \| V \|_{3/2}) \| g \|^2_{H^1}\,.
\end{split}
\end{equation*}
Combining (\ref{eq:es1}) and (\ref{eq:es2}), we obtain from (\ref{eq:ints})
\begin{equation}
\begin{split}
\left\| \left(e^{-i \fh_N t} - e^{i\Delta t} \right) g \right\|^2
\leq \; & 2 \int_0^{N^{-\alpha}} \rd s \; \| V \|_{3/2} (1 + \| V \|_{3/2} )^{1/2} \,
 \| g \|^2_{H^1} \\ &+ 2 \int_{N^{-\alpha}}^t \, \rd s \frac{\| V \|_1 \, \| g \|^2_{H^1}
\, \| W \|_{\infty \to \infty}\, \| W^* \|_{1 \to 1}}{Ns^3} \\ \leq \; &
\Big( C_1 N^{-\alpha} + C_2 N^{2\alpha-1}\Big)\| g \|^2_{H^1}
\end{split}
\end{equation}
for every $t \in \bR$. Choosing $\alpha = 1/3$, we obtain (\ref{eq:equiv}).
\end{proof}

\section{Approximation of the initial data}\label{sec:appro}
\setcounter{equation}{0}

In this section we show how to regularize the initial wave function
$\psi_N$  given in Theorem \ref{thm:main2}.

\begin{proposition}\label{prop:initialdata}
Suppose that $\psi_N \in L^2 (\bR^{3N})$ with $\| \psi_N \| =1$ is a family of
$N$-particle wave functions with \begin{equation}\label{eq:initener}
\langle \psi_N, H_N \psi_N \rangle \leq C N \end{equation} and with one-particle
 marginal density $\gamma^{(1)}_{N}$ such that
\begin{equation}
\label{eq:initcond} \gamma^{(1)}_{N} \to |\ph \rangle \langle \ph | \qquad \text{as }
 N \to \infty \end{equation} for a $\ph \in H^1 (\bR^3)$. For $\kappa >0$ we define
\begin{equation}\label{eq:wtpsi}
\wt \psi_N: = \frac{\chi ( \kappa H_N/N ) \psi_N}{\| \chi (\kappa
H_N/N) \psi_N \|} \, .
\end{equation}
Here $\chi \in C^{\infty}_0 ( \bR )$ is a cutoff function such that
$0\leq \chi\leq 1$, $\chi (s) =1$ for $0 \leq s \leq 1$ and $\chi
(s) =0$ for $s \geq 2$. We denote by $\wt \gamma^{(k)}_{N}$, for $k
=1, \dots, N$, the marginal densities associated with $\wt \psi_N$.
\begin{enumerate}
\item[i)] For every integer $k \geq 1$ we have
\begin{equation}
\langle \wt \psi_N , H_N^k \, \wt \psi_N \rangle \leq \frac{2^k
N^k}{\kappa^k}\,.
\end{equation}
\item[ii)] We have
\[ \sup_N \| \psi_N - \wt\psi_N \| \leq C \kappa^{1/2} \; .
\]
\item[iii)] For $\kappa >0$ small enough and for every
fixed $k \geq 1$ we have
\begin{equation}\label{eq:init-3}
\lim_{N\to\infty}
\tr \; \Big| \wt \gamma^{(k)}_{N} - |\ph \rangle \langle
\ph|^{\otimes k} \Big| = 0\; .
\end{equation}
\end{enumerate}
\end{proposition}

\begin{proof}
For the proof of part i) and ii) see \cite[Proposition 8.1]{ESY}. To prove iii), we
begin by noticing (see (\ref{eq:1tok2})) that it is enough to show that
\[ \lim_{N \to \infty} \tr \Big| \wt \gamma^{(1)}_{N} - |\ph \rangle \langle
\ph| \Big| = 0 \,. \] Moreover, since the limiting density is an orthogonal projection,
 trace-norm convergence is equivalent to weak* convergence. In other words, it is enough
to prove that, for every compact operator $J^{(1)} \in \cK_1$ and for every $\eps >0$
 there exists $N_0 = N_0 (J^{(1)},\eps)$ such that
\begin{equation}\label{eq:iii1} \Big| \tr \; J^{(1)} \left(  \wt \gamma^{(1)}_{N} -
|\ph \rangle \langle \ph| \right) \Big| \leq \eps  \end{equation} for $N > N_0$.
To show (\ref{eq:iii1}), we start by observing that, from (\ref{eq:initcond}),
there exists a sequence $\xi_N^{(N-1)} \in L^2 (\bR^{3(N-1)})$, with $\| \xi_N^{(N-1)} \| = 1$ such that
\begin{equation}\label{eq:ale} \| \psi_N - \ph \otimes \xi_N^{(N-1)} \| \to 0
\qquad \text{as } N \to \infty \, . \end{equation}
This was proven by Alessandro Michelangeli in \cite{M} using the following argument.
Choose an orthonormal basis $\{ f_i \}_{i \geq 1}$ of $L^2 (\bR^3)$ with $f_1 = \ph$.
Choose also an orthonormal basis $\{ g_j \}_{j \geq 1}$ of $L^2 (\bR^{3(N-1)})$. Then one can write
\[ \psi_N = \sum_{ij} \alpha^{(N)}_{ij} f_i \otimes g_j \]
and
\[ |\psi_N \rangle \langle \psi_N| = \sum_{i,j,i',j'} \overline{\alpha}^{(N)}_{i,j}
 \alpha^{(N)}_{i',j'} |f_i \rangle \langle f_{i'}| \otimes |g_j \rangle \langle g_{j'}| \, . \]
This implies that
\begin{equation*}
\begin{split}
\gamma^{(1)}_N = \; &\sum_{j} \left( |\alpha_{1,j}^{(N)}|^2 |\ph \rangle \langle \ph |  +
\alpha_{1,j}^{(N)} \sum_{i \neq 1}  \overline{\alpha}_{i,j}^{(N)} |\ph \rangle \langle f_i| +
\overline{\alpha}_{1,j}^{(N)} \sum_{i \neq 1} \alpha_{i,j}^{(N)} |f_i \rangle \langle \ph| +
\sum_{i,i' \neq 1} \overline{\alpha}_{i,j}^{(N)} \alpha_{i',j}^{(N)} |f_i \rangle \langle f_{i'}|\right)
\end{split}
\end{equation*}
and therefore, using (\ref{eq:initcond}), that
\[ \sum_{j} |\alpha_{1,j}^{(N)}|^2 \to 1 \] as $N \to \infty$.
Thus, putting $\wt\xi^{(N-1)}_N = \sum_j \alpha_{1,j}^{(N)} g_j$, we get
\[ \| \psi_N - \ph \otimes \wt\xi_N^{(N-1)} \|^2 = \sum_j \sum_{i \neq 1} |\alpha_{i,j}^{(N)}|^2 =
 1 - \sum_j |\alpha_{1,j}^{(N)}|^2 \to 0 \] as $N \to \infty$. It is then simple to
 check that $\xi_N^{(N-1)} = \wt \xi_N^{(N-1)} / \| \wt \xi_N^{(N-1)} \|$ satisfies (\ref{eq:ale}).

\medskip

On the other hand, there exists $\ph_* \in H^2 (\bR^3)$ with $\| \ph_* \|=1$ and such that
\[ \| \ph - \ph_* \| \leq \frac{\eps}{32 \| J^{(1)}\|} \, . \]
Let $\Xi = \chi (\kappa H_N/N)$. Then \[ \| (\Xi -1) \psi_N \|^2 \leq \frac{\kappa}{N}
 \langle \psi_N, H_N \psi_N \rangle \leq C \kappa \] independently of $N$. Therefore,
choosing $\kappa >0$ so small that $\| \Xi \psi_N \| \geq 1/2$, we find
\begin{equation}
\begin{split}
\left\| \frac{\Xi \psi_N}{\| \Xi \psi_N \|} -
 \frac{\Xi \left(\ph_* \otimes \xi_{N}^{(N-1)}\right)}{ \|\Xi \left(\ph_*
\otimes \xi^{(N-1)}_N \right)\|} \right\| & \leq \frac{2}{\| \Xi \psi_N \|}
\left\| \Xi \left( \psi_N - \ph_* \otimes \xi^{(N-1)}_{N} \right) \right\| \\ &
\leq 4 \left\| \psi_N - \ph_* \otimes \xi^{(N-1)}_{N} \right\| \\ &
\leq 4 \left\|  \psi_N - \ph \otimes \xi^{(N-1)}_{N} \right\| + 4 \| \ph - \ph_* \| \\
&\leq \frac{\eps}{6 \| J^{(1)} \|}
\end{split}
\end{equation}
for all $N$ sufficiently large. Next we define the Hamiltonian
\begin{equation}
\wh H_N := -\sum_{j=2}^N \Delta_j + \sum_{1<i<j}^N V_N
(x_i-x_j)\,.
\end{equation}
Note that $\wh H_N$ acts only on the last $N-1$ variables. We set
$\wh\Xi := \chi (\kappa \wh H_N /N)$. Then we claim that, if $\eps >0$ is small enough,
\begin{equation}\label{eq:hatXipsi-ph}
\begin{split}
\Big\| \frac{\Xi \psi_N}{\| \Xi \psi_N \|} - \frac{\wh\Xi \left(
\ph_* \otimes \xi^{(N-1)}_{N} \right)}{\|\wh\Xi \left(
\ph_* \otimes \xi^{(N-1)}_{N} \right)\|} \Big\| \leq
\frac{\e}{3 \| J^{(1)}\|}
\end{split}
\end{equation}
for $N$ sufficiently large. The proof of (\ref{eq:hatXipsi-ph}) can be found in
\cite[Proposition 8.1]{ESY}. To get (\ref{eq:iii1}) we define
\[ \wh\psi_{N} := \frac{\wh\Xi
\left( \ph_* \otimes \xi^{(N-1)}_{N} \right)}{\|\wh\Xi
\left( \ph_* \otimes \xi^{(N-1)}_{N} \right)\|} =
\ph_* \otimes \frac{\wh\Xi \xi^{(N-1)}_{N}}{\|\wh\Xi
\xi^{(N-1)}_{N} \|} \] where we used the fact that $\wh \Xi$ acts only on the last $N-1$
variables and the fact that $\| \ph_*\|=1$. Define
\[ \wh \gamma_N^{(1)} (x_1;x'_1) := \int \rd \bx_{N-1} \, \wh
\psi_{N} (x_1, \bx_{N-k}) \overline{\wh\psi}_{N}
(x'_1,\bx_{N-k})\,. \] Note that $\wh \psi_N$ is not symmetric in
all variables, but it is symmetric in the last
$N-1$ variables. In particular, $\wh\gamma_N^{(1)}$ is a density
matrix and clearly $\wh \gamma_N^{(1)} =
|\ph_* \rangle \langle \ph_*|$. Therefore, since
 $\| \wt \psi_N - \wh \psi_{N} \| \leq \e
/(3\| J^{(1)}\|)$ by (\ref{eq:hatXipsi-ph}) and since $\| \ph -
\ph_*\| \leq \e/(32 \|J^{(1)}\|)$, we have
\begin{equation}
\begin{split}
\Big|\tr \; J^{(1)} \left( \wt\gamma^{(1)}_{N} - |\ph \rangle
\langle \ph| \right) \Big| \leq \; &\Big|\tr \; J^{(1)}
\left( \wt\gamma^{(1)}_{N} - |\ph_* \rangle \langle \ph_*|
\right) \Big| + \Big| \tr \; J^{(1)} \left( |\ph_* \rangle
\langle \ph_*| - |\ph\rangle \langle \ph|
\right)\Big| \\ \leq \; &2 \| J^{(1)} \| \, \| \wt \psi_N - \wh
\psi_{N} \| + 2 \| J^{(1)}\| \, \| \ph - \ph_* \| \leq \e
\end{split}
\end{equation}
for $N$ sufficiently large (for arbitrary $\kappa, \e
>0$ small enough). This proves (\ref{eq:iii1}).
\end{proof}

\section{Poincar{\'e}-Sobolev type inequalities}
\setcounter{equation}{0}

In the proof of the convergence we need to estimate potentials converging to a
delta functions, and their difference to a normalized delta-function.
To this end we make use of the following three lemmas.

\begin{lemma}\label{lm:VL1}
Suppose $V \in L^1 (\bR^3)$. Then
\begin{equation}
\begin{split}
\left|\langle \ph, V (x_1 -x_2) \psi \rangle \right| \leq C \| V \|_1 \, \langle
&\psi, \left( (\nabla_1 \cdot \nabla_2)^2 -\Delta_1 - \Delta_2 + 1 \right) \psi \rangle^{1/2} \; \\ &
 \times \langle \ph, \left( (\nabla_1 \cdot \nabla_2)^2 -\Delta_1 - \Delta_2 + 1 \right)
\ph \rangle^{1/2}\end{split}
\end{equation}
for every $\psi,\ph \in L^2 ( \bR^6, \rd x_1 , \rd x_2)$.
\end{lemma}

\begin{proof}
Switching to Fourier space, we find
\begin{equation}
\begin{split}
\langle \ph, V (x_1 -x_2) \psi \rangle
=\; & \int \rd p_1 \rd p_2 \rd q_1 \rd q_2 \; \overline{\widehat\ph (p_1, p_2)}
\widehat{\psi} (q_1, q_2) \, \widehat{V} (q_1 -p_1)\, \delta (p_1 + p_2 - q_1 -q_2) \, .
\end{split}
\end{equation}
Therefore, by a weighted Schwarz inequality,
\begin{equation}
\begin{split}
\Big| \langle \ph, V &(x_1 -x_2) \psi \rangle \Big|  \\
\leq \; & \| \widehat{V} \|_{\infty} \left( \int \rd p_1 \rd p_2 \rd q_1 \rd q_2 \;
\frac{(p_1 \cdot p_2)^2 + p_1^2 + p^2_2 +1}{(q_1 \cdot q_2)^2 + q_1^2 + q^2_2 +1} \,
 |\widehat\ph (p_1, p_2)|^2 \delta (p_1 + p_2 - q_1 -q_2) \right)^{1/2} \\ &
\hspace{1cm} \times \left(\int \rd p_1 \rd p_2 \rd q_1 \rd q_2
\frac{(q_1 \cdot q_2)^2 + q_1^2 + q^2_2 +1}{(p_1 \cdot p_2)^2 + p_1^2 + p^2_2 +1}
 |\widehat{\psi} (q_1, q_2)|^2 \,\delta (p_1 + p_2 - q_1 -q_2)\right)^{1/2} \\
\leq \; & \| V \|_{1} \; \left( \sup_{p} \int \rd q \;
\frac{1}{(q \cdot (p -q))^2 + q^2 + (p - q)^2 +1}\right) \\ & \hspace{1cm} \times
\left\langle \psi, \left((\nabla_1 \cdot \nabla_2)^2 -\Delta_1 - \Delta_2 + 1 \right)
\psi \right\rangle^{1/2} \, \left\langle \ph, \left((\nabla_1 \cdot \nabla_2)^2 -\Delta_1 - \Delta_2 + 1 \right)
\ph \right\rangle^{1/2}\,.
\end{split}
\end{equation}
The lemma will then follow from
\begin{equation}\label{eq:int1}
\sup_{p \in \bR^3} \int \rd q \; \frac{1}{(q \cdot (p -q))^2 + q^2 + (p - q)^2 +1} < \infty \, . \end{equation}
To prove (\ref{eq:int1}), we proceed as follows.
\begin{equation}\label{eq:int2}
\begin{split}
\int \rd q \; \frac{1}{(q \cdot (p -q))^2 + q^2 + (p - q)^2 +1} = \; & \int_{|q-\frac{p}{2}| > |p|}
\rd q \; \frac{1}{\left( \left(q-\frac{p}{2}\right)^2 - \frac{p^2}{4}\right)^2 + q^2 + (p - q)^2 +1} \\ &
+ \int_{|q-\frac{p}{2}| < |p|} \rd q \; \frac{1}{\left( \left(q-\frac{p}{2}\right)^2 -
\frac{p^2}{4}\right)^2 + q^2 + (p - q)^2 +1}\,.
\end{split}
\end{equation}
The first term on the r.h.s. of the last equation is bounded by
\begin{equation}
\begin{split}
\int_{|q-\frac{p}{2}| > |p|} \rd q \; \frac{1}{\left( \left(q-\frac{p}{2}\right)^2 - \frac{p^2}{4}\right)^2
 + q^2 + (p - q)^2 +1} \leq \; & \int_{|q-\frac{p}{2}| > |p|} \rd q \; \frac{1}{\frac{9}{16}
\left|q-\frac{p}{2}\right|^4 +1}
\\ \leq \; & \frac{16}{9} \int_{\bR^3} \rd q \; \frac{1}{|q|^4  +1} < \infty\,,
\end{split}
\end{equation}
uniformly in $p \in \bR^3$. As for the second term on the r.h.s. of (\ref{eq:int2}), we observe that
\begin{equation}
\begin{split}
\int_{|q-\frac{p}{2}| < |p|} \rd q \; &\frac{1}{\left( \left(q-\frac{p}{2}\right)^2 -
\frac{p^2}{4}\right)^2 + q^2 + (p - q)^2 +1} \\ = \; & \int_{|x| < |p|} \rd x \;
\frac{1}{\left( x^2 - \frac{p^2}{4}\right)^2 + \left( x+\frac{p}{2} \right)^2 + \left(x-\frac{p}{2} \right)^2 +1} \\
= \; & 4\pi \int_{0}^{|p|} \rd r \frac{r^2}{\left( r^2 - \frac{|p|^2}{4}\right)^2 +  2r^2 + \frac{|p|^2}{2} +1} \\
\leq \; & C |p|^2 \int_{-|p|/2}^{|p|/2} \rd r \frac{1}{ r^2 \left( r + |p|\right)^2 +
 \left( r+\frac{|p|}{2} \right)^2 + \frac{|p|^2}{4} +1}  \\
\leq\; & C \int_{-|p|/2}^{|p|/2} \rd r \, \frac{1}{ r^2  + 1}  \leq C \int_{\bR} \rd r \, \frac{1}{r^2 + 1} < \infty ,
\end{split}
\end{equation}
uniformly in $p$.
\end{proof}

\begin{lemma}\label{lm:VL12}
Suppose $V \in L^1 (\bR^3)$ with $\int \, V(x) \rd x = 1$. For $\alpha >0$ , let
$V_{\alpha} (x) = \alpha^{-3} V (x/\alpha)$. Then there exists a sequence
$\beta_{\alpha}$ with $\beta_{\alpha} \to 0$ as $\alpha \to 0$ such that
\begin{equation}\label{eq:VL12}
\begin{split}
\left|\langle \ph, \left( V_{\alpha} (x_1 -x_2) - \delta (x_1 -x_2) \right) \psi \rangle
\right| \leq C\beta_{\alpha}  \, \langle &\psi, \left( (\nabla_1 \cdot \nabla_2)^2 -\Delta_1
- \Delta_2 + 1 \right) \psi \rangle^{1/2} \; \\ & \times \langle \ph, \left( (\nabla_1-\nabla_2)^4
+ (\nabla_1 + \nabla_2)^2 +  1 \right) \ph \rangle^{1/2},
\end{split}
\end{equation}
for all $\ph,\psi \in L^2 (\bR^6)$.
\end{lemma}
\begin{proof}
Switching to Fourier space we find
\begin{equation*}
\left\langle \ph, \Big( V_{\alpha} (x_1 -x_2) - \delta (x_1 -x_2) \Big) \psi \right\rangle
=  \text{I} + \text{II} \, ,
\end{equation*}
where we defined
\begin{equation}\label{eq:IandII}
\begin{split}
\text{I} = \; & \int_{| x \cdot (p_1 - q_1)| < \alpha^{-1/2}} \rd p_1 \rd p_2 \rd q_1 \rd q_2
\rd x \, V(x) \, \overline{\widehat{\ph}}  (p_1, p_2) \left( e^{i\alpha x \cdot (p_1 -q_1)} - 1 \right)
 \widehat{\psi} (q_1, q_2) \delta (p_1 +p_2 - q_1 -q_2) , \\
\text{II} = \; & \int_{| x \cdot (p_1 - q_1)| \geq \alpha^{-1/2}} \rd p_1
\rd p_2 \rd q_1 \rd q_2 \rd x \, V(x) \, \overline{\widehat{\ph}} (p_1, p_2)
\left( e^{i\alpha x \cdot (p_1 -q_1)} - 1 \right) \widehat{\psi}  (q_1, q_2)
\delta (p_1 +p_2 - q_1 -q_2)\,.
\end{split}
\end{equation}
To bound the first term we use that $|e^{i\kappa} - 1| \leq |\kappa|$,
for $\kappa \in \bR$, and we observe that
\begin{equation}\label{eq:I1}
\begin{split}
|\text{I}| \leq \; &\alpha^{1/2} \| V \|_1 \int \rd p_1 \rd p_2 \rd q_1 \rd q_2
\frac{\sqrt{|p_1 - p_2|^4 + (p_1 + p_2)^2 + 1}}{\sqrt{(q_1 \cdot q_2)^2 + q_1^2 + q_2^2 +1}} \,
 |\widehat{\ph} (p_1, p_2)| \\ &\hspace{4cm} \times
\frac{\sqrt{(q_1 \cdot q_2)^2 + q_1^2 + q_2^2 +1}}{\sqrt{|p_1 - p_2|^4 + (p_1 + p_2)^2 + 1}}
\, |\widehat{\psi} (q_1, q_2)| \, \delta (p_1 +p_2 - q_1 -q_2)
\end{split}
\end{equation}
With a Schwarz inequality, we obtain that
\begin{equation*}\begin{split}
|\text{I}| \leq \; &\alpha^{1/2} \| V \|_1 \left( \int \rd p_1 \rd p_2 \rd q_1 \rd q_2 \,
 \frac{|p_1 - p_2|^4 + (p_1 + p_2)^2 + 1}{(q_1 \cdot q_2)^2 + q_1^2 + q_2^2 +1} \,
 |\widehat{\ph} (p_1, p_2)|^2 \, \delta (p_1 + p_2 - q_1 - q_2) \right)^{1/2} \\ &
 \times \left( \int \rd p_1 \rd p_2 \rd q_1 \rd q_2  \,
 \frac{\sqrt{(q_1 \cdot q_2)^2 + q_1^2 + q_2^2 +1}}{\sqrt{|p_1 - p_2|^4 + (p_1 + p_2)^2 + 1}}
\,|\widehat{\psi} (q_1, q_2)|^2 \, \delta (p_1 +p_2 - q_1 -q_2)\right)^{1/2} \\
\leq \; &\alpha^{1/2} \| V \|_1 \; \Big\langle \ph, \left( (\nabla_1 - \nabla_2)^4 +
(\nabla_1 + \nabla_2)^2 + 1 \right) \ph \Big\rangle^{1/2} \,  \Big\langle \psi,
\left( (\nabla_1 \cdot \nabla_2)^2 -\Delta_1 - \Delta_2 + 1 \right) \psi \Big\rangle^{1/2} \\ &
\times\left( \sup_{p \in \bR^3} \int \frac{\rd q}{|q-p|^4 + p^2 + 1} \right)^{1/2}
 \,\left( \sup_{q \in \bR^3} \int \frac{\rd p}{(p\cdot (q-p))^2 + p^2 + (q-p)^2+ 1} \right)^{1/2}\,.
\end{split}
\end{equation*}
{F}rom \[  \sup_{p\in \bR^3} \int \frac{\rd q}{|q-p|^4 + p^2 + 1} \leq \int
\frac{\rd q}{|q|^4 + 1} < \infty \] and (\ref{eq:int1}) it follows that
\begin{equation}\label{eq:Ifin} |\text{I}| \leq C \alpha^{1/2} \Big\langle \ph,
\left( (\nabla_1 - \nabla_2)^4 + (\nabla_1 + \nabla_2)^2 + 1 \right) \ph \Big\rangle^{1/2} \,
  \Big\langle \psi, \left( (\nabla_1 \cdot \nabla_2)^2 -\Delta_1 - \Delta_2 + 1 \right)
\psi \Big\rangle^{1/2} \, .\end{equation}

In order to control the second term in (\ref{eq:IandII}), we bound it by
\begin{equation}\label{eq:II1}
\begin{split}
| \text{II}| \leq \; &2 \int_{| x \cdot (p_1 - q_1)| \geq \alpha^{-1/2}} \rd p_1 \rd p_2
\rd q_1 \rd q_2 \rd x \, |V(x)| \, |\widehat\ph (p_1, p_2)| \, | \widehat\psi (q_1, q_2)|
\, \delta (p_1 +p_2 - q_1 -q_2) \\ \leq \; &2 \int_{|x| \geq \alpha^{-1/4}} \rd p_1
\rd p_2 \rd q_1 \rd q_2 \rd x \, |V(x)| \, |\widehat\ph (p_1, p_2)| \,
 |\widehat \psi (q_1, q_2)| \, \delta (p_1 +p_2 - q_1 -q_2) \\ &+ 2 \int_{|p_1-q_1|
\geq \alpha^{-1/4}} \rd p_1 \rd p_2 \rd q_1 \rd q_2 \rd x \, |V(x)| \, |\ph (p_1, p_2)| \,
| \psi (q_1, q_2)| \, \delta (p_1 +p_2 - q_1 -q_2)
\\ \leq \; &\beta_{1,\alpha} \int \rd p_1 \rd p_2 \rd q_1 \rd q_2 \, |\widehat\ph (p_1, p_2)| \,
 |\widehat \psi (q_1, q_2)| \, \delta (p_1 +p_2 - q_1 -q_2) \\ &+2\| V \|_1  \int_{|p_1-q_1| \geq
 \alpha^{-1/4}} \rd p_1 \rd p_2 \rd q_1 \rd q_2 \, |\widehat\ph (p_1, p_2)| \,
 | \widehat\psi (q_1, q_2)| \, \delta (p_1 +p_2 - q_1 -q_2)\\ \leq \; &\beta_{1,\alpha}
 \left\langle \ph , \left((\nabla_1 - \nabla_2)^4 + (\nabla_1 + \nabla_2)^2 +1 \right) \ph \right\rangle^{1/2}
 \, \left\langle \psi , \left((\nabla_1 \cdot \nabla_2)^4 - \Delta_1 - \Delta_2 +1 \right)
\ph \right\rangle^{1/2}  \\ &+2\| V \|_1  \int_{|p_1-q_1| \geq \alpha^{-1/4}} \rd p_1
\rd p_2 \rd q_1 \rd q_2 \, |\widehat\ph (p_1, p_2)| \, | \widehat\psi (q_1, q_2)| \, \delta (p_1 +p_2 - q_1 -q_2)\, ,
\end{split}
\end{equation}
where we defined \[ \beta_{1,\alpha} = 2 \int_{|x| \geq \alpha^{-1/4}} |V (x)| \, , \]
and we bounded the first integral analogously as we did with the integral in (\ref{eq:I1}).
 Note that $\beta_{1,\alpha} \to 0$ as $\alpha \to 0$, because $V \in L^1 (\bR^3)$.
We still need to control the last integral, on the r.h.s. of the last equation. To this end, we observe that
\begin{equation}\label{eq:II2}
\begin{split}
\int_{|p_1-q_1| \geq \alpha^{-1/4}} \rd p_1 &\rd p_2 \rd q_1 \rd q_2 \,
|\widehat\ph (p_1, p_2)| \, |\widehat \psi (q_1, q_2)| \, \delta (p_1 +p_2 - q_1 -q_2) \\
 \leq \; & 2 \int_{|q_1| \geq \alpha^{-1/4}/8} \rd p_1 \rd p_2 \rd q_1 \rd q_2 \,
 |\widehat\ph (p_1, p_2)| \, |\widehat \psi (q_1, q_2)| \, \delta (p_1 +p_2 - q_1 -q_2) \\ &+
\int_{|q_2| \geq \alpha^{-1/4}/8} \rd p_1 \rd p_2 \rd q_1 \rd q_2 \, |\widehat\ph (p_1, p_2)| \,
 |\widehat \psi (q_1, q_2)| \, \delta (p_1 +p_2 - q_1 -q_2) \\ &+\int_{\stackrel{|p_1| \geq
 \alpha^{-1/4}/2}{|q_1 + q_2| \leq \alpha^{-1/4}/4}} \rd p_1 \rd p_2 \rd q_1 \rd q_2 \,
|\widehat\ph (p_1, p_2)| \, |\widehat \psi (q_1, q_2)| \, \delta (p_1 +p_2 - q_1 -q_2)\,.
\end{split}
\end{equation}
The first two terms can be bounded by
\begin{equation}\label{eq:bdq}
\begin{split}
\int_{|q_j| \geq \alpha^{-1/4}/8} &\rd p_1 \rd p_2 \rd q_1 \rd q_2 \, |\widehat\ph (p_1, p_2)| \,
 | \widehat\psi (q_1, q_2)| \, \delta (p_1 +p_2 - q_1 -q_2) \\ \leq \; &C \alpha^{1/12}
\left\langle \ph , \left((\nabla_1 - \nabla_2)^4 + (\nabla_1 + \nabla_2)^2 +1 \right)
\ph \right\rangle^{1/2} \, \left\langle \psi , \left((\nabla_1 \cdot \nabla_2)^4 -
\Delta_1 - \Delta_2 +1 \right) \ph \right\rangle^{1/2}\,
\end{split}
\end{equation}
which holds for both $j=1,2$, and for a universal constant $C$, independent of $\alpha,
\ph,\psi$. To show (\ref{eq:bdq}) note that, proceeding as in (\ref{eq:I1}) (for example
 for $j=1$), we have
\begin{equation}
\begin{split}
&\int_{|q_1| \geq \alpha^{-1/4}/8} \rd p_1 \rd p_2 \rd q_1 \rd q_2 \, |\widehat\ph (p_1, p_2)|
\, |\widehat \psi (q_1, q_2)| \, \delta (p_1 +p_2 - q_1 -q_2) \\ &\hspace{.1cm} \leq \;
\left\langle \ph , \left((\nabla_1 - \nabla_2)^4 + (\nabla_1 + \nabla_2)^2 +1 \right) \ph
\right\rangle^{1/2} \, \left\langle \psi , \left((\nabla_1 \cdot \nabla_2)^4 -
\Delta_1 - \Delta_2 +1 \right) \ph \right\rangle^{1/2} \\ &\hspace{.3cm} \times
\left( \sup_{q \in \bR^3} \int \frac{\rd p}{(p\cdot (q-p))^2 + p^2 + (q-p)^2 + 1}
\right)^{1/2} \, \left(\sup_{p \in \bR^3} \int_{|q| \geq \alpha^{-1/4}/8}
 \frac{\rd q}{|q-p|^4 + p^2 + 1}\right)^{1/2}
\end{split}
\end{equation}
and thus (\ref{eq:bdq}) follows from (\ref{eq:int1}) and
\begin{equation}
\begin{split}
\sup_{p\in \bR^3} \int_{|q| \geq \alpha^{-1/4}/8} \frac{\rd q}{|q-p|^4 + p^2 + 1} &
\leq  (8\alpha^{1/4})^{1/3} \sup_{p\in \bR^3} \int \rd q \frac{ |q|^{1/3}}{|q-p|^4 + p^2 + 1} \\ &
 \leq (8\alpha^{1/4})^{1/3} \left( \sup_{q,p \in \bR^3}
\frac{|q+p|^{1/3}}{(|q|^4 + p^2 + 1)^{1/6}} \right) \, \int \frac{\rd q}{(|q|^4 + 1)^{5/6}} \\ &
\leq C \alpha^{1/12}\,.
\end{split}
\end{equation}
As for the last term on the r.h.s. of (\ref{eq:II2}), we note that
\begin{equation}
\begin{split}
&\int_{\stackrel{|p_1| \geq \alpha^{-1/4}/2}{|q_1 + q_2| \leq \alpha^{-1/4}/4}} \rd p_1
 \rd p_2 \rd q_1 \rd q_2 \, |\widehat\ph (p_1, p_2)| \, | \widehat\psi (q_1, q_2)| \,
\delta (p_1 +p_2 - q_1 -q_2) \\& \leq \left\langle \ph , \left((\nabla_1 - \nabla_2)^4 +
 (\nabla_1 + \nabla_2)^2 +1 \right) \ph \right\rangle^{1/2} \, \left\langle \psi ,
\left((\nabla_1 \cdot \nabla_2)^4 - \Delta_1 - \Delta_2 +1 \right) \ph \right\rangle^{1/2} \\ &
\hspace{.2cm} \times \left( \sup_{p\in \bR^3} \int \frac{\rd q}{|q-p|^4 + p^2 + 1} \right)^{1/2}
\left( \sup_{q} \int_{\stackrel{|p| \geq 2 |q|}{|p| \geq \alpha^{-1/4}/2}}
\frac{\rd p}{(p \cdot (q-p))^2 + p^2 + (q-p)^2 + 1} \right)^{1/2}\,.
\end{split}
\end{equation}
Since
\begin{equation}
\begin{split}
\int_{\stackrel{|p| \geq 2 |q|}{|p| \geq \alpha^{-1/4}/2}} &
\frac{\rd p}{(p \cdot (q-p))^2 + p^2 + (q-p)^2 + 1} \\ & \leq
(2\alpha^{1/4})^{1/3} \int_{|p| \geq 2 |q|} \rd p \, \frac{|p|^{1/3}}{\left(\left(p-\frac{q}{2} \right)^2
- \frac{q^2}{4} \right)^2 + p^2 + (q-p)^2 + 1} \\ & \leq C \alpha^{1/12} \int_{|p| \geq |q|} \rd p \,
\frac{\left|p+ \frac{q}{2} \right|^{1/2}}{\left(p^2 - \frac{q^2}{4} \right)^2 + p^2 +
\frac{q^2}{4} + 1} \\ & \leq C \alpha^{1/12} \int \rd p \frac{|p|^{1/2}}{\frac{9}{16} |p|^4 + 1}
 \\ & \leq C \alpha^{1/12}\, ,
\end{split}
\end{equation}
it follows that the last term on the r.h.s. of (\ref{eq:II2}) is bounded by
\begin{equation}
\begin{split}
&\int_{\stackrel{|p_1| \geq \alpha^{-1/4}/2}{|q_1 + q_2| \leq \alpha^{-1/4}/2}} \rd p_1
\rd p_2 \rd q_1 \rd q_2 \, |\widehat\ph (p_1, p_2)| \, |\widehat \psi (q_1, q_2)| \,
\delta (p_1 +p_2 - q_1 -q_2) \\& \leq C \alpha^{1/12} \, \left\langle \ph ,
\left((\nabla_1 - \nabla_2)^4 + (\nabla_1 + \nabla_2)^2 +1 \right) \ph \right\rangle^{1/2} \,
\left\langle \psi , \left((\nabla_1 \cdot \nabla_2)^4 - \Delta_1 - \Delta_2 +1 \right) \ph
\right\rangle^{1/2}\,.
\end{split}
\end{equation}
{F}rom the last equation, (\ref{eq:bdq}), (\ref{eq:II2}), and (\ref{eq:II1}),
it follows that
\begin{equation*}
\begin{split}
|\text{II}| \leq \; &C (\beta_{1,\alpha} + \alpha^{1/12}) \\ &\hspace{.5cm} \times \left\langle \ph ,
 \left((\nabla_1 - \nabla_2)^4 + (\nabla_1 + \nabla_2)^2 +1 \right) \ph \right\rangle^{1/2} \,
\left\langle \psi , \left((\nabla_1 \cdot \nabla_2)^4 - \Delta_1 - \Delta_2 +1 \right)
\ph \right\rangle^{1/2}\,.
\end{split}
\end{equation*}
This together with (\ref{eq:Ifin}), implies (\ref{eq:VL12}) with $\beta_{\alpha} =
 C (\beta_{1,\alpha} + \alpha^{1/12} + \alpha^{1/2})$.
\end{proof}

When dealing with the limit points $\gamma^{(k)}_{\infty,t}$, for which we have stronger
 a-priori estimates, we will make use of the following lemma, whose proof can be found in
\cite{ESY2} (Lemma 8.2).
\begin{lemma}\label{lm:sobsob}   Suppose that
$\delta_\alpha(x)$ is  a function satisfying $0 \leq \delta_{\alpha}
(x) \leq C \alpha^{-3} {\bf 1} (|x| \leq \alpha)$ and $\int
\delta_{\alpha} (x) \rd x=1$ (for example $\delta_{\alpha} (x) =
\alpha^{-3} g (x/\alpha)$, for a bounded probability density $g(x)$
supported in $\{ x : |x| \leq 1\}$). Moreover, for $J^{(k)} \in
\cK_k$, and for $j=1,\dots ,k$, we define the norm
\be\label{eq:Jnorm} \tri J^{(k)} \tri_{j} := \sup_{\bx_k, \bx'_k}
\la x_1 \ra^4 \dots \la x_k \ra^4 \la x'_1 \ra^4 \dots \la x'_k
\ra^4  \left( |J^{(k)} (\bx_k ; \bx'_k)| + |\nabla_{x_j} J^{(k)}
(\bx_k;\bx'_k)| + |\nabla_{x'_j} J^{(k)} (\bx_k;\bx'_k)| \right) \,
\ee and $S_j = (1-\Delta_{x_j})$ (here $\la x \ra^2 := 1+ x^2$).
 Then if $\gamma^{(k+1)} (\bx_{k+1};\bx'_{k+1})$ is the
kernel of a density matrix on $L^2 (\bR^{3(k+1)})$, we have, for any
$j\leq k$,
\begin{multline}\label{eq:gammaintbound}
\Big| \int \rd \bx_{k+1} \rd \bx'_{k+1} \, J^{(k)} (\bx_k ; \bx'_k)
\left(\delta_{\alpha_1} (x_{k+1} - x'_{k+1}) \delta_{\alpha_2} (x_j
-x_{k+1}) - \delta (x_{k+1} -x'_{k+1}) \delta (x_j -
x_{k+1})\right)\\ \times \gamma^{(k+1)} (\bx_{k+1} ; \bx'_{k+1})
\Big| \\ \leq C_k \, \tri J^{(k)} \tri_j \left( \alpha_1 +
\sqrt{\alpha_2}\right) \, \tr \, | S_j S_{k+1} \gamma^{(k+1)} S_j
S_{k+1}|\;.
\end{multline}
The same bound holds if $x_j$ is replaced with $x_j'$ in
(\ref{eq:gammaintbound}) by symmetry.
\end{lemma}

\appendix

\section{Properties of the cutoff function $\theta_i^{(n)}$}
\setcounter{equation}{0}

Recall that the cutoff functions
$\Theta_k^{(n)}=\Theta_k^{(n)}(\bx)$ defined for $k=1,\dots,N$ and
$n \in \bN$, in Eq. (\ref{eq:thetan}). In the following lemma, whose proof can be found in
\cite[Appendix A]{ESY}, we collect some of their important properties which were used in the
energy estimate, Proposition \ref{prop:hk}.

\begin{lemma}\label{lm:theta}
\begin{itemize}
\item[i)] The functions $\Theta_k^{(n)}$ are monotonic in both
indices, that is for any $n, k \in \bN$,
\[ \Theta_{k+1}^{(n)}  \leq \Theta_k^{(n)}  \leq 1\; ,\qquad
 \Theta_k^{(n+1)}  \leq \Theta_k^{(n)}  \leq 1 \; .
 \]
Moreover,  $\Theta^{(n)}_k$ is permutation symmetric  in the first
$k$ and the last $N-k$ variables.
\item[ii)] For every $k =1,\dots ,N$, $n \in\bN$ we have
\begin{equation}\label{eq:lmthetaiv}
\begin{split}
\sum_{j=1}^N &\frac{\left| \nabla_j \Theta_k^{(n)}
\right|^2}{\Theta_k^{(n)}} \leq C \ell^{-2} \Theta_k^{(n-1)}
\end{split}
\end{equation}
\item[iii)] For every fixed $k =1,\dots,N$ and $n \in \bN$ we have
\begin{equation}\label{eq:lmthetavi}
\begin{split}
\sum_{i,j} \left| \nabla_i \nabla_j \Theta_k^{(n)} \right| \leq C
\ell^{-2} \Theta_k^{(n-1)}\,.
\end{split}
\end{equation}
\end{itemize}
\end{lemma}

\section{Removal of the assumption on derivatives of $V$}
\label{app:nablaV}

The goal of this appendix is to explain how the assumption
\begin{equation}\label{eq:assV2}
 |\nabla^{\alpha} V (x)| \leq C \qquad \text{for all } x \in \bR^3, \; |\alpha|\leq 2
\end{equation}
in Theorem \ref{thm:main} can be removed. The main observation is that (\ref{eq:assV2}) is only
used in the proof of the higher order energy estimate, Proposition \ref{prop:hk}, in the form
 $\|\nabla V_N\|_\infty\leq CN^3$, $\|\nabla^2 V_N\|_\infty \leq CN^4$. More precisely, the
 estimate on $\|\nabla V_N\|_\infty$ is first used in the study of the third term on the r.h.s.
of (\ref{eq:Tk+1}) (the third term on the r.h.s. of (\ref{eq:for}) in the case $k=2$); namely
 the term containing the commutator $[T^{1/2}, D_k] = [(H_N+N)^{1/2}, \nabla_1 \dots \nabla_k ]$.
  Bounds on the first and second derivatives are also used in the proof of Lemma \ref{lemma:exp}.
However, in both cases, the final estimates turn out to be subexponentially small in $N$ (see
(\ref{eq:nablaVterm}) and  (\ref{eq:commm})). For this reason, the proof of Proposition
\ref{prop:hk} remains unchanged if, instead of (\ref{eq:assV2}), we allow $V= V^{(N)}$
to depend on $N$ and only assume
\be
\| \nabla^\alpha V^{(N)}\|_\infty \leq e^{cN^\kappa}, \quad |\alpha|\leq 2,
\label{eq:sube}
\ee
for some sufficiently small $\kappa>0$.

\medskip

More precisely, suppose that the potential $V\geq 0$ satisfies $V(x) \leq C
\langle x \rangle^{-\sigma}$ for some $\sigma >5$, with no assumptions on the derivatives
$\nabla^{\alpha} V$, for $|\alpha| \geq 1$. Then consider the evolution $\psi_{N,t} =
 e^{-iH_N t} \psi_N$ of an initial $N$-body wave function $\psi_N$ satisfying the two assumptions
 (\ref{eq:assH1}) and (\ref{eq:asscond}), with respect to the evolution generated by the Hamiltonian
\[ H_N = \sum_{j=1}^N -\Delta_j + \sum_{i<j}^N V_N (x_i -x_j) \] with $V_N (x) = N^2 V (Nx)$.
As in Theorem \ref{thm:main} we claim that, for every fixed $t \in \bR$, and every $k \geq 1$,
the $k$-particle marginal $\gamma^{(k)}_{N,t}$ associated with $\psi_{N,t}$ is such that
\be\label{eq:clapp} \gamma_{N,t}^{(k)} \to |\ph_t \rangle \langle \ph_t|^{\otimes k} \end{equation}
as $N \to \infty$, w.r.t. to the trace-norm topology.

\medskip

To prove (\ref{eq:clapp}) we can assume, without loss of generality, that the initial data $\psi_N$ is such that
\begin{equation}\label{eq:psiN2} \langle \psi_N, H_N^k \psi_N \rangle \leq C^k N^k \, .
\end{equation}
In fact, if this is not the case, we can use the argument outlined in Section \ref{sec:outline}
(in the proof of Theorem \ref{thm:main}) and based on the analysis of Section \ref{sec:appro}
(which does not use any assumption on the derivatives of $V$) to approximate $\psi_N$.

\medskip

{F}rom Theorem \ref{thm:compactness} it follows that the sequence $\Gamma_{N,t}^{(k)} =
 \{\gamma^{(k)}_{N,t} \}_{k \geq 1}$ is compact with respect to the product topology
$\tau_{\text{prod}}$ defined in Section \ref{sec:outline}. If we could prove,
similarly to Theorem \ref{thm:aprik}, that an arbitrary limit point $\Gamma_{\infty,t} =
\{\gamma_{\infty,t}^{(k)} \}_{k\geq 1}$ satisfies the a-priori estimates (\ref{eq:aprik}),
it would follow from Theorem \ref{thm:conv} that $\Gamma_{\infty,t}$ is a solution to the
infinite hierarchy (\ref{eq:BBGKYinf}) and, by the uniqueness result of Theorem~\ref{thm:uniqueness},
we could conclude the proof of (\ref{eq:clapp}) using the same strategy outlined in Section
\ref{sec:outline}.

\medskip

To prove that every limit point $\Gamma_{\infty,t}$ satisfies the a-priori estimates
(\ref{eq:aprik}), we introduce a potential $\wt V^{(N)} = V * \nu^{(N)}$, where $\nu^{(N)} (x)
 = (\const) \,  e^{3N^{\kappa}/2} \exp (- e^{N^{\kappa}} x^2)$ with some sufficiently small
 $\kappa>0$ (here the constant is chosen so that $\int \rd x \, \nu^{(N)} (x) =1$), and
we consider the evolution $\wt \psi_{N,t} = e^{-i\wt H_N t} \psi_N$ of the initial data
$\psi_N$ with respect to the modified Hamiltonian
\[
\wt H_N = -\sum_{j=1}^N \Delta_j + \sum^N_{i<j} N^2 \wt V^{(N)}
(N (x_i -x_j)) \,.
\]
The potential $\wt V^{(N)}$ satisfies the bounds $\|\nabla^{\alpha}\wt V^{(N)}\|_{L^{\infty}}
\leq C e^{|\alpha| N^{\kappa}/2}$ for all $|\alpha| \leq 2$. As we remarked above, this very
weak control on the $L^{\infty}$ norm of $\nabla^{\alpha} \wt V^{(N)}$ is enough to prove
Proposition \ref{prop:hk}. Therefore, it follows from Theorem \ref{thm:aprik} that for any
fixed $t\in \bR$ every limit point $\wt \Gamma_{\infty,t} =
\{ \wt \gamma^{(k)}_{\infty,t} \}_{k\geq 1}$ of the sequence $\wt \Gamma_{N,t} =
 \{ \wt \gamma_{N,t}^{(k)} \}_{k=1}^N$ (w.r.t. to the product of the weak* topologies)
satisfies the bound
\be\label{eq:clapp3} \tr \; (1-\Delta_1) \dots (1- \Delta_k) \wt \gamma_{\infty,t}^{(k)}
\leq C^k \ee for all $k \geq 1$. To show that a limit point $\Gamma_{\infty,t} =
 \{\gamma_{\infty,t}^{(k)} \}_{k\geq 1}$ of the original sequence $\Gamma_{N,t} =
\{ \gamma_{N,t}^{(k)} \}_{k \geq 1}$ also satisfies this bound, it is therefore
enough to prove that, for every fixed $t \in \bR$, \be \label{eq:clapp2}
\| \psi_{N,t} - \wt \psi_{N,t} \| \to 0 \ee as $N \to \infty$; in fact (\ref{eq:clapp2})
immediately implies that every limit point $\gamma_{\infty,t}^{(k)}$ of $\gamma_{N,t}^{(k)}$
is also a limit point of the sequence $\wt \gamma_{N,t}^{(k)}$ and therefore satisfies (\ref{eq:clapp3}).

\medskip

To verify (\ref{eq:clapp2}), we observe that
\begin{equation}
\begin{split}
\frac{\rd}{\rd t} \Big\| \psi_{N,t} - \wt \psi_{N,t} \Big\|^2 = \; &2 \, \text{Im} \,
\langle \psi_{N,t},
\left(H_N - \wt H_N \right) \wt \psi_{N,t} \rangle \\ = \; & N^3(N-1) \, \text{Im} \,
\langle \psi_{N,t},
\left( V * ( \delta - \nu_N) \right) (N(x_1 - x_2)) \wt \psi_{N,t} \rangle \\ = \; &N^3(N-1)\,
\text{Im} \, \int \rd \bx \rd y \; \overline{\psi}_{N,t} (\bx)  \, V (y) (\delta -
\nu^{(N)}) (N(x_1-x_2)-y) \wt \psi_{N,t} (\bx) \\
= \; &(N-1)\, \text{Im} \, \int \rd y \; V (y) \int \rd \bx \; \overline{\psi}_{N,t} (\bx)
 \, (\delta - \nu_N) (x_1-x_2-y/N) \wt \psi_{N,t} (\bx)\,
\end{split}
\end{equation}
where we defined $\nu_N (x) = N^3 \nu^{(N)} (Nx)$ (this implies that $\nu_N (x) = (\const) \,
N^3 e^{3N^{\kappa}/2} e^{-N^2 \, e^{N^{\kappa}} x^2}$ is still normalized with $\| \nu_N \|_1=1$).
 Therefore
\begin{equation}\label{eq:diffe}
\begin{split}
\Big| \frac{\rd}{\rd t} \Big\| \psi_{N,t} - \wt \psi_{N,t} \Big\|^2 \Big| \leq \; &C N \| V \|_1 \;
\sup_{y \in \bR^3} \Big| \langle \psi_{N,t}, \left(\delta_y - \nu_{N,y} \right) (x_1 -x_2)
 \wt \psi_{N,t} \rangle \Big|
\end{split}
\end{equation}
with $\delta_y (x) = \delta (x-y)$ and $\nu_{N,y} (x) = \nu_{N} (x-y)$. It is simple to check that
\begin{equation}\label{eq:poin}
\begin{split}
\sup_{y\in \bR^3} \Big|\langle \psi_{N,t}, &\left(\delta_y - \nu_{N,y} \right) (x_1 -x_2)
\wt \psi_{N,t} \rangle \Big|
 \\ &\leq C e^{-\frac{N^{\kappa}}{8}} \langle \psi_{N,t}, (1 -\Delta_1)(1- \Delta_2)
\psi_{N,t} \rangle^{1/2}
\langle \wt \psi_{N,t}, (1-\Delta_1)(1- \Delta_2) \wt\psi_{N,t} \rangle^{1/2} \,.\end{split}
\end{equation}
In fact, (\ref{eq:poin}) can be proven using the Fourier representation
\begin{equation}
\begin{split}
\langle \psi_{N,t}, (\delta_y - &\nu_{N,y}) (x_1 -x_2) \wt \psi_{N,t}\rangle \\ =
\; \int &\rd \bp_{N-2} \, \rd p_1 \rd p_2
\rd q_1 \rd q_2 \; \delta (p_1 + p_2 - q_1 -q_2) e^{i \, y\cdot (p_1 - q_1)}
\left( 1 - e^{- N^{-2} e^{-N^{\kappa}} \, (p_1 - q_1)^2/4} \right) \\ &\times
\frac{\sqrt{(p_1^2 + 1)(p_2^2 + 1)}}{\sqrt{(q_1^2 + 1) (q_2^2 + 1)}} \,
\overline{\widehat{\psi}}_{N,t} (p_1,p_2, \bp_{N-2})
\frac{\sqrt{(q_1^2 + 1) (q_2^2 + 1)}}{\sqrt{(p_1^2 + 1)(p_2^2 + 1)}} \,
\widehat{\wt\psi}_{N,t} (q_1, q_2, \bp_{N-2}) \,
\end{split}
\end{equation}
with $\bp_{N-2} = (p_3, \dots ,p_N)$. Using that $|1-e^{-a}|\leq C a^{1/4}$ for $a >0$, applying a Schwarz inequality, and changing variables $q \to p$ in $\widehat{\wt \psi}_{N,t}$, we obtain
\begin{equation}
\begin{split}
\sup_{y\in \bR^3}\Big| \langle \psi_{N,t}, (\delta_y - \nu_{N,y}) &(x_1 - x_2)
\wt \psi_{N,t} \rangle \Big| \\ \leq \;
C e^{-\frac{N^{\kappa}}{8}} \int &\rd \bp_{N-2} \rd p_1 \rd p_2 \rd q_1 \rd q_2 \;
 \delta (p_1 + p_2 - q_1 -q_2) \frac{|q_1|^{1/2}+1}{(q_1^2 + 1) (q_2^2 + 1)} \\
 &\times (p_1^2 + 1)(p_2^2 + 1)
\left(\beta |\widehat{\psi}_{N,t} (p_1,p_2, \bp_{N-2})|^2 + \beta^{-1}
|\widehat{\wt \psi}_{N,t} (p_1,p_2,\bp_{N-2})|^2 \right)
\end{split}
\end{equation}
for every $\beta >0$. This implies (\ref{eq:poin}) because
\be \label{eq:trivv}
 \sup_{p \in \bR^3} \int \rd q \, \frac{|q|^{1/2}+1}{(1+q^2)(1+(q-p)^2)} < \infty \,.
\ee

To bound the expectations of $(1-\Delta_1)(1-\Delta_2)$ on the r.h.s. of (\ref{eq:poin})
we observe that, as an operator inequality on $L^2_s (\bR^{3N})$, we have
\begin{equation}\label{eq:HN2bd}
\begin{split}
(H_N + N)^2 = \; &\left( \sum_{j=1}^N (1-\Delta_j) + \sum_{i<j} V_N (x_i -x_j) \right)^2  \\
 \geq \; &\frac{1}{2} \left( \sum_{j=1}^N (1-\Delta_j) \right)^2 - 2 \left( \sum_{i<j}
V_N (x_i -x_j) \right)^2 \\ \geq \; & N(N-1) (1-\Delta_1)(1-\Delta_2) - N^8 \| V \|_{\infty}
\end{split}
\end{equation}
which, by (\ref{eq:psiN2}), implies that
\begin{equation}\label{eq:HN2bd2}
\begin{split}
\langle \psi_{N,t}, (1 -\Delta_1)(1- \Delta_2) \psi_{N,t} \rangle \leq \; &
\left\langle \psi_{N,t}, \left( \frac{(H_N+N)^2+ C N^8}{N(N-1)}  \right) \psi_{N,t}
\right\rangle \\ = \; & \left\langle \psi_{N}, \left( \frac{(H_N+N)^2+ C N^8}{N(N-1)}  \right)
\psi_{N} \right\rangle \leq C N^6\,.
\end{split}
\end{equation}
Using (\ref{eq:psiN2}) and a Schwarz inequality similar to (\ref{eq:HN2bd}) to compare $H_N^2$ and $\wt H_N^2$, it is simple to check that
\[ \langle \psi_N, \wt H_N^2 \psi_N \rangle \leq C N^8 \] and therefore, proceeding
analogously to (\ref{eq:HN2bd}) and (\ref{eq:HN2bd2}), we also obtain that
\begin{equation}\label{eq:HN2bd3} \langle \wt \psi_{N,t}, (1-\Delta_1)(1- \Delta_2)
\wt\psi_{N,t} \rangle \leq C N^6 \,.\end{equation}
Inserting (\ref{eq:HN2bd2}) and (\ref{eq:HN2bd3}) into (\ref{eq:poin}), and using
 (\ref{eq:diffe}), we find
\[ \Big| \frac{\rd}{\rd t} \Big\| \psi_{N,t} - \wt \psi_{N,t} \|^2 \Big| \leq C e^{- c N^{\kappa}} \]
 for some $c >0$, which implies that
\begin{equation}
\Big\| \psi_{N,t} - \wt \psi_{N,t} \Big\| \leq C t^{1/2} e^{-c N^{\kappa}} \to 0
\end{equation}
as $N \to \infty$, for every fixed $t \in \bR$. This completes the proof of (\ref{eq:clapp2}).

\end{document}